\documentclass[11pt,letterpaper,a4paper]{article}
\pdfoutput=1
\usepackage{jheppub}
\usepackage{xspace}
\usepackage{multirow}

\newcommand{\SVirtual}{\mbox{SVirtual }}

\newcommand{\mrm}[1]{\mathrm{#1}}

\newcommand{\ttt}[1]{\texttt{#1}}

\newcommand{\pT}[1][]{\ensuremath{p_{\perp #1}}\xspace}
\newcommand{\TeV}{\,\mbox{Te\kern-0.2exV}}
\newcommand{\GeV}{\,\mbox{Ge\kern-0.2exV}}
\newcommand{\MeV}{\,\mbox{Me\kern-0.2exV}}
\newcommand{\keV}{\,\mbox{ke\kern-0.2exV}}
\newcommand{\eV}{\,\mbox{e\kern-0.2exV}}
\newcommand{\qbar}{\ensuremath{\bar{q}}}      
\renewcommand{\d}[1]{\ensuremath{\mrm{d}#1}}
\newcommand{\dA}{\delta\hspace*{-0.12ex}A}

\newcommand{\smax}{\ensuremath{s_{\max}}}

\newcommand{\gq}{\ensuremath{g\bar{q}}}
\newcommand{\qg}{\ensuremath{qg}}
\newcommand{\sgq}{\ensuremath{s_{g\bar{q}}}}
\newcommand{\sqg}{\ensuremath{s_{qg}}}
\newcommand{\zmax}{\ensuremath{z_{\max}}}
\newcommand{\zmin}{\ensuremath{z_{\min}}}
\newcommand{\ximin}{\ensuremath{\xi_{\min}}\xspace}

\newcommand{\Li}[1][2]{\ensuremath{\mrm{Li}_{#1}}}

\newcommand{\PS}{\ensuremath{\Phi}}
\newcommand{\dPS}[1]{\d{\PS_{#1}}}

\def\Re{\mathop{\rm Re}\nolimits}

\newcommand{\eps}{\epsilon}

\newcommand{\appRef}[1]{appendix~\ref{#1}\xspace}

\newcommand{\eqRef}[1]{eq.~\eqref{#1}\xspace}

\newcommand{\eqsRef}[1]{eqs.~\eqref{#1}\xspace}

\newcommand{\secRef}[1]{section~\ref{#1}\xspace}

\newcommand{\tabRef}[1]{tab.~\ref{#1}\xspace}
\newcommand{\TabRef}[1]{Tab.~\ref{#1}\xspace}
\newcommand{\figRef}[2][]{figure~\ref{#2}#1\xspace}

\newcommand{\cname}[1]{\textsc{#1}}

\newcommand{\Ar}{\cname{ariadne}\xspace}

\newcommand{\Fj}{\cname{fastjet}\xspace}
\newcommand{\Fw}{\cname{mc@nlo}\xspace}
\newcommand{\Hw}{\cname{herwig}\xspace}
\newcommand{\Hp}{\cname{herwig++}\xspace}
\newcommand{\Js}{\cname{jetset}\xspace}

\newcommand{\Pw}{\cname{powheg}\xspace}
\newcommand{\Py}{\cname{pythia}\xspace}
\newcommand{\Pp}{\cname{pythia~8}\xspace}

\newcommand{\Vc}{\cname{vincia}\xspace}
\newcommand{\Vr}{\cname{vinciaroot}\xspace}

\definecolor{lgreen}{rgb}{0.1,1.0,0.1}
\definecolor{lygreen}{rgb}{0.65,0.97,0.0}
\definecolor{yell}{rgb}{0.95,0.95,0.0}
\newcommand{\gbox}[1]{\hspace*{-1.7mm}\colorbox{lgreen}{#1}\hspace*{-1.7mm}}
\newcommand{\gybox}[1]{\hspace*{-1.7mm}\colorbox{lgreen}{#1}\hspace*{-1.7mm}}
\newcommand{\ybox}[1]{\hspace*{-1.7mm}\colorbox{yell}{#1}\hspace*{-1.7mm}}

\title{Antenna Showers with One-Loop Matrix Elements}
\author[a]{L.~Hartgring,}
\author[a,b,c]{E.~Laenen,}
\author[d]{and P.~Skands}
 
\affiliation[a]{Nikhef Theory Group, Science Park 105, Amsterdam, The Netherlands}
\affiliation[b]{ITFA, University of Amsterdam, Science Park 904, Amsterdam, The Netherlands}
\affiliation[c]{ITF, Utrecht University, Leuvenlaan 4, Utrecht, The Netherlands}
\affiliation[d]{Theoretical Physics CERN,\\ CH-1211 Geneva 23, Switzerland}

\emailAdd{lhartgri@nikhef.nl}
\emailAdd{Eric.Laenen@nikhef.nl}
\emailAdd{peter.skands@cern.ch}

\abstract{We consider the probability for a colour-singlet
$q\bar{q}$ pair to emit a gluon, in strongly and smoothly ordered
antenna showers. We expand to second order in
$\alpha_s$ and compare to the second-order QCD 
matrix elements for $Z \to 3$ jets, neglecting terms suppressed by $1/N_C^2$.
We give a prescription that corrects the shower to the
matrix-element result at this order for both soft and hard emissions,
thereby explicitly reducing its dependence on evolution- and
renormalization-scale choices. We confirm that the choice
of $p_\perp$ for both of these scales absorbs all logarithms through ${\cal
  O}(\alpha^2_s)$, and contrast this with various
alternatives. 
We include these corrections in the \Vc shower generator and study the
impact on LEP event-shape and fragmentation
observables. An uncertainty estimate is provided for each event, in
the form of a vector of alternative weights.  }


\preprint{CERN-PH-TH/2013-038\\NIKHEF-2013-004\\ITF-UU-13-02}

\arxivnumber{1303.4974}

\begin{document}

\maketitle
\flushbottom

\section{Introduction}
\label{sec:introduction}

Modern QCD descriptions of hard-scattering events at particle colliders can
be roughly divided into two broad categories. In the first, 
fixed-order descriptions, matrix elements are computed for all allowed initial
states with a given final state, $F$, plus a limited number
of additional partons. The leading-order (LO) description has 
the minimal number (often zero) of additional partons. For improved
accuracy, one includes matrix elements with one extra parton beyond
leading order and one loop correction (next-to-leading order) and so
forth. The squared 
matrix elements are numerically integrated over the allowed phase
space, after accounting for any divergences. Given the accuracy,
i.e.\ order of the description, the possible number of additional final-state
particles is in essence predetermined, and can take one (LO), or 
two (NLO) etc, values. 
In the second, parton-shower descriptions, one also starts from 
matrix elements for the desired hard process, $F$, but additional radiation
is now generated stochastically via a shower algorithm, which is
essentially Markovian. 
This is a unitary process, with
probability one, and therefore does not change the probability of the
underlying hard process to occur. The number of
final-state partons is now not predetermined, and can take an infinity of
different values. 
The two approaches have complementary strengths and weaknesses (for
a review, see e.g.~\cite{Buckley:2011ms}). When hard extra emissions
(e.g., hard jets) are important to model well, one looks to descriptions of
the first category. However, the calculation is then unpredictive for
near-collinear and soft radiation (e.g., jet substructure and soft
wide-angle jets). The obverse holds for the second category. 

Even from this very cursory summary, it is clear that methods to  
unite the two --- combining strengths and eliminating weaknesses ---
are very important. Two  
longstanding and very successful approaches for combining one-loop
matrix elements with parton showers are 
\Fw~\cite{Frixione:2002ik,Frixione:2003ei,Hirschi:2011pa,Frederix:2009yq,Frederix:2011ss}
and \Pw~\cite{Nason:2004rx,Frixione:2007vw,Alioli:2010xd}. 
An important restriction of both of these is that only the
spectra of the basic hard-scattering partons 
are corrected to NLO precision, while those of additional QCD emissions
are not. Removing this limitation, fully or partially, has been the
focus of much recent
effort~\cite{Lavesson:2008ah,Platzer:2011bc,Hoeche:2012yf,Gehrmann:2012yg,Frederix:2012ps,Lonnblad:2012ng,Platzer:2012bs,Alioli:2012fc,Lonnblad:2012ix,Hamilton:2012rf},
and is 
also among the main goals of this paper.
While most other approaches employ parton-shower algorithms which are
based on $1 \rightarrow 2$ splitting kernels, we develop an
approach for matching NLO descriptions to showers based on $2
\rightarrow 3$ splittings~\cite{Gustafson:1987rq}. 
The equivalent of the $1 \rightarrow 2$ splitting
kernels are, for us, dipole-antenna functions
\cite{Gustafson:1987rq,Kosower:1997zr,GehrmannDeRidder:2005cm}. At the
practical level, our approach is in the context of the \Vc framework
\cite{Giele:2007di,Giele:2011cb}. 
Whatever the splitting kernel, the parton-shower approaches rest upon
the factorization  
of both phase space and matrix element when the splitting is either
soft or collinear, or both. 
A technical advantage of our approach is
that the $(n+3)$-particle phase-space factorizes exactly
into a $(n+2)$-particle phase space times a $2\rightarrow 3$ phase
space with all momenta on-shell,  
without need for momentum reshuffling~\cite{Kosower:2003bh}. 
Phase-space factorization and the antenna-based matrix-element
factorization are important to our approach in about equal measure.

The essential bottleneck in such combinations of fixed order and
parton shower is how to avoid double counting both real emissions as well as
virtual effects. A key aspect of this is how well the NLO emissions are 
mimicked by parton-shower emissions. 
Emissions generated by a parton-shower Markov chain in fact produce 
approximations to tree-level matrix elements up to arbitrary
numbers of legs, while the no-emission Sudakov factors 
generate the equivalent all-orders loop corrections\footnote{For an
  introduction to such chains 
and a description of their properties, see e.g.\
\cite{Buckley:2011ms,Skands:2011pf}.}.
This all-orders resummation of contributions is ordered in a
measure of jet resolution, called the 
evolution scale, which we denote $Q_E$. It is typically chosen to be a 
measure of transverse momentum \cite{Gustafson:1987rq} or invariant
mass. Its fundamental role is to separate resolved from unresolved
emissions, in analogy to a jet-clustering measure. The different
evolution variables each have their strengths, depending on the
context. As part of our present study, we judge them by how well their 
fixed-order expansions approximate the NLO matrix elements.

The main purpose of this paper is to define, for $e^+e^-$ initial
states, an antenna-based shower algorithm that incorporates multileg NLO
corrections for both soft and hard emissions, and to study
the quality of the matching for a variety of evolution variables. We strive for
next-to-leading  logarithmic (NLL) accuracy, in a way we shall detail
below.  

The leading-logarithmic (LL) structure of antenna showers was
discussed in \cite{Nagy:2009re,Skands:2009tb}, with 
explicit comparisons of various
algorithms to tree-level ${\cal   O}(\alpha_s^2)$ matrix elements
presented in \cite{Andersson:1991he,Skands:2009tb,Giele:2011cb}. 
A prescription for matching the showers
to reproduce tree-level matrix elements exactly (over all of phase
space) was developed in~\cite{Giele:2011cb}, with uncertainty
variations provided in the form of vectors of alternative weights for
each event. In \cite{LopezVillarejo:2011ap} and
\cite{Larkoski:2013yi}  
substantial speedups of the matching algorithm were obtained by dividing
phase space into so-called sectors, and by deriving a
formalism for using individual helicity amplitudes to correct the
shower evolution, respectively.
To further probe the structure of antenna showers, at the
subleading-logarithmic level, we shall here consider the expansion of
exclusive $2\to 3$ splitting probabilities to ${\cal O}(\alpha_s^2)$,
comparing these to one-loop matrix elements \cite{Ellis:1980wv} and
to corresponding second-order antenna functions 
\cite{GehrmannDeRidder:2004tv,GehrmannDeRidder:2005cm}. 

We shall compare six different types of ordering
criteria for the shower 
evolution: 1) strong ordering in
transverse momentum, 2) strong ordering in dipole virtuality, 
3 \& 4) strong ordering in two variants of emission energy (mostly intended as 
cross-checks), and 5 \& 6) so-called smooth ordering in $p_\perp$ and
in dipole virtuality, as defined by \cite{Giele:2011cb}. 
We also consider several different choices for
the renormalization scale $\mu_R$ used in the tree-level antenna
functions and discuss how to systematically 
absorb contributions proportional to the $\beta$-function by this choice, elaborating on earlier
arguments~\cite{Amati:1980ch,Catani:1990rr}.  

Finally, we will present a prescription for how to systematically
incorporate the second-order (one-loop) $q\qbar\to qg\qbar$ antenna into
the shower evolution, for each of the studied
evolution variable choices. 
This will essentially constitute the NLL accuracy mentioned above.
The  resulting shower algorithm, whose $q\qbar\to qg\qbar$ splitting
probability  
should therefore be correct to ${\cal O}(\alpha_s^2)$ over all of
phase space, has been
implemented in the publicly available \Vc plug-in~\cite{Giele:2007di} 
to the \Pp event generator~\cite{Sjostrand:2007gs}. 

We have organized the paper as follows. In section
\ref{sec:notat-incl-vs} we discuss introductory aspects of (antenna)
shower algorithms, define the various evolution 
variables, and the implementation of an ordering prescription that rules the 
shower evolution. In section \ref{sec:match-antenna-show} we present
our matching 
prescription in detail, initially for 3-parton final states in
$Z$-decay, then generally 
for $n$ partons. In section \ref{sec:integrals} we discuss details of
the Sudakov 
integrals required in the matching prescription and compare the
infrared limits of those integrals to those of the one-loop matrix elements.
In section \ref{sec:results} we combine one-loop and tree-level
corrections in a single algorithm, perform speed benchmarking, and
study the impact on LEP observables, especially in the context of
$\alpha_s(m_Z)$ extractions. We conclude in section
\ref{sec:conclusions} and elaborate on technical aspects in the
appendices.  

\section{Antenna Showers}
\label{sec:notat-incl-vs}

In this section, we recap the basic antenna-shower formalism, as used
in the \Vc shower algorithm. This also serves to introduce the 
basic notation and conventions that will be used in later sections. 

\subsection{The Formal Basis of Antenna Showers}

Antenna showers are based on the factorization of (squared) colour-ordered 
QCD amplitudes  in soft and collinear limits, which can be
expressed as follows 
\begin{eqnarray}
|M(\ldots,p_i,p_j,\ldots)|^2 &\stackrel{\mbox{$i||j$}}{\to}&
g_s^2 \ {\cal C} \ \frac{P(z)}{s_{ij}} \
|M(\ldots,p_i+p_j,\ldots)|^2 \\
|M(\ldots,p_i,p_j,p_k,\ldots)|^2 &\stackrel{\mbox{$j_g$ soft}}{\to}&
g_s^2 \ {\cal C} \ A_{g}(s_{ij},s_{jk},s_{ijk}) \
|M(\ldots,p_i,p_k,\ldots)|^2 ~,
\end{eqnarray}
with $g_s^2=4\pi\alpha_s$ the strong coupling and the subscript $g$ in the
second line emphasizing that the soft limit is only relevant for
gluons. 

In the collinear limit (first line), 
$P(z)$ are the Altarelli-Parisi splitting kernels~\cite{Altarelli:1977zs}, 
$z$ is the energy fraction taken by parton $i$ (with a 
fraction $(1-z)$ going to parton $j$), and $\cal C$ is a colour
factor, which we discuss below. This limit forms the
basis for traditional parton showers, such as those in the \Py
generator~\cite{Bengtsson:1986hr}. 

In the soft-gluon limit (second line), the function $A$ has dimension
$\mrm{GeV}^{-2}$, and is called an antenna function. 
For unpolarized massless partons\footnote{In
the context of massive particles, replace $s_{ab}$ by
$2p_a\cdot p_b$ in all expressions. For a more complete treatment, 
see~\cite{GehrmannDeRidder:2011dm}.}, its leading term is the
so-called eikonal or dipole factor, 
\begin{equation}
A_\mrm{Eik}(s_{ij},s_{jk},s_{ijk}) = \frac{2s_{ik}}{s_{ij}s_{jk}}~, \label{eq:eik}
\end{equation}
where $s_{ik} = s_{ijk} - s_{ij} - s_{jk}$ for massless partons. It
was found early on that this factor can be reproduced by a traditional 
parton shower by imposing the requirement of angular ordering of
subsequent emissions~\cite{Marchesini:1983bm}. 
This gave rise to the angular-ordered
showers \cite{Marchesini:1987cf,Gieseke:2003rz} 
in the \Hw and \Hp generators~\cite{Corcella:2000bw,Bahr:2008pv} 
as well as the imposition of an angular-ordering constraint
\cite{Bengtsson:1986hr,Bengtsson:1986et} in the 
\Js and \Py generators~\cite{Sjostrand:2006za,Sjostrand:2007gs}.   

In fixed-order calculations, dipole \cite{Catani:1996vz} 
and antenna
\cite{Kosower:1997zr,Kosower:2003bh,GehrmannDeRidder:2005cm} 
functions are frequently used to define subtraction
terms. These functions include additional subleading terms, beyond the
eikonal one, which are necessary to correctly describe 
both soft and collinear limits in all regions of phase space. In the
parametrization we shall use, their most general forms, for the
branching process $IK \to ijk$, are
\begin{eqnarray}
A_{\mrm{Emit}}(s_{ij},s_{jk},m_{IK}^2) & = &\frac{1}{m_{IK}^2}\left(
\frac{2y_{ik}}{y_{ij}y_{jk}}  
 + \frac{y_{jk}(1-y_{jk})^{\delta_{ig}}}{y_{ij}} 
 + \frac{y_{ij}(1-y_{ij})^{\delta_{kg}}}{y_{jk}} 
 + F_{\mrm{Emit}} \right) \label{eq:aEmit}\\
A_{\mrm{Split}}(s_{ij},s_{jk},m^2_{IK}) & = &\frac{1}{m_{IK}^2}\left(
\frac{y_{jk}^2+y_{ik}^2}{2y_{ij}}
+ F_{\mrm{Split}}\right) ~,\label{eq:aSplit}
\end{eqnarray}
for gluon-emission and gluon-splitting processes, respectively, 
with the parent antenna invariant mass, $m_{IK}^2 =
(p_I+p_K)^2 = (p_i+p_j+p_k)^2$ and 
the scaled invariants, 
\begin{eqnarray}\label{eq:20}
y_{ij} = \frac{s_{ij}}{m^2_{IK}} & ; & y_{jk} =
\frac{s_{jk}}{m^2_{IK}}~, \label{eq:yDef}
\end{eqnarray}
and we use the notation $\delta_{ig} = 1$ if parton $i$ is a gluon and zero
otherwise. The functions $F_\mrm{Emit}$ and $F_\mrm{Split}$ 
allow for the presence of non-singular
terms, which are in principle arbitrary. 
A logical choice would be $F=0$, but this would not be
invariant under reparametrizations of the antenna functions across the
gluon-collinear singular limits~\cite{Giele:2011cb}. Since the $F$
functions can anyway be made useful in the context of uncertainty 
estimates~\cite{Giele:2007di,Giele:2011cb}, 
we therefore leave them as functions whose forms we are free to 
choose.

In the soft-gluon limit, the eikonal factor is reproduced by the first term
in in \eqRef{eq:aEmit}. In the collinear $q\to qg$ limit, the AP
splitting kernel also is reproduced. For collinear $g\to gg$ and $g\to
q\bar{q}$ branchings, one must sum over the contributions from 
two neigbouring antennae, which together reproduce the AP splitting
kernel. Limits that are both soft and collinear are also correctly
reproduced \cite{GehrmannDeRidder:2005cm}. 

In the antenna context, 
the colour factors are $2C_F$ for $q\bar{q} \to
qg\bar{q}$, $C_A$ for $gg\to ggg$\footnote{Note that in a process like $H^0\to
gg$, there are two $gg$ antennae at the Born level,  
and hence the antenna approximation to $H^0\to ggg$ is twice as large
as the single $gg\to ggg$ antenna. Likewise, the collinear limit of a
gluon is obtained by summing over the contributions from both of the
dipoles/antennae it is connected to. One must also include a 
sum over permutations of the final-state gluons, if comparing to a 
summed matrix-element expression.}, and $2T_R$ for gluon splitting to
$q\bar{q}$, again using the normalization convention adopted in
\cite{Giele:2011cb}. However, for $qg\to qgg$ there is a genuine subleading
ambiguity whether to prefer, say, $2C_F$, $C_A$, or something interpolating
between them~\cite{Gustafson:1992uh}. At fixed order, 
the question of subleading colour can in fact be dealt with quite
elegantly, by using $C_A$ for all antennae and then including an
additional $q\bar{q}$ antenna with a negative colour factor,
$-C_A/N_C^2$, spanned between the two endpoint quarks, for each 
$qg\ldots \bar{q}$ chain~\cite{Berends:1988zn}. In the
context of an 
antenna-based shower, however,
it is desirable to use only positive-definite antenna functions,
and a prescription for absorbing the negative one into the
positive ones was given in~\cite{Giele:2011cb}. In the context of this work,
however, we shall largely ignore subleading-colour aspects and, unless
explicitly stated otherwise, assign a colour factor $C_A$
to the $qg\to qgg$ antenna function, thereby overcounting the
collinear limit in the quark direction by a factor $C_A/(2C_F) \simeq
1+1/N_C^2$.

The renormalization scale used to evaluate the strong coupling in the
antenna function, $g_s^2 = 4\pi \alpha_s(\mu_{\mrm{PS}})$, is typically
chosen proportional to $p_\perp$ (following
\cite{Amati:1980ch}). As alternatives, we shall also consider 
 $\mu^2_{\mrm{PS}}\propto m^2_D =
2\mrm{min}(s_{ij},s_{jk})$, and, as an extreme case which connects with
fixed-order 
calculations, the invariant mass of the antenna,
$\mu_\mrm{PS}^2 \propto m_{IK}^2$.

A final aspect concerns the phase-space factorization away from the
collinear limit. Within the framework of collinear factorization (and
hence, in traditional parton showers), the momentum fraction, $z$, is
only uniquely defined in the exactly collinear limit; outside that
limit, the choice of $z$ is not unique. In addition, 
a prescription must be adopted for ensuring overall momentum conservation, 
leading to the well-known ambiguities concerning recoil strategies
(see e.g. \cite{Buckley:2011ms}). In antenna showers, on the other hand,
the antenna function is defined in terms of the unique branching
invariants, $s_{ij}$ and $s_{jk}$, over all of phase space, and the
phase space itself has an exact Lorentz-invariant and
momentum-conserving factorization, 
\begin{equation}
\dPS{n} = \dPS{n-1} \ \times \ \dPS{\mrm{Ant}}~, \label{eq:PSfac}
\end{equation}
with 
\begin{equation}
\dPS{\mrm{Ant}} \ = \ \frac{1}{16\pi^2 m_{IK}^2} \ \d{s_{ij}}
\d{s_{jk}} \frac{\d{\phi}}{2\pi}\label{eq:PhiAnt} 
\end{equation}
for massless partons (for massive ones, see
\cite{GehrmannDeRidder:2011dm}), with the  
$\phi$ angle parametrizing rotations around the antenna axis, in
the CM of the antenna. Note the equality signs; no approximation is
involved in this step. The only remaining phase-space ambiguity,
outside the singular limits, is present when
specifying how the post-branching momenta are related to the
pre-branching ones. This is defined by a kinematics map, the antenna
equivalent of a recoil strategy, which we here take to be of the class
defined by~\cite{Kosower:1997zr,Giele:2007di}.

\subsection{Constructing a Shower Algorithm \label{sec:shower}}

In a shower context, the amplitude and phase-space factorizations
above imply that we can interpret the radiation functions (AP splitting
kernels or dipole/antenna functions)
as the probability for a radiator (parton or dipole/antenna) 
to undergo a branching, per unit
phase-space volume,  
\begin{equation}
\frac{\d{P(\Phi)}}{\d{\Phi}} \ = \ g_s^2 \ {\cal C} \ A(\Phi) ~,
\end{equation}
where we use $\Phi$ as shorthand to denote a phase-space point. (If
there are several partons/dipoles/antennae, the total probability for
branching of the event as a whole is obtained as a sum of such terms.)  

An equally fundamental object in both analytical resummations and in parton
showers is the Sudakov form factor, which defines the probability for
a radiator \emph{not} to emit anything, as a function of the shower 
evolution parameter, $Q$ (i.e., similarly to a jet veto, with
$Q$ playing the role of the jet clustering scale; we return to the
choice of functional form for the shower evolution scale 
in \secRef{sec:ordering}). 
In the all-orders shower construction, these factors 
generate the sum over virtual amplitudes plus unresolved real
radiation, and hence their first-order expansions 
play a crucial role in matching to next-to-leading order matrix
elements. 
We here recap some basic properties.
The Sudakov factor, giving the no-emission probability between two
values of the shower evolution parameter, $Q_1$ and $Q_2$ (with $Q_1 >
Q_2$), is defined by 
\begin{equation}
\Delta(Q^2_1,Q^2_2) \ = \ \exp\left(-
\int_{Q_2^2}^{Q_1^2} \frac{\d{P(\Phi)}}{\d{\Phi}} \ \dPS{}
\right) \ = \ 
\exp\left(-\int_{Q_2^2}^{Q_1^2} g_s^2 \ {\cal C} \ A(\Phi) \ \dPS{}\right)~,
\label{eq:I}
\end{equation}
where it is understood that the integral boundaries must be
imposed either as step functions on the integrand or by a suitable
transformation of integration variables, accompanied by Jacobian
factors.  

This description has a very close analogue in the simple process of 
nuclear decay, in which the probability for a nucleus to undergo a
decay, per unit time, is given by the nuclear decay constant, 
\begin{equation}
\frac{\d{P(t)}}{\d{t}} \ = \ c_N~.
\end{equation}
The probability for a nucleus existing at time $t_1$ to remain
undecayed before time $t_2$, is 
\begin{equation}
\Delta(t_2,t_1) = \exp\left(-\int_{t_1}^{t_2} c_N \ \d{t} \right)~=~
\exp\left(-c_N \ \Delta t\right)~.
\end{equation}
This case is especially simple,
since the decay probability per unit time, $c_N$, is constant. By
conservation of the total number of nuclei (unitarity), the activity
per nucleon at time $t$, equivalent to the ``resummed'' decay
probability per unit time, is minus the derivative of $\Delta$,
\begin{equation}
\frac{\d{P_{\mrm{res}}(t)}}{\d{t}} \ = \ -\frac{\d{\Delta}}{\d{t}} \ = \ c_N
\ \Delta(t,t_1)~.
\end{equation} 

In QCD, the emission probability varies over phase space, hence the
probability for an antenna not to emit has the more elaborate integral
form of \eqRef{eq:I}. By unitarity, the resummed branching probability
is again minus the derivative of the Sudakov factor, 
\begin{equation}
\frac{\d{P_{\mrm{res}}(\Phi)}}{\d{\Phi}} \ = \ g_s^2 \ {\cal C}
\ A(\Phi) \ \Delta(Q_1^2, Q^2(\Phi))~, \label{eq:Pres}
\end{equation}
with $Q^2(\Phi)$ the shower evolution scale (typically chosen as a
measure of invariant mass or transverse momentum, see 
\secRef{sec:ordering}), evaluated at the phase-space point $\Phi$. 
 
In shower algorithms, branchings are generated with this distribution,
starting from a uniformly distributed random number ${\cal R}\in[0,1]$, by 
solving the equation,
\begin{equation}
{\cal R} = \Delta(Q_1^2,Q^2) \label{eq:R}
\end{equation}
for $Q^2$. 
For an initial distribution of ``trial''
branching scales, we do not use the full antenna function,
\eqRef{eq:aEmit}, as the evolution kernel, but only its
leading singularity, 
\begin{equation}
A_{T} \ = \ \frac{2m_{IK}^2}{s_{ij}s_{jk}} \ = \ \frac{2}{p_{\perp
    A}^2}~, \label{eq:aTrial}
\end{equation}
where $p_{\perp A}$ is the \Ar definition of transverse 
momentum~\cite{Lonnblad:1992tz}, which is also the one used in \Vc.
This reflects the universal $1/p_\perp^2$ behaviour of soft-gluon emissions. 
In addition to the trial scale, $Q$, two complementary 
phase-space variables are also generated (which we usually label $\zeta$ and
$\phi$~\cite{Giele:2011cb}),  
according to the shape of $A_T$ over a phase-space contour of 
constant $Q$. From these, the model-independent 
set of trial phase-space variables ($s_{ij}$,
$s_{jk}$, $\phi$) are determined by inversion of the defining
relations $Q(s_{ij},s_{jk})$ and $\zeta(s_{ij},s_{jk})$, and 
the full kinematics (i.e., four-momenta) 
of the trial branching can then be constructed~\cite{Giele:2007di}. 

To decide whether to accept the trial or not, we note that 
the function $A_T$  differs from the eikonal in
\eqRef{eq:eik} by the replacement of $s_{ik}$  in the numerator by
$m^2_{IK}$. By accepting the trial scales generated by $A_T$
with the probability 
\begin{equation}
P_\mrm{eik}~=~\frac{A_\mrm{eik}}{A_T}~=~\frac{s_{ik}}{m_{IK}^2}~\le~1~,
\end{equation}
the eikonal approximation can be recovered, by virtue of the veto
algorithm~\cite{Sjostrand:1985xi,Sjostrand:1987su,Buckley:2011ms}. 
Crucially, any 
other function that has the eikonal as its soft-collinear limit could
equally well be imprinted on the trial distribution by a similar veto. Two
particularly relevant choices are the full physical antenna function,
\eqRef{eq:aEmit} (which includes additional collinear-singular terms
in addition to the eikonal)
and the GKS-corrected antenna
function (which also incorporates a multiplicative factor that
represents tree-level matching in \Vc), 
\begin{eqnarray}
P^\mrm{LL}_\mrm{accept} & = & \frac{A_\mrm{Emit}}{A_T}~,\label{eq:PLL}\\[2mm]
P^\mrm{LO}_\mrm{accept} & = & P^\mrm{LL}_\mrm{accept}\, R_n~, \label{eq:PLO}
\end{eqnarray}
with $A_\mrm{Emit}$ given in \eqRef{eq:aEmit}
and $R_n$ the $n$-parton tree-level GKS matching
factor~\cite{Giele:2011cb}, to which we return in
\secRef{sec:LO}. 

Note that, for gluon-splitting antenna functions
($Xg\to Xq\bar{q}$), we use $Q=m_{q\bar{q}}$, with 
a trial function $\propto 1/m_{q\bar{q}}^2$, 
and again implement the physical antenna function, \eqRef{eq:aSplit},
and LO matching corrections by vetos. 
We also include the so-called \Ar
factor, $P_\mrm{Ari}$, defined by
\begin{equation}
A_\mrm{Split}~\to~P_\mrm{Ari} \, A_\mrm{Split}~=~\frac{2s_N}{s_P+s_N}\,A_\mrm{Split}~,\label{eq:Ariadne}
\end{equation}
with $s_N$ the invariant mass squared of the colour neighbour on the other side
of the splitting gluon and $s_P=m_{IK}^2$ the invariant mass squared of
the parent (splitting) antenna. This does not modify the singular
behavior (as will be elaborated upon below), and was shown to give
significantly better agreement with the $Z\to qq\bar{q}\bar{q}$ matrix
element in~\cite{LopezVillarejo:2011ap}.

Explicit solutions to \eqRef{eq:R} using the trial function defined by
\eqRef{eq:aTrial} were presented in \cite{Giele:2011cb}, for fixed and
first-order running couplings. In the context of the present
work, two-loop running has been implemented using a simple numerical
trick, as follows: given a value of $\alpha_s(M_Z)$, we determine the
corresponding two-loop value of
$\Lambda^\mrm{2-loop}_\mrm{QCD}$. We then use that $\Lambda$ value 
in the one-loop solutions in~\cite{Giele:2011cb}, and correct the
resulting distribution by inserting an additional trial accept veto: 
\begin{equation}
P_\mrm{accept}^\mrm{2-loop} \ =
\ \frac{\alpha^\mrm{2-loop}_s(Q,\Lambda^\mrm{2-loop}_\mrm{QCD})} 
{\alpha^\mrm{1-loop}_s(Q,\Lambda^\mrm{2-loop}_\mrm{QCD})}~.
\end{equation}
Due to the faster pace of 2-loop running,
$\alpha_s^\mrm{2-loop}(Q,\Lambda) < \alpha_s^\mrm{1-loop}(Q,\Lambda)$,
hence this accept probability is guaranteed to be smaller than or
equal to unity.  

A final point concerns if 
there are several ``competing'' radiators (equivalent to several
competing nuclei, and/or several competing available decay channels
for each nucleus). In this case, the trial with the highest value of $Q$ is
selected (corresponding to the one happening at the earliest time,
$t$), and consideration of any other branchings (decays) are postponed
temporarily. 
After a branching, any partons involved in that branching
are replaced by the post-branching ones, and any postponed trial
branchings involving those partons are deleted. The evolution is then
restarted, from the scale $Q$ of the new configuration, 
until there are no radiators left with trial branching scales larger than a
fixed, lower, cutoff, normally identified with the hadronization
scale, $Q_\mrm{had} \sim 1\,\mrm{GeV}$.  

\subsection{Evolution and Ordering \label{sec:ordering}}

In order to solve \eqRef{eq:R} we need to specify the form of
\eqRef{eq:I}, which takes us 
from one scale $Q_1^2$ to a lower scale $Q_2^2$. We change
variables to parametrize the integral by the ordering variable, $Q$, and
another, complementary (but otherwise arbitrary), phase-space variable
which we denote by $\zeta$. The generic  
evolution integral now reads
\begin{align}
\mathcal{A}\left(Q_1^2,Q_2^2 \right) = \int_{Q_2^2}^{Q_1^2} g_s^2\, \mathcal{C}\, \d Q^2 \d\zeta \,|J| \,A(Q^2,\zeta)
\label{eq:AntPS}
\end{align}   
with $|J|$ denoting the Jacobian of this transformation. For branchings involving gluon emission, we consider 
three possible choices for the ordering
variable: 
 dipole virtuality $m_D$, transverse momentum, 
and the energy of the emitted parton, $E_j^*$
 (in the CM of the parent antenna), with the
following definitions, 
\begin{align}
Q^2_{E1} \ &= m_D^2 &\hspace*{-1cm} &=  2 m^2_{IK}
\min(y_{ij},y_{jk})~, &\hspace*{-1cm} &  \\[2mm]
Q^2_{E2} \ &= 4\pT^2 & \hspace*{-1cm}&=  4 m^2_{IK} \, y_{ij} y_{jk}~, &\hspace*{-1cm} & \\[2mm]
Q^2_{E3} \ &= 4E_j^{*2}& \hspace*{-1cm}&=  m^2_{IK} \,
(y_{ij}+y_{jk})^2 &\hspace*{-1cm} & =
\ x_j^2 \, m^2_{IK}~,
\end{align}
with the energy fraction $x_j = 2E^*_j/m_{IK}$. 

\begin{figure}[t!]
\centering
\begin{tabular}{cccc}
& \ Dipole Virtuality-Ordering & \ $\pT$-ordering & \ Energy-Ordering\\
& \ $(m_\mrm{min}^2)$ 
& \ $(\left<m^2\right>_{\mrm{geometric}})$ 
& \ $(\left<m^2\right>_{\mrm{arithmetic}})$\\
\rotatebox{90}{\hspace*{1.5cm}
  Linear in $y$} 
& 
\includegraphics*[scale=0.48]{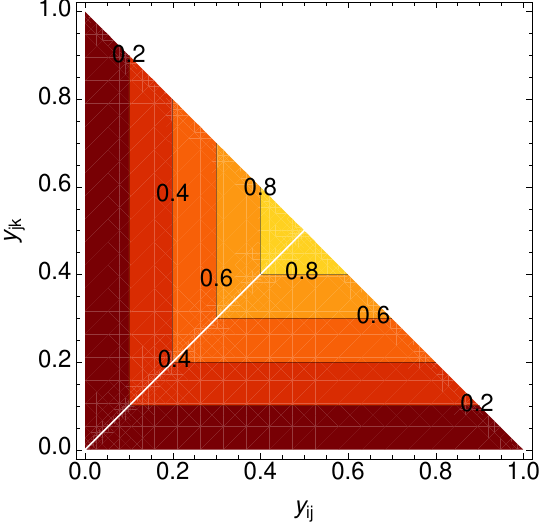}\label{fig:mDlin}
& 
\includegraphics*[scale=0.48]{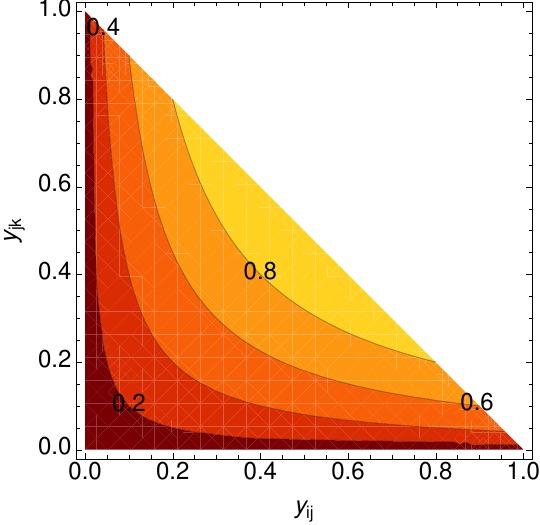}\label{fig:pTlin}
&
\includegraphics*[scale=0.48]{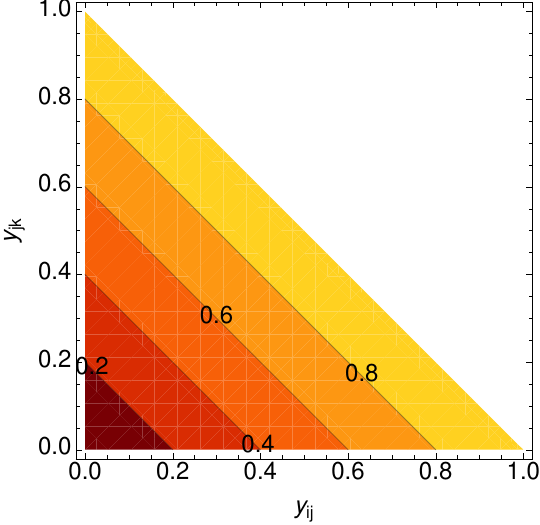}\label{fig:Enlin}
\\[-1mm] $Q_E^2:$ 
& (a) $m_D^2=2\min(y_{ij},y_{jk})s$
&
 (b) $2\pT\sqrt{s}=2\sqrt{y_{ij}y_{jk}}s$
&
(c) $2E^*\sqrt{s}=(y_{ij}+y_{jk})s$
\end{tabular} \\[5mm]
\begin{tabular}{cccc}
\centering
\rotatebox{90}{\hspace*{1.5cm} Quadratic in $y$}
&
\includegraphics*[scale=0.48]{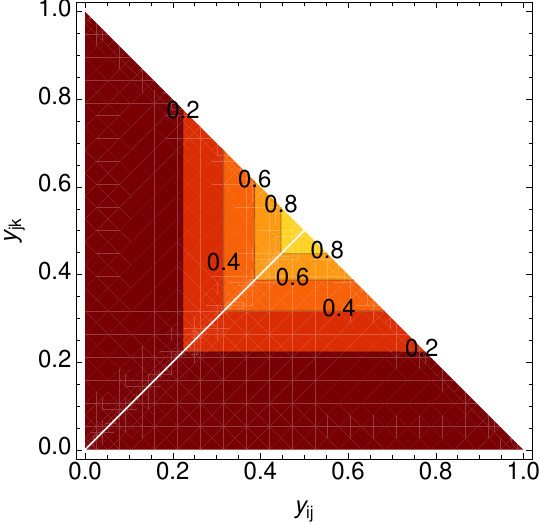}\label{fig:mDsp}
&
\includegraphics*[scale=0.48]{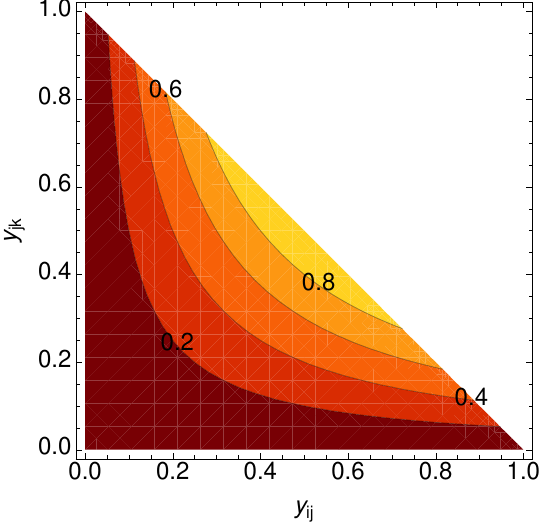}\label{fig:pTsq}
&
\includegraphics*[scale=0.48]{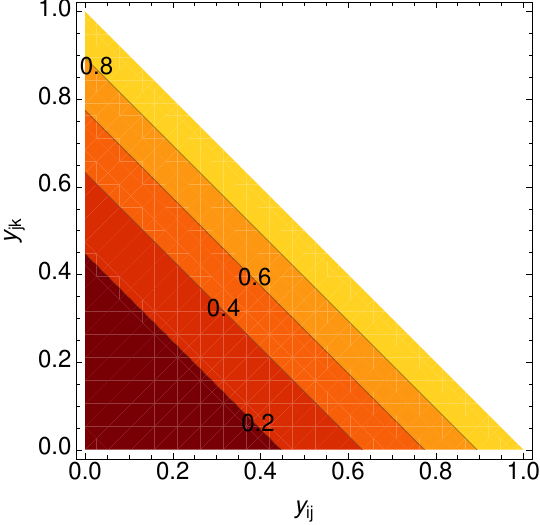}\label{fig:Ensq}\\[-1mm]
$Q_E^2:$
& (d) $\frac{m_D^4}{s}=4\min(y_{ij}^2,y^2_{jk})s$\hspace{2ex}
& (e) $4\pT^2=4y_{ij}y_{jk}s$
& (f) $4E^{*2}=(y_{ij}+y_{jk})^2s$
\end{tabular}\\
\caption{Contours of constant value of $y_E = Q_E^2/m_{IK}^2$ for evolution
  variables  
  linear ({\sl top}) and quadratic ({\sl bottom}) in the branching invariants, 
  for virtuality-ordering ({\sl left}), \pT-ordering ({\sl middle}), and
  energy-ordering ({\sl right}). Note that the energy-ordering variables 
 intersect 
  the phase-space boundaries, where the antenna functions are
  singular, for finite values of the evolution variable. They can
  therefore only be used as evolution variables together with a
  separate infrared regulator, such as a cut in invariant mass, not shown
  here.  \label{fig:ev} }
\end{figure}
All three
options are available as ordering variables in the \Vc shower Monte Carlo.
They are illustrated in \figRef{fig:ev}, where contours of constant 
value of $y_E=Q_E^2/m^2_{IK}$ are shown for each variable, as a function
of $y_{ij}$ and $y_{jk}$. 
For completeness, we show both the case of
a linear (top row) and quadratic (bottom row) dependence on the
branching invariants, for each variable. Since the ordering variable
raised to any positive power will result in the same relative ordering
of emissions within a given antenna, the distinction between linear
and quadratic forms does not affect  individual antenna Sudakov
factors. It does, however, affect  the ``competition'' between
different antennae, and the choice of restart scale for subsequent
evolution after a branching has taken place,  as will be 
discussed further below.

In labeling the columns in
\figRef{fig:ev}, we have also emphasized that
mass-ordering, as defined here, corresponds to choosing the smallest of
the daughter antenna masses as the ``resolution scale'' of the
branching, whereas $\pT$ and energy correspond to using the geometric
and arithmetic means of the daughter invariants,
respectively. Naively, each of these could be taken as a plausible 
measure of the resolution scale of a given phase-space point. We shall
see below which ones lead to better agreement with the one-loop matrix
elements. 

We consider two possible definitions for the complementary phase-space
variable $\zeta$,
\begin{eqnarray} 
\zeta_1 & = & \frac{y_{ij}}{y_{ij}+y_{jk}} \label{def:zeta1} \\
\zeta_2 & = & y_{ij}~. \label{def:zeta2}
\end{eqnarray}
We emphasize that the choice of $\zeta$ has no physical
consequences, it merely serves to reparametrize the Lorentz-invariant
phase space. We may therefore let the choice be governed purely by
convenience, and, for each antenna integral, select 
whichever of the above definitions give the simplest final
expressions.  
The corresponding Jacobian factors, for each of the evolution-variable 
choices we shall consider, are listed in
\tabRef{tab:jacobians}.
\begin{table}[t]
\centering
\begin{tabular}{lcccccccc}
\hline
\multicolumn{1}{c}{$y_E = \frac{Q^2}{s_{ijk}}$} & = & $\frac{m_{jk}^2}{s}$ &
$\frac{m_D^2}{s}$ & $\frac{m_D^4}{s^2}$ 
& $\frac{2\pT}{\sqrt{s}}$ & $\frac{4\pT^2}{s}$ &
$\frac{2E^*}{\sqrt{s}}$ & $\frac{4E^{*2}}{s}$\\
\hline
 $|J(y_E,\zeta_1)|$ & = & $\frac{y_E}{(1-\zeta_1)^2}$ & $\frac{y_E}{4(1-\zeta_1)^2}$ & $\frac{1}{8(1-\zeta_1)^2}$ &
$\frac{y_E}{4\zeta_1(1-\zeta_1)}$ & $\frac{1}{8\zeta_1(1-\zeta_1)}$ & $y_E$ & $\frac12$\\[2mm]
$|J(y_E,\zeta_2)|$ & = & $1$ &
$\frac12$ & $\frac{1}{4\sqrt{y_E}}$ & $\frac{y_E}{2\zeta_2}$ &
$\frac{1}{4\zeta_2}$ & $1$ 
& $\frac{1}{2\sqrt{y_E}}$
\\ 
\end{tabular}\\
Note : $|J(Q^2,\zeta)| = s_{ijk} |J(y_E,\zeta)|$
\caption{Jacobian factors for all combinations of evolution variables
  and $\zeta$ choices. 
\label{tab:jacobians}}
\end{table}

Note that, for the special case of the $m_D^2$ and $m_D^4$ variables,
which contain the non-analytic function $\min(y_{ij},y_{jk})$,
the $\zeta$ definitions in \eqsRef{def:zeta1} and
\eqref{def:zeta2} apply to the branch with $y_{ij}>y_{jk}$. For the
other branch, $y_{ij}$ and $y_{jk}$ should be interchanged. With this
substitution, the Jacobians listed in \tabRef{tab:jacobians} apply to
both branches\footnote{This corresponds to replacing $y_{ij}$ by
$\max(y_{ij},y_{jk})$ in the numerator of \eqRef{def:zeta1} and in 
\eqRef{def:zeta2}.}.

For branchings involving gluon
splitting, $g\to q\bar{q}$, we restrict our attention to two
possibilities, ordering in $\pT$, defined as above, 
and ordering in gluon virtuality,
defined as 
\begin{equation}
Q_{E4}^2 \ = \ m_{g^*}^{2} \ = \ m^2_{q\bar{q}} ~~~~~~\mbox{(for gluon splitting)}~.
\end{equation}
Note that the normalization factors for the ordering variables 
have in all cases been chosen such that the maximum value of the ordering
variable is $m^2_{IK}$.

Since the phase space for subsequent branchings is limited both by the
scale $Q_E$ of the previous branching (according to strong ordering) and by
the invariant mass of the antenna $m_j$, the effective ``restart scale'',
after a branching in a strongly ordered shower, is given by
\begin{equation}
Q^2_{Rj} = \min(Q^2,m^2_{j})~, \label{eq:QR}
\end{equation}
for each antenna $j$. 

Depending on the choice and value of $Q$, one or both daughter
antennae after a splitting may have a non-trivial
restriction on the phase space available for subsequent branching. 
 Conversely, if $Q>m_j$, there is no such restriction. 
 Physically, we distinguish between the case
in which the strong-ordering condition implies a non-trivial
constraint on the evolution of the produced antennae, eating into the
phase-space that would otherwise be accessible, 
and the case in which the strong-ordering condition does
not imply such a constraint.

\begin{figure}[t!]
\centering\scalebox{0.95}{
\begin{tabular}{cccc}
& \ Mass-Ordering & \ $\pT$-ordering & \ Energy-Ordering\\
\rotatebox{90}{\hspace*{1.5cm}
  Linear in $y$} 
& 
\includegraphics*[scale=0.48]{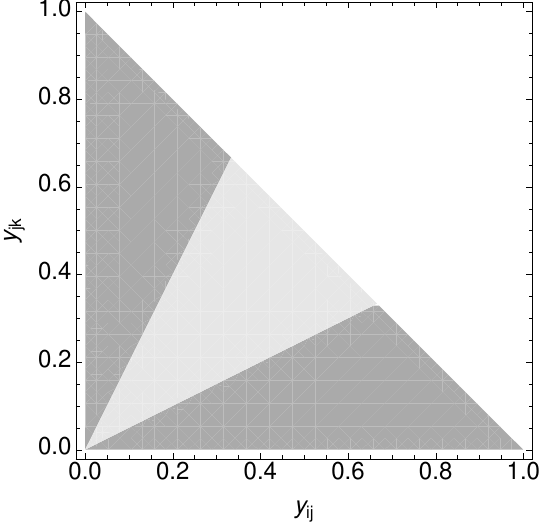}\label{fig:mdlinOrd}
& 
\includegraphics*[scale=0.48]{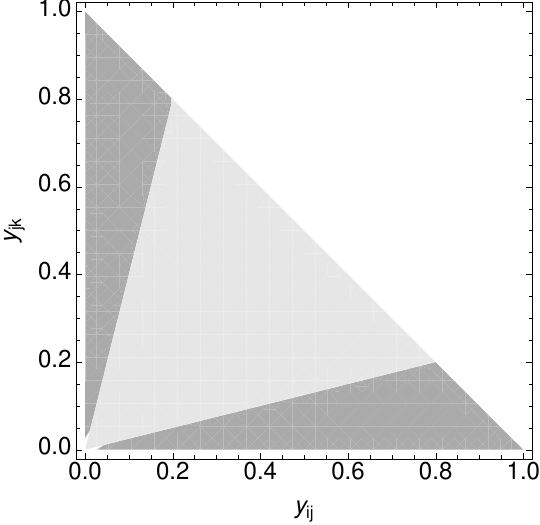}\label{fig:mdsqOrd}
&
\includegraphics*[scale=0.48]{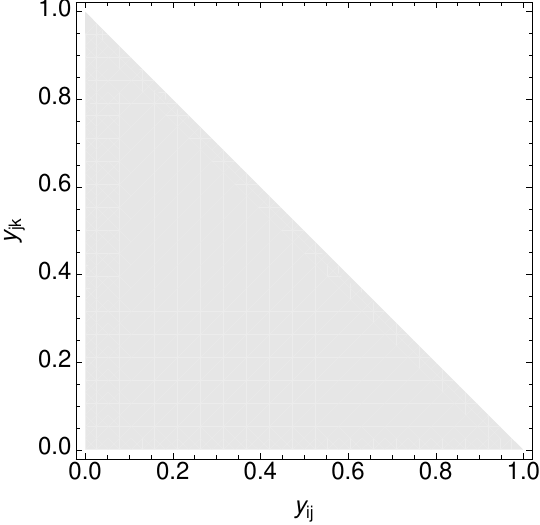}\label{fig:pTlinOrd}
\\[-1mm] $Q_E^2=$
& (a) $m_D^2=2\min(y_{ij},y_{jk})s$
& (b) $2\pT\sqrt{s}=2\sqrt{y_{ij}y_{jk}}s$
& (c) $2E^*\sqrt{s}=(y_{ij}+y_{jk})s$ \\[5mm]
\rotatebox{90}{\hspace*{1cm} Quadratic in $y$}
&
\includegraphics*[scale=0.48]{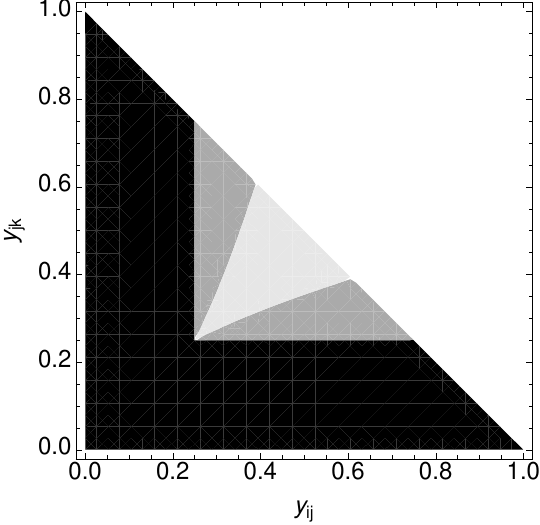}\label{fig:pTsqOrd}
&
\includegraphics*[scale=0.48]{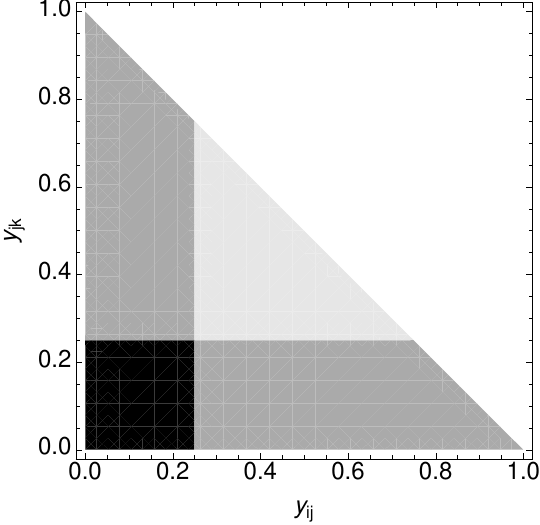}\label{fig:EnlinOrd}
&
\includegraphics*[scale=0.48]{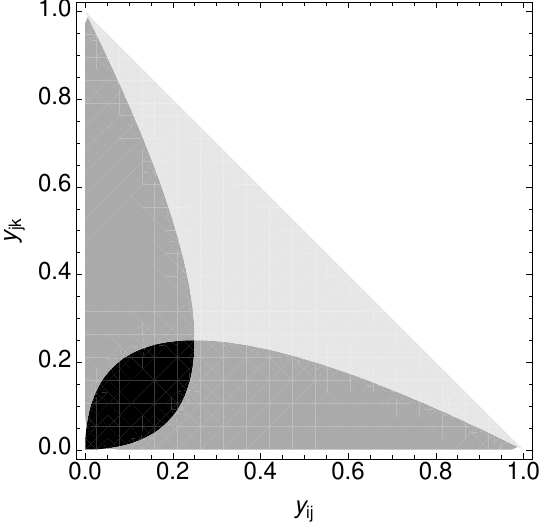}\label{fig:EnsqOrd} \\[-1mm]
$Q_E^2=$
& (d) $\frac{m_D^4}{s}=4\min(y_{ij}^2,y^2_{jk})s$
& (e) $4\pT^2=4y_{ij}y_{jk}s$
& (f) $4E^{*2}=(y_{ij}+y_{jk})^2s$
\end{tabular}}\\
\caption{Illustration of the regions of 3-parton phase space in which
  the subsequent evolution of the $qg$ and $g\bar{q}$ antennae is
  restricted (from above) by the strong-ordering condition. See the
  text for further clarification of this plot. {\sl Black:} both
  antennae restricted. {\sl Dark Gray:} one antenna restricted, the other
  unrestricted. {\sl Light Gray:} both antennae unrestricted. {\sl
    Top/Bottom:} $Q^2$ linear/quadratic in the branching invariants,
  for mass-ordering ({\sl left}), \pT-ordering ({\sl middle}), and
  energy-ordering ({\sl right}).\label{fig:psOrd}}
\end{figure}
The regions of $q\bar{q} \to qg\bar{q}$ phase space in
which either zero, one, or both of the daughter antennae ($qg$ and
$g\bar{q}$ respectively) are constrained by the
ordering condition are illustrated in
\figRef{fig:psOrd}, for each of the choices of evolution variable
under consideration. The black shaded areas correspond to regions in
which both the $qg$ and $g\bar{q}$ antennae are restricted, 
by having $Q < m_j$. The
darker gray shaded areas show regions in which only one of the antennae is
restricted, while the other will still be allowed to evolve over its full
phase space. In the light-gray shaded areas, both
of the antennae are allowed to evolve over all of their available
phase spaces, equivalent to the ordering condition imposing no
constraint on the subsequent evolution. We recall that we are here
discussing the \emph{upper} boundary on the subsequent evolution,
hence the infrared\footnote{Note: we use the word 
  infrared to refer collectively to soft and/or collinear regions of
  phase space.}
 (IR) poles are not affected.

To further clarify the meaning of the plots in \figRef{fig:psOrd}, 
let us discuss panel (e) as an
example. The coordinates, $(y_{ij}, y_{jk})$, represent the 3-parton 
state before it evolves to a 4-parton state, and each point corresponds to
a specific value of the evolution variable at
hand, cf.~\figRef{fig:ev}. Assuming ordering in $p_\perp$ and 
using subscript $(3)$
for quantities evaluated in the 3-parton state, the value of the
evolution variable for a specific $(y_{ij},y_{jk})$ point is $Q^2_{E(3)} =
4p_{\perp(3)}^2 = 4 y_{ij} y_{jk} s$, with $s=m_Z^2$ at the $Z$ pole.
The further evolution of the shower, from a 3- to a 4-parton state, involves a 
sum over all possible branchings of the $qg$ and $g\bar{q}$
antennae. Consider the $qg$ one. Its branchings can again be
characterized by two 
invariants $(s_1, s_2)$, both of which will be smaller than
$m^2_{qg}$. However, depending on the value of $m^2_{qg}$ (or,
equivalently,  $y_{ij}$) the $p_\perp$ of the new configuration, 
$4p_{\perp(4)}^2=4 s_1 s_2/m^2_{qg}$ may actually be \emph{larger} than 
$4p_{\perp(3)}^2$. In a strongly ordered shower, such
configurations are not allowed, and would be discarded. Whether this
situation can occur or not, for one or both of the $qg$ and $g\bar{q}$
antennae, as a function of $(y_{ij},y_{jk})$, is what
figure~\ref{fig:psOrd} reveals, for each type of ordering variable. 

The mathematical consequence is that, in the dark- and black-shaded
regions, respectively, the upper boundary of one or both of 
the $qg$ and $g\bar{q}$ antenna integrals is set by the evolution variable,
rather than by phase space. This creates an important difference
between the integrals generated by a shower
algorithm and those used for IR subtractions in traditional 
fixed-order applications for which the integrals 
often run over all of phase space, although some subtraction schemes 
feature parameters that allow restrictions on the phase space for the
subtraction terms \cite{Frixione:1995ms,Nagy:2003tz}.
In particular, we see that the strong-ordering
condition will generate additional logarithms involving 
$s_{ij}/Q^2_{E(3)}$ as argument. For a ``good'' choice of evolution
variable, these logarithms should explicitly 
cancel against ones present in the one-loop matrix elements,
a question we
shall return to in detail in \secRef{sec:integrals}.

Several interesting structures can be seen in
\figRef{fig:psOrd}. Firstly, the linearized variables imply less
severe constraints on the subsequent evolution than the quadratic
ones. This is easy to understand given that the linearized variables,
$Q_\mrm{lin}$, are related to the quadratic ones,
$Q_\mrm{qdr}$, by 
\begin{equation}
Q_\mrm{lin}^2 =  Q_\mrm{qdr}\, m_{IK}~,
\end{equation}
and hence $Q_\mrm{lin} > Q_\mrm{qdr}$,
implying a higher absolute restart scale for the
linearized ordering variables.

It is also apparent that, for a given choice of linearity,  
mass-ordering reduces the phase-space for further evolution more than 
\pT-ordering does, which in turn is more constraining than
energy-ordering. In this comparison, however, 
it becomes important to recall that the
traditional ordering variables used, e.g., in \Vc, are the
\emph{linearized} mass-ordering and the \emph{quadratic} \pT and
energy-ordering variables\footnote{This distinction comes about from
  using quantities that are similar to a squared mass, squared
  transverse momentum, and squared energy, respectively.}. 
Within that group, \pT-ordering appears to
be the most restrictive, followed by energy-ordering, with
(traditional, linearized) mass-ordering leading to the most open phase
space for the subsequent evolution. 

\begin{table}[tp]
\centering 
\scalebox{0.95}{
\begin{tabular}{l l | | c | c c | c } 
\multicolumn{2}{l||}{Ordering type} & $Q_E^2$ & $\zeta_{\min}(Q_E^2)$ & $\zeta_{\max}(Q_E^2)$ &  $3\to 4$ restriction\\ \hline
\multirow{2}{*}{$p_\perp$-ordering} & linear & $2m^2_{IK} \sqrt{y_{ij}y_{jk}}$   & \multicolumn{2}{c|}{$\frac{1\mp \sqrt{1-Q_E^4/m^4_{IK}}}{2}$}   &  $\theta \left( m^2_\mrm{ant} - 2 \sqrt{s_{ij}s_{jk}}  \right)$  \\
                                                          & squared & $4 m^2_{IK} \,y_{ij}y_{jk}$ & \multicolumn{2}{c|}{$\frac{1\mp \sqrt{1-Q_E^2/m^2_{IK}}}{2}$} & $\theta \left( m^2_\mrm{ant} - 4\frac{s_{ij}s_{jk}}{s}  \right)$ \\ 
\multirow{2}{*}{$m_D$-ordering} & linear  &$2m^2_{IK}\min(y_{ij},y_{jk})$ & $\frac{Q_E^2}{2m^2_{IK}}$ & $1-\frac{Q_E^2}{2m^2_{IK}}$ & $\theta \left( m^2_\mrm{ant} - 2\min(s_{ij},s_{jk})  \right)$ \\
                                                            &  squared & $4m^2_{IK}\min(y^2_{ij},y^2_{jk})$ & $\sqrt{\frac{Q_E^2}{4m^2_{IK}}}$ & $1-\sqrt{\frac{Q_E^2}{4m^2_{IK}}}$ & $\theta \left( m^2_\mrm{ant} - 4\frac{\min(s_{ij}^2,s_{jk}^2)}{s}  \right)$  \\
\multirow{2}{*}{$E^*$-ordering} & linear  & $m^2_{IK}\left( y_{ij} + y_{jk} \right)$          &  $0$ & $1$ & 1 \\
                                                         & squared &  $m^2_{IK}\left( y_{ij} + y_{jk} \right)^2$ & $0$ & $1$ &  $\theta \left( m^2_\mrm{ant} - \frac{(s_{ij}+s_{jk}  )^2}{s}\right)$\\ \hline
\end{tabular}}
\caption{Boundaries corresponding to the ordering variables portrayed
  in \figRef{fig:ev}, with $m^2_\mrm{ant}$ corresponding to the active
  $3\to 4$ dipole $\sqg$ or $\sgq$, and $s=m_Z^2$ at the $Z$ pole. We
  have chosen $\zeta_2$ as the 
  energy sharing variable for $m_D$ and $p_\perp$ ordering and $\zeta_1$
  for $E^*$ ordering, with $\zeta$ defined as in \eqRef{def:zeta1} and
  \eqRef{def:zeta2}. The energy variable will lead to infinities if
  the hadronization scale is not imposed as a cut-off. }  
\label{tab:QELimits}
\end{table}

We are now able to fully specify the boundaries of the evolution
integrals in \eqRef{eq:AntPS}. For each $Q_E$ contour (see
\figRef{fig:ev}), the integration limits in $\zeta$ are listed in 
\tabRef{tab:QELimits}. Combined with a $Q_E$ interval and an antenna
function, these boundaries account for the integrated tree-level
splitting probability when going from one scale $Q_1^2$ to another
$Q_2^2$, as expressed by \eqRef{eq:AntPS}. 
The last column in \tabRef{tab:QELimits} tells when the $3\to 4$ ordering
condition is active. In \figRef{fig:psOrd} this corresponds to a
region darkening due to the restriction, with its shade determined by
the amount of restricted dipoles.   

Finally, we note that the dependence on $Q$ in \eqRef{eq:QR} 
causes explicit non-Markovian behavior at the $4$-parton level and beyond, 
since the value of $Q$ then depends
explicitly on which branching was the last to occur. A more
strictly  Markovian variant of this is obtained by 
letting the $\min()$ function act on all possible $Q$ values
(corresponding to all possible colour-connected clusterings) 
of the preceding topology. In that case, a single $Q$ value can be
used to characterize an entire $n$-parton topology, irrespective of which
branching was the last to occur. Since the distinction between
Markovian and non-Markovian shower restart conditions only enters 
starting from the $4\to 5$ parton evolution step, it will not affect
our discussion of the second-order $2\to 3$ branching process. For
completeness, we note that the strongly ordered showers in \Vc are
of the ordinary non-Markovian type, while the smoothly ordered ones
are Markovian. 

\subsection{Smooth Ordering}\label{sec:SmOrd}
In addition to traditional (strongly ordered) showers, 
we shall also consider so-called smooth ordering \cite{Giele:2011cb}: 
applying the ordering criterion as a smooth
  dampening factor instead of as a step
  function. This is not as radical as it may seem at first. Applying a
    jet algorithm to any set of events will in general result in some small
    fraction of unordered clustering sequences. This is true even if
    the events were 
    generated by a strongly ordered shower algorithm, if the jet
    clustering measure is not strictly identical to the shower
    ordering variable. An example of this, for strong ordering
    in $p_\perp$ and in $m_D$, clustered with the $k_T$ algorithm, can
    be found in \cite{lvs2011}. 

  In smooth ordering, the only quantity which must still be strictly
  ordered are the antenna invariant masses, a condition which follows
  from the 
  nested antenna phase spaces and momentum conservation. Within each
  individual antenna, and between competing ones, the 
  measure of evolution time is still provided by the ordering
  variable, which we therefore typically refer to as 
  the ``evolution variable'' in this context (rather than the ``ordering
  variable''), in order to prevent
  confusion with the strong-ordering case. The evolution variable 
  can in principle 
  still be chosen to be any of the possibilities given
  above, though we shall typically use $2p_\perp$ for gluon emission and
  $m_{q\bar{q}}$ for gluon splitting. 
  
\begin{figure}[t]
\centering
\includegraphics*[scale=0.78]{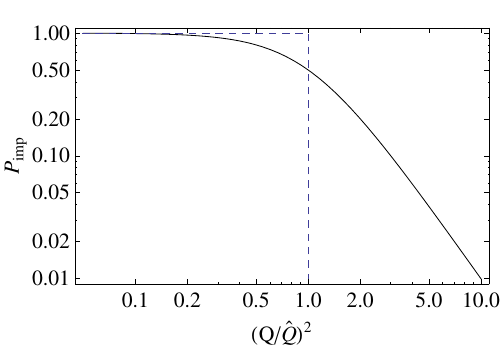}
\caption{The smooth-ordering factor ({\sl solid}) compared to a
  strong-ordering $\Theta$ function ({\sl dashed}). \label{fig:smooth}}
\end{figure}

In terms of an arbitrary evolution variable, $Q$, 
the smooth-ordering factor is \cite{lvs2011}
\begin{equation}
P_\mrm{imp} \ = \ \frac{\hat{Q}^2}{\hat{Q}^2 + Q^2}~,
\end{equation}
where $Q$ is the evolution scale associated with the current
branching, and $\hat{Q}$ measures the scale of the parton
configuration before branching. A comparison to the strong-ordering
step function is given in \figRef{fig:smooth}, on a log-log scale. 

In the strongly-ordered region of phase-space, $Q \ll \hat{Q}$, 
we may rewrite the $P_\mrm{imp}$ factor as 
\begin{equation}
P_\mrm{imp} 
  \ = \ 
    \frac{1}{1 + \frac{Q^2}{\hat{Q}^2}} 
  \ \stackrel{Q<\hat{Q}}{=} \ 
    1 - \frac{Q^2}{\hat{Q}^2} + \ldots ~. \label{eq:pimpO}
\end{equation}
Applying this to the $2\to3$ antenna function whose leading singularity,
\eqRef{eq:aTrial}, is proportional to $1/Q^2$, 
we see that the overall correction arising from the $Q^2/\hat{Q}^2$
and higher terms is finite and of order $1/\hat{Q}^2$; a power
correction. The LL singular behaviour is therefore not
affected. However, at the multiple-emission level, the $1/\hat{Q}^2$
terms do modify the \emph{subleading}
logarithmic structure, starting from ${\cal O}(\alpha_s^2)$,  
as we shall return to below. 

In the \emph{unordered} region of phase-space, $Q > \hat{Q}$, we rewrite the
$P_\mrm{imp}$ factor as 
\begin{equation}
P_\mrm{imp} 
  \ = \ 
    \frac{\hat{Q}^2}{Q^2} \ \frac{1}{1 +
    \frac{\hat{Q}^2}{Q^2}} 
  \ \stackrel{Q>\hat{Q}}{=} \ 
    \frac{\hat{Q}^2}{Q^2} \left(
    1 - \frac{\hat{Q}^2}{Q^2} + \ldots \right)~,\label{eq:pimpU}
\end{equation}
which thus effectively modifies the leading 
singularity of the LL $2\to 3$ function from
$1/Q^2$ to $1/Q^4$, removing it from the LL counting. The only
effective terms $\propto 1/Q^2$ arise from finite terms in the
radiation functions, if any such are present, multiplied by the
$P_\mrm{imp}$ factor. Only a matching to the full tree-level 
$2\to 4$ functions would enable a precise control over these terms. Up to any
given fixed order, this can effectively be achieved by matching to
tree-level matrix elements, as will be discussed in 
\secRef{sec:LO}. Matching beyond the 
fixed-order level is beyond the scope of this paper. We thus restrict
ourselves to the observation that, at the LL level, smooth ordering is
equivalent to strong ordering, with differences only
appearing at the subleading level.  

The effective $2\to 4$ probability in the shower arises from a sum
over two different $2\to 3 \otimes 2\to 3$ histories, each 
of which has the tree-level form 
\begin{equation}
\hat{A} \ P_\mrm{imp} \ A \ \propto \ \frac{1}{\hat{Q}^2} \
  \frac{\hat{Q}^2}{\hat{Q}^2 + Q^2} \ \frac{1}{Q^2} \ = \ \frac{1}{\hat{Q}^2
    + Q^2} \ \frac{1}{Q^2}~,
\end{equation}
thus we may also perceive the combined effect of the modification as 
the addition of a mass term in the denominator of the propagator
factor of the previous splitting. In the strongly ordered region, this
correction is negligible, whereas in the unordered region, there is a
large suppression from the necessity of the propagator in the previous
topology having to be very off-shell, which is reflected by the
$P_\mrm{imp}$ factor. 
Using the expansion for the unordered region, \eqRef{eq:pimpU}, we
also see that the effective $2\to 4$ radiation function, obtained from
iterated $2\to 3$ splittings, is modified as follows,
\begin{equation}
P_{2\to 4} 
  \ \propto \ 
    \frac{1}{\hat{Q}^2}\frac{\hat{Q}^2}{Q^2} \ \frac{1}{Q^2} 
  \ \to \
    \frac{1}{Q^4} + {\cal O}(...)~,
\end{equation}
in the unordered region. That is, the intermediate low scale
$\hat{Q}$, is \emph{removed} from 
the effective $2\to 4$ function, by the application of the
$P_\mrm{imp}$ factor. 

\begin{figure}[t]
\centering
\includegraphics[scale=0.85]{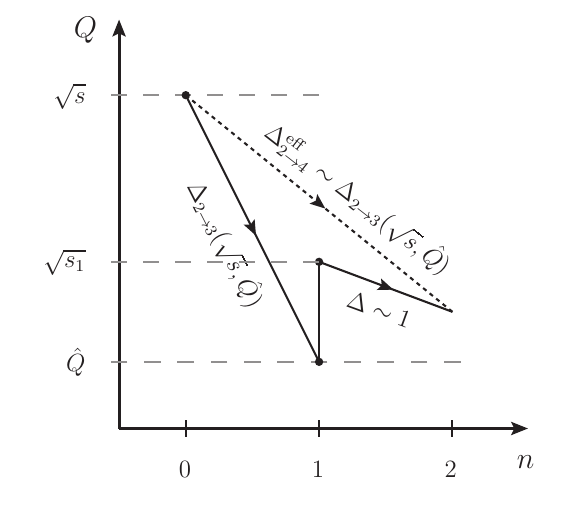}
\caption{Illustration of scales and Sudakov factors involved in an
  unordered sequence of two $2\to 3$ branchings, representing the
  smoothly ordered shower's approximation to a hard $2\to 4$
  process. 
\label{fig:ordg}}
\end{figure}
Finally, to illustrate what happens to the Sudakov factors, we
illustrate the path through phase space taken by an unordered shower
history 
in \figRef{fig:ordg}. An antenna starts showering at a
scale equal to its invariant mass, $\sqrt{s}$, 
and a first $2\to 3$ branching occurs at the
evolution scale $\hat{Q}$. This produces the overall Sudakov factor
$\Delta_{2\to 3}(\sqrt{s},\hat{Q})$. A daughter antenna, produced by the
branching, then starts showering at a scale equal to its own invariant
mass, labeled $\sqrt{s_1}$. However, for all scales larger than $\hat{Q}$,
the $P_\mrm{imp}$ factor suppresses the evolution in this new dipole
so that no leading logs are generated. To leading approximation, 
the effective Sudakov factor for the combined $2\to 4$ splitting is
therefore given by 
\begin{equation}
\Delta_{2\to 4}^{\mrm{eff}} \ \sim \ \Delta_{2\to 3}(\sqrt{s},\hat{Q})~,
\end{equation}
in the unordered region. Thus, we see that a dependence on the
intermediate scale $\hat{Q}$ still remains in the effective Sudakov
factor generated by the smooth-ordering procedure. Since $\hat{Q}<Q$
in the unordered region, the effective Sudakov suppression of these
points might be ``too strong''.  The smooth ordering therefore allows
for phase space occupation in regions corresponding to dead zones in a
strongly ordered shower, but it does suggest that a correction to the
Sudakov factor may be desirable, in the unordered region.  
 A study
of $Z\to 4$ jets at one loop would be required to shed further light
on this question.

Having presented introductory aspects of (antenna) showers, we now
turn to a detailed discussion of how we match them to higher
fixed-order 
calculations. 

\section{Matched Antenna Showers}
\label{sec:match-antenna-show}

\subsection{Tree-Level Matching \label{sec:LO}} 

The strategy for matching to tree-level matrix elements used in \Vc
was defined by GKS in  \cite{Giele:2011cb} 
and is tightly related to the veto
algorithm outlined above. The philosophy is to view the shower
produced by the smoothly ordered antenna functions as generating 
an all-orders approximation to QCD, which can be systematically
improved, order by order, by including one more factor in the accept
probability, called the matrix-element correction. For a
given trial branching, the full
trial accept probability, up to the highest matched number of partons,
is then obtained as a product of the ordinary trial-accept probability
in the shower, multiplied by this extra correction factor. 

Since the shower is
already correct in the soft and collinear limits, the matrix-element
correction factor will
tend to unity in those limits, but it can deviate on either side of 
unity outside those limits. As long as the combined accept probability
is still smaller than unity, a probabilistic accept/reject step can
still be applied, without having to worry about reweighting the 
events (which would be required if the total accept probability should
exceed unity). It is also important to define the factor only in terms
of physical cross sections (here represented by LO matrix elements), 
which guarantees positivity. (Again, if
it were allowed to become negative, one would have to introduce
negative-weight events, but this is avoided in the GKS strategy as
defined in  \cite{Giele:2011cb}). 

As we have seen, the shower furnishes an all-orders approximation to
QCD. The aim is, for each resolved parton/jet multiplicity, to match
the answer provided by the shower 
to an, ideally, all-orders exact expression, by applying a
multiplicative correction factor,
schematically~\cite{Bengtsson:1986hr,Giele:2011cb}
\begin{equation}
\mbox{Matched} \ = \ \mbox{Approximate}
\ \frac{\mbox{Exact}}{\mbox{Approximate}}~. \label{eq:philosophy}
\end{equation}
At tree level, we in fact know only the first term in the expansion of
the numerator, and we therefore expand the shower approximation to the
same level. For $n$ partons, assuming the approximation
has already (recursively) been matched to the preceding order, 
\begin{eqnarray}
\mbox{Exact} & \to & |M_n|^2 \\[2mm]
\mbox{Approximate} & \to & \sum_{j} g_{Tj}^2 \ {\cal C}_{Tj} \ A_{Tj}
\ P_{\mrm{accept}} \ |M_{n-1_j}|^2 \ = \ \sum_j g_j^2 \ {\cal C}_j \ P_\mrm{imp}
  \ P_\mrm{Ari} \ A_j \ |M_{n-1_j}|^2~, \hspace*{1.5cm} \label{eq:approximate}
\end{eqnarray}
where the subscript ``T'' indicates trial quantities
(cf.\ \secRef{sec:shower}), 
we have suppressed the dependence on phase-space points, $\Phi$,
and the subscript $j$ in the $(n-1)$-parton matrix element indicates
the configuration obtained by 
performing the inverse shower step that removes parton $j$ from the
$n$-parton state.  

The factors in \eqRef{eq:approximate} are easy to calculate 
if a tree-level matrix-element (ME) generator is available to provide the
$|M|^2$ factors. The total ME-corrected accept probability is
thus simply \eqRef{eq:PLO},
\begin{equation}
P_{\mrm{accept}}^\mrm{LO} \ = \ P_{\mrm{accept}} \ R_n \ = \ 
 P_{\mrm{accept}} \ 
\frac{ |M_n|^2}
     {\sum_j g_j^2 \ {\cal C}_j \ P_\mrm{imp} \ P_\mrm{Ari}
       \ A_j \ |M_{n-1_j}|^2}~. \label{eq:PME}
\end{equation}
As mentioned above, this factor should be positive and smaller than
unity, in order to avoid having to reweight any events. In practice,
we have found the trial function defined in \eqRef{eq:aTrial} to
guarantee this for all processes we have so far considered, mainly
consisting of $Z\to n$ and $H\to n$ partons. As shown in  \cite{Giele:2011cb},
it is also possible to absorb subleading-colour corrections into this
matching factor in a positive-definite way, but since subleading colour
goes beyond the scope of our study we do not reproduce the arguments
here. 

The fact that these factors change the distribution of the 
final set of generated events to the correct matrix-element answer 
can be explained by following the steps of the algorithm and summing
over all possible branching histories. 
We start from Born-level
matrix-element events, and generate trial 
shower branchings, for a trial approximation to the
$(B+1)$-parton matrix element of:
\begin{equation}
|M^\mrm{Trial}_{B+1}|^2 = \sum_i g_{Ti}^2 \, {\cal C}_{Ti} \, A_{Ti} \,
|M^{\mrm{LO}}_{B_i}|^2~,
\end{equation}
with $i$ running over all possible single-parton clusterings that
correspond to allowed shower branchings.
Applying the LO accept probability, \eqRef{eq:PME}, changes this to
\begin{eqnarray}
 & \to & \sum_i g_{Ti}^2 \, {\cal C}_{Ti} \, A_{Ti} \,
  P_{\mrm{accept}}^{\mrm{LO}}\, |M^{\mrm{LO}}_{B_i}|^2~ \nonumber\\[2mm]
& = & \sum_i g_{i}^2 \, {\cal C}_{i} \, P_\mrm{imp} \, P_\mrm{Ari} \, A_{i} \,
\frac{ |M^\mrm{LO}_{B+1}|^2}
     {\sum_j g_{sj}^2 \ {\cal C}_j \, P_\mrm{imp} \, P_\mrm{Ari} \, 
      A_j \, |M^\mrm{LO}_{B_j}|^2} |M^{\mrm{LO}}_{B_i}|^2~\nonumber\\[2mm]
& = & |M^\mrm{LO}_{B+1}|^2~.
\end{eqnarray}
That is, summed over shower histories, numerators and
denominators are designed to cancel exactly, leaving only the LO
matrix element for B+1 partons, as desired. Due to the full phase-space
coverage and explicitly Markovian nature of the smoothly ordered
shower algorithm, this procedure is straightforward to iterate 
for Born + 2, 3, etc partons\footnote{That is not the case for
  ordinary strongly ordered frameworks, due to the presence of dead
  zones in phase space 
and to the generally non-Markovian shower restart conditions. For such
algorithms, addition of events in the dead zones~\cite{Seymour:1994df},
with CKKW-like Sudakov-factor prescriptions for multi-leg
matching~\cite{Catani:2001cc,Lonnblad:2001iq}, 
would presumably be necessary.}.

To provide a connection with antenna subtraction, which will be useful
when we extend to NLO matching below, we can rewrite the ratio in
\eqRef{eq:philosophy} by a trivial rearrangement, 
\begin{equation}
\mbox{Matched} \ = \ \mbox{Approximate} \left( 1 + 
\ \frac{\mbox{Exact}-\mbox{Approximate}}
       {\mbox{Approximate}}\right)~. \label{eq:philosophySub}
\end{equation}
The numerator in this equation is very similar to a standard
antenna-subtracted matrix element, though we emphasize that our
antennae are of course modified by the presence of the $P_\mrm{imp}$ and
$P_\mrm{Ari}$ factors. 

Let us finally re-emphasize that since we apply the correction factor
to the antenna functions themselves, thereby modifying the probability for a
branching to occur, the probability for a branching \emph{not} to
occur is also modified. These corrections will therefore 
also be present in the Sudakov factors generated by the
corrected shower evolution. The fact  
that the correction factor, $R_{n}$, 
is unity in all LL singular limits (since the
shower functions are guaranteed to match the matrix-element
singularities there) implies that 
the LL properties of the Sudakov factors are not affected by this
modification. However, the tower of subleading logarithms \emph{is}
changed, for better or worse. While it is known that finite terms do
not exponentiate our corrections here also
include a more subtle aspect, namely a resummation of the subleading
logarithms present in the higher-order matrix elements. At this level,
however, we cannot be sure that this procedure produces the
\emph{correct} subleading logarithms of a formally higher-order
resummation. Therefore, we view it at present merely as an
interesting, and hopefully beneficial, side-effect of unitarity-based
matching. The examination of formally subleading terms carried out in
this paper is a first step towards a more rigorous study of
these aspects. 

\subsection{One-Loop Matching at the Born Level \label{sec:Born-nlo}}

For the Born level, at NLO, the GKS matching strategy
\cite{Giele:2011cb} 
reduces to the
\Pw one~\cite{Nason:2004rx,Frixione:2007vw,Alioli:2010xd}. We
nonetheless begin by recapitulating the steps 
used to derive the one-loop correction to the Born-level matrix element, 
in our notation. We then continue to higher multiplicities. 

As our basis for one-loop matching we take the tree-level 
strategy described in \secRef{sec:LO}.
 Since the corrections are applied as
modifications to the branching probabilities, without adding,
subtracting, or reweighting events, the total inclusive rate after
tree-level matching to any number of partons, is still just the
leading-order, Born-level one. 
By the same token, after one-loop matching, at the integrated level, the total
NLO correction to the inclusive rate must therefore just be the 
NLO ``$K$-factor'',
\begin{equation}
K^\mrm{NLO} \ = \ \frac{\sigma_\mrm{inc}^\mrm{NLO}}{\sigma_\mrm{inc}^\mrm{LO}}~.
\end{equation}
For processes like $Z$ decay, where the NLO correction has no
dependence on the Born-level kinematics, this is trivial to implement as
an overall reweighting factor on the Born-level events,
\begin{equation}
K^\mrm{NLO}_Z \ = \ 1 + V_Z \ = \ 1 + \frac{\alpha_s}{\pi}~, \label{eq:KZ}
\end{equation}
where we have introduced the notation $V$ for the NLO correction term,
anticipating a similar notation for the multileg case below. 
Note that one could equally trivially normalize 
to NNLO or to data, as
desired for the application at hand (we note though that such a
normalization choice does not, by itself, ensure NNLO precision for any
quantity besides the total inclusive rate).

However, when the amount of final state particles exceeds two, the NLO
correction depends on the Born-level kinematics, therefore it is worth
illustrating the general procedure for deriving a fully differential
K-factor, for each phase-space point. This also serves as a useful 
warm-up exercise for the multi-leg case below. 

At NLO, we may distinguish
between inclusive and exclusive rates for the first time. Either can
in principle be used to derive matching equations between showers and
fixed-order calculations, but the exclusive one is
best suited for deriving expressions at the fully differential level. 
We recall that 
the exclusive $n$-jet cross section is defined as the cross section
for observing $n$ \emph{and only} $n$ jets, while the inclusive $n$-jet cross
section counts the number of events with $n$ \emph{or more} jets. One therefore has
the trivial relation
\begin{equation}
\label{eq:5}
\sigma_n^\mrm{incl}(Q) = \sum_{k\ge n} \sigma_{k}^\mrm{excl}(Q)~.
\end{equation}
with $Q$ the resolution scale of whatever (IR safe)
 algorithm is used to define the jets. 

\subsubsection{Inclusive Born}

The total inclusive rate produced by the tree-level matched shower 
is just the Born-level matrix element, 
\begin{equation}
\mbox{Approximate} \ \to \ |M^{0}_2|^2~,\label{eq:sigLO}
\end{equation}
where the subscript indicates the parton multiplicity (2 for
 $Z\to q\bar{q}$ decay) and the superscript
indicates the loop order beyond the Born level (0 indicates the
Born loop order). Because cancellation of real and virtual corrections is exact
in both the unmatched shower as well as in the tree-level matching
scheme described above, there are no further corrections to consider
for the inclusive rate. In other words, 
the total integrated cross section produced by the shower
is obtained merely by
integrating \eqRef{eq:sigLO} over all of the Born-level phase space. 
We now seek a correction term, $V_2$, such that
\begin{equation}
\mbox{Matched} \ \to \ (1+V_{2Z}) \ |M^{0}_2|^2
\end{equation}
gives the correct inclusive NLO rate. From \eqRef{eq:KZ}, we know that
the correction term for $Z$ decay is 
\begin{equation}
V_{2Z} \ = \ \frac{\alpha_s}{\pi}~.\label{eq:V0}
\end{equation}

A systematic way of deriving this result, which can be applied to
arbitrary processes, is provided by considering the cross section at
the exclusive level.  

\subsubsection{Exclusive Born}

The shower expression for the exclusive $Z\to q\bar{q}$ rate (defined at the
hadronization cut-off, which is the lowest meaningful resolution scale
in the perturbative shower) is
\begin{equation}
|M_2^0|^2 \ \Delta(s,Q_\mrm{had}^2) \ = \ 
  |M_2^0|^2 \left(1 - \int_{Q^2_\mrm{had}}^s \hspace*{-2mm} \dPS{\mrm{ant}} \ g_s^2
  \ {\cal C} \ A_{g/q\bar{q}} + {\cal O}(\alpha_s^2) \right)~, \label{eq:DShad}
\end{equation}
where we have expanded the Sudakov factor $\Delta$ to first
order. Due to the presence of the hadronization scale, this expression
is IR finite and can be defined in 4 dimensions.

We remark here on the validity of this expansion in $\alpha_s$ for the
exclusive cross section. For the purpose of constructing the matching
factor to order $\alpha_s$ the expansion is a parametric one. In the
ratio of the exact and approximate exclusive cross section, since the
singularities match to the shower accuracy, divergences or large
logaritms (depending on whether one choose zero or finite resolution
scale) cancel and the resulting factor has a well-behaved expansion in
$\alpha_s$.

The colour factor for
$q\bar{q} \to qg\bar{q}$ is
\begin{equation}
{\cal C}_{g/q\bar{q}} = 2C_F~,
\end{equation}
and we assume that the
antenna function, $A$, is either the one derived from $Z$ decay 
 \cite{Gustafson:1987rq} or has been matched to it, using LO
 matching. That is,  
\begin{equation} 
g_s^2 \ {2C_F} \ A_{g/q\bar{q}} \ = \ \frac{|M_3^0|^2}{|M_2^0|^2}~. \label{eq:A3}
\end{equation}
We first consider the limit $Q_\mrm{had}\to 0$, in which case the
expression becomes 
\begin{equation}
|M_2^0|^2 \ \Delta(s,0) \ = \ 
  |M_2^0|^2 \left(1 - \int_{0}^s \dPS{\mrm{ant}} \ g_s^2
  \ 2C_F \ A_{g/q\bar{q}} + {\cal O}(\alpha_s^2) \right)~, \label{eq:D0}
\end{equation}
which can only be defined in the presence of an IR regularization
scheme. We shall here use dimensional regularization, working in
$d=4-2\epsilon$ dimensions. Below, we rederive the
matching equations in 4 dimensions, for $Q_\mrm{had} \ne 0$, 
and show that the \emph{same} final matching factors are obtained in
both cases.

At NLO, the exclusive $Z\to
q\bar{q}$ rate at ``infinite'' perturbative resolution is
\begin{equation}
|M_2^0|^2 \ + \ 2\Re[M_2^0{M_2^1}^*] ~=~|M_2^0|^2 \left(1 +
\frac{2\Re[M_2^0{M_2^1}^*]}{|M_2^0|^2}\right)~, \label{eq:2V}
\end{equation}
where we have written the right-hand side in a form similar to
\eqRef{eq:D0}, in $d$
dimensions. Because the resolution scale has been taken to zero, there
are no unresolved 3-parton configurations to include. 
The virtual matrix element is
\begin{equation}
\frac{2\Re[M_2^0{M_2^1}^*]}{|M_2^0|^2} \ = \ 
   \frac{\alpha_s}{2\pi} \ 2C_F \ \left(2I_{q\bar{q}}(\epsilon,\mu^2/s)
   - 4\right)~,
\end{equation}
with the function $I_{q\bar{q}}$ used to classify the $\epsilon$
divergences
\cite{GehrmannDeRidder:2004tv,GehrmannDeRidder:2005cm,GehrmannDeRidder:2007jk}. Note
that we have modified the definition of $I$ to make it explicitly
dimensionless, see \appRef{app:I}. On the shower side,
the integral of the $Z\to qg\bar{q}$ antenna in \eqRef{eq:D0} is
\cite{GehrmannDeRidder:2005cm} 
\begin{equation}
\int_{0}^s \hspace*{-2mm} \dPS{\mrm{ant}} 
  \ 2C_F \ g_s^2 \ A_{g/q\bar{q}} \ = \ \frac{\alpha_s}{2\pi} \ 2C_F \ \left(
  -2I_{q\bar{q}}(\epsilon,\mu^2/s) + \frac{19}{4}\right)~, \label{eq:Aint}
\end{equation}
and, not surprisingly, the difference comes out to be exactly 
$\alpha_s/\pi \times |M_2^0|^2$. 
Writing this correction as a multiplicative $K$-factor,
we obtain \eqRef{eq:KZ}.

As a cross-check, we now repeat the derivation in 4 dimensions, 
reinstating the hadronization scale. The fixed-order side is then
\begin{equation}
|M_2^0|^2 \left(1 +
\frac{2\Re[M_2^0{M_2^1}^*]}{|M_2^0|^2} \ + \ 
\int_0^{Q_\mrm{had}^2}   \dPS{\mrm{ant}} \ g_s^2
  \ {\cal C} \ A_{g/q\bar{q}} \right)
 \label{eq:2VexcHad}~,
\end{equation}
where the integral that has been added corresponds to unresolved
3-parton configurations, with $A$ again given by \eqRef{eq:A3}. Though
\eqRef{eq:DShad} is now 
defined entirely in 4 dimensions, we still need dimensional
regularization to regulate the two last terms in the fixed-order
expression. In principle, the integral in the last term could be
carried out explicitly, but it is simpler to rewrite it as
\begin{equation}
\int_0^{Q_\mrm{had}^2}   \dPS{\mrm{ant}} \ g_s^2
  \ {\cal C} \ A_{g/q\bar{q}}  \ = \ 
\int_0^{m_Z^2}   \dPS{\mrm{ant}} \ g_s^2
  \ {\cal C} \ A_{g/q\bar{q}} \ - \int_{Q_\mrm{had}^2}^{m_Z^2} \dPS{\mrm{ant}} \ g_s^2
  \ {\cal C} \ A_{g/q\bar{q}}  
\end{equation}
where the first term is just the full antenna integral,
\eqRef{eq:Aint}, and the second term is identical to the
one appearing in \eqRef{eq:DShad}, with which it cancels completely,
cf.\ the definition of the tree-level matching, \eqRef{eq:A3}. 
The final correction term  is therefore 
again exactly equal to $\alpha_s/\pi \times |M_2^0|^2$. 

Note that the scale and
scheme dependence of the $\alpha_s/\pi$ correction is not specified
since its ambiguity is formally of order $\alpha_s^2$. For
definiteness we take the renormalization scale for this correction 
to be proportional to the invariant mass of the system, $\mu_R = k^\mrm{inc}_{\mu}
\sqrt{\hat{s}}$ (so 
that $\mu_R = k^\mrm{inc}_{\mu} m_Z$ at the $Z$ pole), with
$k^\mrm{inc}_{\mu}$ thus representing the free
parameter that governs the choice of renormalization scale for the total 
inclusive rate for $Z\to$ hadrons. 
We shall consider both one-loop and
two-loop running options. The value of $\alpha_s(m_Z)$ will be
determined from LEP data in \secRef{sec:results}.

\subsection{One-Loop Matching for Born + 1 Parton}

The approximation to the $3$-parton exclusive rate produced by a
shower matched to (at least) NLO for the $2$-parton inclusive rate and
to LO for the 3-parton one, is 
\begin{equation}
\mbox{Approximate} \ \to \ 
(1+V_2) \ |M_3^0|^2 
  \  \Delta_{2}(m_Z^2,Q_3^2)
  \ \Delta_3(Q_{R3}^2,Q_\mrm{had}^2)~, \label{eq:ps3excl} 
\end{equation}
where $M^0_3$ is the tree-level $Z\to qg\bar{q}$ matrix element and
$Q_{R3}$ denotes 
the ``restart scale''. For strong ordering, $Q_{R3}$ is equal to $Q_3$, 
while, for smooth ordering, it is given by the nested antenna phase
spaces, i.e., by the successive antenna invariant masses.
\begin{figure}[t]
\centering
\includegraphics*[scale=0.65]{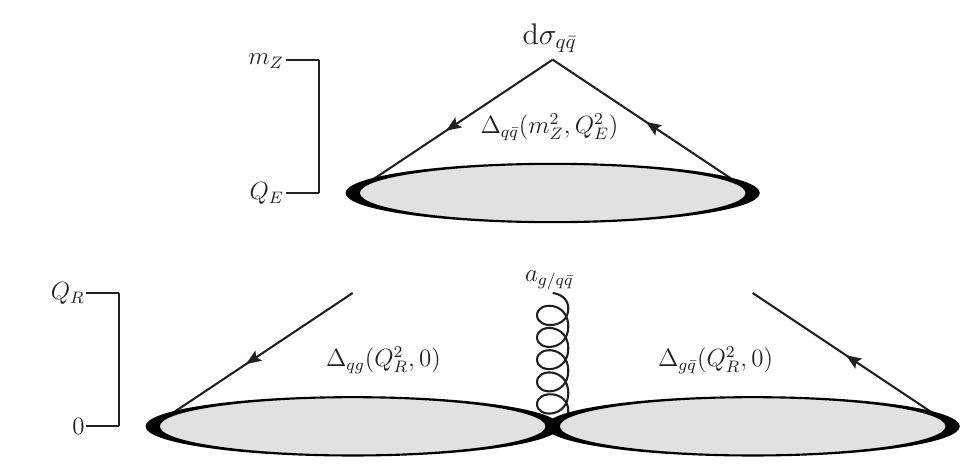}
\caption{Illustration of the evolution scales and Sudakov factors 
appearing in the exclusive 3-jet
  cross section, \eqRef{eq:ps3excl}. \label{fig:ps3excl}}
\end{figure}
The subscripts on the two Sudakov factors $\Delta_2$ and
$\Delta_3$ make it explicit that
they refer to the event as a whole, see the illustration in
\figRef{fig:ps3excl}. Again, we have the choice whether we wish to
work in 4 dimensions, with a non-zero hadronization scale,
$Q_\mrm{had}$, or in $d$ dimensions with the hadronization scale taken
to zero. We have maintained the hadronization scale
in \eqRef{eq:ps3excl}, though we shall see below 
that the dependence on it does indeed cancel in the final
result. 

The 2-parton Sudakov factor, $\Delta_2$, 
is generated by the (matched) evolution from 2 to 3
partons, 
\begin{equation}
\Delta_2(m_Z^2,Q_3^2) \ = \ 
1 -  \int_{Q_3^2}^{m_Z^2} \dPS{\mrm{ant}} \ g_s^2
  \ {2C_F} \ A_{g/q\bar{q}} + {\cal O}(\alpha_s^2) ~,
\end{equation}
with $A_{g/q\bar{q}}$ again defined by \eqRef{eq:A3}. Notice that 
the integral  only runs from the starting scale, $m_Z^2$, to the
3-parton resolution scale, $Q_3^2$, hence this integral is IR finite, 
though it does contain logarithms. In the remainder of this paper, we
shall work only with the leading-colour part of the Sudakov and
matrix-element expressions, hence from now on we replace $2C_F$ in the
above expression by $C_A$, 
\begin{equation}
\Delta_2^{\mrm{LC}}(m_Z^2,Q_3^2) \ = \ 
1 -  \int_{Q_3^2}^{m_Z^2} \dPS{\mrm{ant}} \ g_s^2
  \ {C_A} \ A_{g/q\bar{q}} + {\cal O}(\alpha_s^2) ~.
\end{equation}

The 3-parton Sudakov factor, $\Delta_3$, imposes exclusivity and is given by 
\begin{equation}
\Delta_3(Q_{R3}^2,Q_\mrm{had}^2) \ = \ 
1 - \sum_{j=1}^2 \int_{Q_{\mrm{had}}^2}^{Q_{R3}^2} \dPS{\mrm{ant}} \ g_s^2 \ 
\left(C_A \ A_{Ej} \ + \  
  2T_R \ A_{Sj}\right) + {\cal O}(\alpha_s^2) ~,
\end{equation}
where the index $j$ runs over the $qg$ and $g\bar{q}$ antennae, and we
use subscripts $E$ and $S$ for gluon emission and gluon splitting,
respectively. We have implicitly assumed smooth ordering here, which
implies that the upper boundaries on the integrals are given by the
respective dipole invariant masses (squared), $s_j$. 
Note also that we must take into account all modifications that are
applied to the LL antenna functions, including $P_\mrm{imp}$, $P_\mrm{Ari}$, and
LO matrix-element matching factors. (We do not write out these factors
here, to avoid clutter.) I.e., the antenna functions in the
above expression must be the ones actually generated by the shower 
algorithm, including the effect of any
modifications imposed by vetos. 

For strong
ordering, there are no $P_\mrm{imp}$ factors, and the upper 
integral boundary is instead $\min(Q_3^2,s_j)$, 
\begin{equation}
\Delta_3(Q_3^2,Q_\mrm{had}^2) \ = \ 
1 - \sum_{j=1}^2 \int_{Q_{\mrm{had}}^2}^{\min(Q_3^2,s_{j})} \! \dPS{\mrm{ant}} \ g_s^2 \ 
\left(C_A \ A_{Ej} \ + \  
  2T_R \ A_{Sj}\right) + {\cal O}(\alpha_s^2) ~.
\end{equation}
However, since strong ordering is not able to fill the entire 4-parton
phase space~\cite{Andersson:1991he,Giele:2011cb}, 
full NLO matching can only be obtained for the smoothly
ordered variant. It is nonetheless interesting to examine both types of
shower algorithms, since even in the strongly ordered case, we may
compare the Sudakov logarithms arising at ${\cal O}(\alpha_s^2)$ to
those present in the fixed-order calculation.

On the fixed-order side, the expression for the 3-parton exclusive
rate is simply
\begin{equation}
\mbox{Exact} \ \to \ |M_3^0|^2 \ + \ 2\Re[M_3^0M_3^{1*}] \ + \ 
\int_0^{Q_\mrm{had}^2} \frac{\dPS{4}}{\dPS{3}} |M_4^0|^2
~,
\label{eq:me3excl}
\end{equation}
where the last term represents 4-parton configurations in which a
single parton is unresolved with respect to the hadronization
scale. For $Z$ decay, $d$-dimensional expressions for the
virtual matrix element have been available since long 
\cite{Ellis:1980nc,Ellis:1980wv,GehrmannDeRidder:2004tv,GehrmannDeRidder:2005cm}.   
Details  on the calculation and in particular its renormalization, are given in
\appRef{app:FO}, in a notation convenient for our purposes.

We now seek a fully differential matching factor, $K_3 = 1+V_3$, such that
the expansion of 
\begin{equation}
\mbox{Matched} \ = \ (1+V_3) \ \mbox{Approximate} ~,
\end{equation}
reproduces the exact expression, \eqRef{eq:me3excl}, to one-loop
order. (``Approximate''
here stands for the tree-level matched shower approximation, 
\eqRef{eq:ps3excl}.) 
Trivial algebra yields
\begin{eqnarray}
V^{\mrm{LC}}_3 & = & 
  \left[\frac{2\Re[M_3^0 M_3^{1*}]}{|M_3^0|^2} \right]^{\mrm{LC}} \ - \ V_2\nonumber\\[2mm]
  & & \ + \ \int_{Q_3^2}^{m_Z^2} \dPS{\mrm{ant}} \ g_s^2 \ C_A \ A_{g/q\bar{q}} 
  \ + \ \sum_{j=1}^2 \int_{0}^{s_{j}} \dPS{\mrm{ant}} \ g_s^2 \ 
\left(C_A \ A_{Ej} \ + \  
  2T_R \ A_{Sj}\right) 
  \nonumber\\[2mm]
 & & \ + \ \int_0^{Q_\mrm{had}^2} \frac{\dPS{4}}{\dPS{3}}
  \frac{|M_4^0|^2}{|M_3^0|^2} \ - \ 
  \sum_{j=1}^2 \int_{0}^{Q_\mrm{had}^2} \dPS{\mrm{ant}} \ g_s^2 \ 
\left(C_A \ A_{Ej} \ + \  
  2T_R \ A_{Sj}\right) ~,
\end{eqnarray}
where we have reinstated $d$-dimensional forms of the
  one-loop matrix element and of the divergent $3\to 4$ terms. For a
  shower matched to $|M_4^0|^2$ at leading order, 
  the last two terms will cancel, by definition of the matched
  antenna functions (even for an unmatched shower, the difference
  could at most be a finite power correction in the hadronization
  scale, since the matrix element and the shower antenna functions
  have the same singularities), yielding:
\begin{eqnarray}
V^{\mrm{LC}}_{3Z} & = & 
  \left[\frac{2\Re[M_3^0 M_3^{1*}]}{|M_3^0|^2} \right]^{\mrm{LC}} \ - \ V_{2Z}\nonumber\\[2mm]
  & & \ + \ \int_{Q_3^2}^{m_Z^2} \dPS{\mrm{ant}} \ g_s^2 \ C_A \ A_{g/q\bar{q}} 
  \ + \ \sum_{j=1}^2 \int_{0}^{s_{j}} \dPS{\mrm{ant}} \ g_s^2 \ 
\left(C_A \ A_{Ej} \ + \  
  2T_R \ A_{Sj}\right) 
~.\label{eq:V3formal}
\end{eqnarray} 

Rewriting the remaining 
  integrals in terms of a set of standardized antenna subtraction
  terms, writing out the ordering functions for 
  gluon emission and gluon splitting, $O_E$ and $O_S$, explicitly, 
  and denoting the \Ar factor for gluon splitting by $P_A$, 
  we arrive at 
  the following master equation for the second-order correction to the
  3-jet rate: 
\begin{eqnarray}
V^{\mrm{LC}}_{3Z} & = & 
\left[  \frac{2\Re[M_3^0 M_3^{1*}]}{|M_3^0|^2}\right]^{\mrm{LC}} -
\ V_{2Z} \ + \ 
\sum_{j=1}^2 \int_{0}^{s_{j}} \dPS{\mrm{ant}} \ g_s^2 \ 
    \left(C_A A^{\mrm{std}}_{Ej} \ + \  
    n_F A^{\mrm{std}}_{Sj}\right) \label{eq:V3}\\
& &  \ + \  \int_{Q_3^2}^{m_Z^2} \dPS{\mrm{ant}} \ g_s^2 \ C_A
    \ A^\mrm{std}_{g/q\bar{q}} \ + \ \int_{Q_3^2}^{m_Z^2}
    \dPS{\mrm{ant}} \ g_s^2 \ C_A \ \dA_{g/q\bar{q}} 
   \nonumber\\
& & \ - \ \sum_{j=1}^2  \int_{0}^{s_{j}} \dPS{\mrm{ant}} \ g_s^2 \ 
\left(C_A \ (1-O_{Ej}) \ A_{Ej}^\mrm{std}
\ + \  
  n_F \ (1-O_{Sj}) \ P_{A_j}\ A^\mrm{std}_{Sj}\right) \nonumber\\
& &  
  \ + \sum_{j=1}^2 \int_{0}^{s_{j}} \dPS{\mrm{ant}} \ g_s^2 \ 
\left(C_A \ \dA_{Ej} \ + \  
  n_F \ \dA_{Sj}\right) - \sum_{j=1}^2 \int_0^{s_j} \dPS{\mrm{ant}} g_s^2 n_F
  \left( 1- P_{A_j} \right) A^\mrm{std}_{Sj} ~,\nonumber 
\end{eqnarray}
with the standardized Gehrmann-Gehrmann-de Ridder-Glover (GGG) 
subtraction terms defined by \cite{GehrmannDeRidder:2005cm}:
\begin{equation}
\begin{array}{rclcrcl}
\displaystyle A^\mrm{std}_{g/q\bar{q}} & = & a_3^0 \ ( \ = A_3^0) 
 & ,& \displaystyle  \int_0^s
\dPS{\mrm{ant}} \ g_s^2 \ A^\mrm{std}_{g/q\bar{q}} 
   & = & \displaystyle 
\frac{\alpha_s}{2\pi}
  \left( -2 I^{(1)}_{q\bar{q}}(\epsilon,\mu^2/s) + \frac{19}{4} \right)   
\\[4mm]
\displaystyle A^\mrm{std}_{g/qg} & = & d_3^0 
 & ,& \displaystyle  \int_0^s
\dPS{\mrm{ant}} \ g_s^2 \ A^\mrm{std}_{g/qg} 
   & = & \displaystyle \frac{\alpha_s}{2\pi} 
 \left( -2 I^{(1)}_{qg}(\epsilon,\mu^2/s) + \frac{17}{3} \right) 
\\[4mm]
\displaystyle A^\mrm{std}_{\bar{q}/qg} & = & e_3^0 \ ( \ = \frac12
E_3^0)
 & ,& \displaystyle  \int_0^s
\dPS{\mrm{ant}} \ g_s^2 \ A^\mrm{std}_{\bar{q}/qg} 
   & = & \displaystyle 
\frac{\alpha_s}{2\pi}
  \left( -2 I^{(1)}_{qg,F}(\epsilon,\mu^2/s) - \frac{1}{2}
  \right)  
\end{array} \label{eq:AGGG}
\end{equation}
whose IR limits and integrated pole structures 
were examined thoroughly in
\cite{GehrmannDeRidder:2004tv,GehrmannDeRidder:2005cm,GehrmannDeRidder:2007jk},
though we have here rewritten the IR singularity operators $I^{(1)}$
in explicitly dimensionless forms, see \appRef{app:I}. 
(The alphabetical labeling in \eqsRef{eq:AGGG} corresponds to the 
notation used in \cite{GehrmannDeRidder:2005cm}.)

The first line combined with the first term on the second line in \eqRef{eq:V3} represent a standard antenna-subtracted
one-loop matrix element, normalized to the Born level, with the
standardized subtraction terms tabulated in \eqRef{eq:AGGG}, and the
additional finite term $V_{2Z}$ originating from the NLO matching at
the preceding order; see \secRef{sec:Born-nlo}, \eqRef{eq:V0}. 

The subsequent terms express the difference between the simple fixed-order
subtraction carried out in the first line and the actual terms that are
generated by a matched Markovian antenna shower. 
Physically, these terms represent the difference between the
evolution of a single dipole (the original $q\bar{q}$ system) and evolution
of two dipoles (the post-branching $qg\bar{q}$ system), with a
transition occuring at the branching scale $Q_3$. As mentioned above, the
$O_{Ej}$ and $O_{Sj}$ factors in the third line 
represent the ordering criterion imposed
in the evolution, either strong or smooth. For smooth ordering, they
are 
\begin{eqnarray}
1-O_{Ej} & = & 1-\frac{Q_{3}^2}{Q_{Ej}^2 + Q_{3}^2}~, \label{eq:pimpGluEm}\\ 
1-O_{Sj} & = & 1-\frac{Q_3^2}{m_{q\bar{q}}^2 + Q_3^2}~,\label{eq:pimpGluSpl}
\end{eqnarray}
with $Q_{Ej}$ the evolution variable used for gluon emissions, while
for strong ordering, the factor $(1-O_j)$ can be removed if  
the integral boundaries are replaced by $[Q_3^2,s_j]$ (note: this
replacement should only be done in the
third line).  

The last term 
in \eqRef{eq:V3} is an artifact of the 
\Ar factor, $P_\mrm{Ari}$, which was introduced in \secRef{sec:shower}
and is applied to
gluon-splitting antennae in \Vc. Summed over the two ``sides'' of
the splitting gluon, this produces the same collinear singularities as
the standard gluon-splitting antenna, but in highly asymmetric
configurations in which the splitting gluon is near-collinear to a
neighbouring colour line, the \Ar factor produces a strong
suppression, which improves the agreement with the tree-level 4-parton matrix
element \cite{LopezVillarejo:2011ap}, and which then generates an additional
logarithm.

Notice that all but the $\delta A$ terms are defined in terms of standarized
antenna functions, and the corresponding integrals can be carried out
analytically, once and for all. We give explicit forms for each of
these terms, for each choice of evolution variable, in the following
section. 

The only terms of \eqRef{eq:V3} that need to be integrated numerically
are thus the $\delta A$ terms, which express the 
difference between the standardized antenna functions
and those generated by the actual (matched) shower
evolution, which may have different finite terms and/or be matched to
the LO 4-parton matrix element. 
Nonetheless,
since the previous lines already contain most of the structure, we
expect these functions to be relatively well-behaved and numerically
sub-leading. 
Specifically, the $\delta A$ terms for
gluon emission and gluon splitting, respectively, are
defined by 
\begin{eqnarray}
\dA^{\mrm{LC}}_{Ej} & = & O_{Ej} \ \left( R_{4E}^{\mrm{LC}}
A^{\mrm{LL}}_{Ej} \ - \ A_{Ej}^\mrm{std}\right)~,  \label{eq:dAE} \\
\dA^{\mrm{LC}}_{Sj} & = & O_{Sj} \ P_{Aj} \ \left( R_{4S}
A^{\mrm{LL}}_{Sj} \ - \ A_{Sj}^\mrm{std}\right)~,  \label{eq:dA}
\end{eqnarray}
with $A^\mrm{LL}$ the unmatched shower antenna function (as defined in 
\cite{GehrmannDeRidder:2011dm,LopezVillarejo:2011ap}) and 
the second-order LO matching factors, $R_{4E}$ and $R_{4S}$ 
(for $Z\to qgg\bar{q}$ and $Z\to q\bar{q}'q'\bar{q}$, respectively),
defined as in \eqRef{eq:PME}, but including only the leading-colour
terms in $R^\mrm{LC}_{4E}$. 
For strong ordering, 
similarly to above, the $O_j$ factors can be removed by changing the
integration boundaries of the $\delta A$ terms to $[0,Q_3^2]$. 

Finally, we note that one could in principle 
equally well have defined \eqRef{eq:V3} without
the terms on the third line. The $\dA$ terms in \eqsRef{eq:dAE} and
\eqref{eq:dA}  
would then likewise have to be defined without $P_\mrm{imp}$ and
$P_{\mrm{Ari}}$ factors. However, while this would give a seemingly cleaner
relation with standard fixed-order subtraction, the behaviour of the
(numerical) integrals over the $\dA$ terms would be more difficult, 
due to over-subtraction in the unordered regions. (Showers without
either a strong-ordering condition or a smooth-ordering suppression
greatly overestimate the real-radiation matrix elements in the
unordered region~\cite{Skands:2009tb,Giele:2011cb,LopezVillarejo:2011ap}.)  
Numerically, it is therefore more convenient 
to integrate the contributions represented by the third 
line in \eqRef{eq:V3} analytically, leaving only the suppressed 
terms in \eqRef{eq:dA} to be integrated over numerically. 

To be specific, the numerical integration over the $\dA$ terms is performed by
rewriting the $\dA$ integrals in dimensionless MC form, as:
\begin{eqnarray}
\frac{\alpha_s}{2\pi} C_A \sum_{j=1}^2 \frac{1}{4} \frac{1}{N}
\sum_{i=1}^N \left(\,s_j\,\delta\! A_j(\Phi_i)\,\right)
\label{eq:dAnum}~,
\end{eqnarray}
and similarly for the gluon-splitting terms, with $\Phi_i$ a
number of random (uniformly distributed) antenna phase-space points. 
The common factor 1/4
arises from combining the prefactor $8\pi^2$ above with the area of
the phase-space triangle, $1/2$, and the 
factor $1/(16\pi^2)$ from the phase-space factorization, $\dPS{\mrm{ant}}$.
For smooth
$p_\perp$-ordering with an arbitrary normalization factor $N_\perp$
(so $Q_E^2 = N_\perp p_\perp^2$),
the ordering factors, $O_j$, reduce to: 
\begin{eqnarray}
O_{E}(q_ig_j,\bar{q}_k \to q_ag_bg_c,\bar{q}_k) 
  & = & \frac{y_{jk}}{y_{jk} + x_{ab}x_{bc}} ~,\\[2mm]
O_{E}(q_i,g_j\bar{q}_k \to q_i,g_ag_b\bar{q}_c) & = & 
  \mbox{same with $i\leftrightarrow k$}~,\\[2mm]
O_{S}(q_ig_j,\bar{q}_k \to q_a\bar{q}'_bq'_c,\bar{q}) & = & 
  \frac{N_\perp y_{jk}}{N_\perp y_{jk} + x_{bc}}~,\\[2mm]
O_{S}(q_i,g_j\bar{q}_k \to q_i,\bar{q}_c'q'_b\bar{q}_a) & = & 
  \mbox{same with $i\leftrightarrow k$}~,
\end{eqnarray}
where we have used $y$ with $ijk$ indices for the scaled invariants in
the original $qg\bar{q}$ topology and $x$ with $abc$ indices for the
integration variables in the antenna phase space. 
Note also that the $y$
values are normalized to the full 3-parton CM energy (squared), while
the $x$ values are normalized to their respective dipole CM energies
(squared).

\subsection{The Renormalization Term}
\label{sec:renormalization-term}

A further ingredient to be discussed is the choice of
renormalization scale on both the fixed order and parton shower sides of the calculation,
as these scales are in general chosen differently in both 
sides. Hence a translation term arises at second
order, which must account for this difference, keeping in
mind that, as the scale evolves from one to the other, flavour thresholds
are passed. Our aim is to have the flexibility to use fixed order matrix elements renormalized according
to their usual scheme, while maintaining the freedom to make a
different choice on the shower side. 

The fixed order calculations for $Z$-decay to jets to which we match are 
customarily renormalized in a version of the $\overline{\mathrm{MS}}$ scheme
called the Zero-Mass Variable Flavour Number Scheme (ZM-VFNS). In this scheme
the bare QCD coupling is renormalized as
\begin{equation}
\label{eq:35}
   g_b = \mu^\epsilon g(\mu_R^2)\left[1 + \frac{\alpha_s(\mu_R^2)}{8\pi}
\left\{\left(-\frac{1}{\epsilon}+\gamma_E - \ln 4\pi + \ln\frac{\mu_R^2}{\mu^2}\right)\beta_0 \right\}\right]
\end{equation}
with $\beta_0 = (11 C_A-2n_F)/3 \equiv \beta_0^F $ and $n_F$ is the number of light flavours.
One thus ignores flavours that are heavier
than the scale of the calculation, both in the virtual and
in the real calculations. Once all the UV poles are cancelled, 
one has a running coupling that depends on the number of light
flavours for the scale $\mu_R$ at hand. One then changes the flavour number
when a threshold is crossed. For our present case of $Z$ boson decay to jets
we take $n_F=5$ for $\mu_R$ not too different from the $Z$-boson mass.


Let us be more specific about the matching of $\alpha_s$ across
flavour thresholds. At one loop, 
\begin{equation}
  \label{eq:29}
  \alpha_s^{(n_F)}(\mu_R) = \frac{4\pi/\beta_0^F}{\ln(\mu_R^2/\Lambda_F^2)}\,.
\end{equation}
The value of $\Lambda_F$ depends on the number of active flavours, as follows. 
When passing flavour thresholds the following one-loop matching conditions are imposed
\begin{equation}
  \label{eq:30}
  \alpha_s^{(5)}(m_b) = \alpha_s^{(4)}(m_b), \qquad  \alpha_s^{(4)}(m_c) = \alpha_s^{(3)}(m_c)\,.
\end{equation}
These conditions can be satisfied if  $\Lambda_F$ obeys the matching
conditions
\begin{equation}
\label{eq:38}
 \ln\frac{\Lambda^2_F}{\Lambda^2_{F+1}}= \frac{2}{3\beta_0^F} \ln\frac{m^2_{F+1}}{\Lambda^2_{F+1}} \,.
\end{equation}
With these conditions one can also express $\alpha_s$ values for different flavour numbers into eachother.
E.g. if  $m_c < \mu_R < m_b$, one can express $\alpha_s^{(4)}(\mu_R)$ in terms of 
$\alpha_s^{(5)}(\mu_R)$ by the relation
\begin{equation}
  \label{eq:33}
  \alpha_s^{(4)}(\mu_R) =   \alpha_s^{(5)}(\mu_R) \frac{1}{\frac{\beta_0^4}{\beta_0^5} + 
\left(1-\frac{\beta_0^4}{\beta_0^5} \right) \frac{ \alpha_s^{(5)}(\mu_R)}{ \alpha_s^{(5)}(m_b)}}\,.
\end{equation}
For completeness we briefly review how this $n_F$-dependent UV singularity occurs in the
context of the (inclusive) 3-jet rate, in the case where we only consider massless quarks \cite{Ellis:1980nc,Ellis:1980wv}.
In the virtual contribution, 
the only one-loop diagram for $Z\rightarrow q\bar{q}g$ that is sensitive to the number of flavours is
the quark self-energy correction on the external gluon.  The self-energy diagram itself, being scaleless, is zero in
dimensional regularization. However, renormalization of the coupling
amounts to adding a $n_F$ counterterm on the exteral gluon line proportional to
  \begin{equation}
\label{eq:10}
 n_F \frac{2}{3\epsilon} \left(\frac{\mu^2}{\mu_R^2}\right)^\epsilon \,.
  \end{equation}
The real contribution contributes a $n_F$ dependent (collinear)
$1/\epsilon$ pole as well, from gluon-splitting 
  \begin{equation}
\label{eq:16}
-  n_F \frac{2}{3\epsilon} \left(\frac{\mu^2}{s}\right)^\epsilon \,.
  \end{equation}
In the sum over real and virtual contributions the poles cancel, as
guaranteed by the KLN theorem, leaving a logarithm of the form
\begin{equation}
  \label{eq:15}
 n_F \frac{2}{3} \ln\left(\frac{s}{\mu_R^2}\right) \,.
\end{equation}

On the shower side a related prescription is used, in which the running coupling is 
evaluated at a shower scale $\mu_{\mrm{PS}}$, such that the scale again depends on the number of 
flavours. Depending on the value of $\mu_{\mrm{PS}}$, a corresponding value of $n_F$ is chosen, 
as well as of the QCD scale $\Lambda_F$. This is often different from that for a fixed
order calculation. To give a specific example, matrix elements
will typically be renormalized at a scale characteristic of the total CM
energy, i.e., $\mu_\mrm{ME}^2=s$ an event-independent value, while resummation
arguments imply one best chooses a running scale, such as $\mu_\mrm{PS} = \pT$,
for shower applications~\cite{Amati:1980ch,Catani:1990rr}, which can differ per
event. 

Shifting to a different scale for $\alpha_s$ of a given flavour number
is quite straightforward. Translating from a shower scale
$\mu_{\mrm{PS}}$ to a matrix-element scale $\mu_\mrm{ME}$ amounts to replacing,
for an antenna function 
\begin{equation}
\left.a_{g/q\bar{q}}\right\vert_{\mu_R=\mu_\mrm{PS}} \to
\left(1+\frac{\alpha_s}{2\pi}\frac{11N_C-2n_F}{6}\ln\left(\frac{\mu^2_\mrm{ME}}{\mu^2_\mrm{PS}}\right)
+ {\cal
  O}(\alpha_s^2)\right)\,\left.a_{g/q\bar{q}}\right\vert_{\mu_R=\mu_\mrm{ME}}
~. \label{eq:aSrun}
\end{equation}

A further aspect is that shower Monte Carlos normally
switch to 4-flavour (3-flavour) running for scales $\mu<m_b$ ($\mu <
m_c$), matching the $\alpha_s$ value across the thresholds to obtain a
continuous running. For a consistent treatment, such thresholds must
be taken into account when translating $\alpha_s$ from the shower
scale to the matrix-element one. At one-loop order (which is all that 
is relevant for the NLO expansion), 
this can be done by inserting an
additional term for each flavour threshold in the region
$[\mu_\mrm{PS},\mu_{\mrm{ME}}]$,
\begin{equation}
  + \frac{\alpha_s}{2\pi}\frac{1}{3} 
\ln\left(\frac{m^2_\mrm{thres}}{\mu^2_\mrm{PS}}\right)~,\label{eq:aSthres}
\end{equation}
with $m_\mrm{thres}$ the flavour threshold. 
Physically, \eqRef{eq:aSrun} expresses running with $n_F$ flavours all
the way from $\mu_{\mrm{PS}}$ to $\mu_\mrm{ME}$. The correction term,
\eqRef{eq:aSthres}, 
expresses that the number of flavours was effectively lower 
below each flavour threshold passed on the way.
Note that this can also be
used to account for a larger number of
flavours in the shower calculation, e.g., at scales $\mu_\mrm{PS} >
m_t$, with the sign change of the correction then automatically reflected
by the logarithm.

For coherent parton-shower models, the arguments presented in
\cite{Catani:1990rr} also 
motivate a change to a ``Monte Carlo'' scheme for $\alpha_s$, in which
$\Lambda_{\mrm{QCD}}$ is rescaled, for each $n_F$, by the so-called
CMW factor $\sim 
1.5$ (with some mild flavour dependence), relative to its
$\overline{\mrm{MS}}$ value. If the shower model being matched employs
this scheme, then a further rescaling of the renormalization-scale
argument, $\mu_{\mrm{PS}} \to \mu_{\mrm{PS}}/k_{\mrm{CMW}}$, should be
used in \eqRef{eq:aSrun}, with 
\begin{equation}
k_{\mrm{CMW}} 
\ = \ 
 \exp\left(
  \frac{67-3\pi^2-10n_F/3}{2(33-2n_F)}
 \right)
\ = \ 
 \left\{ 
  \begin{array}{ll}
    1.513 & n_F = 6 \\
    1.569 & n_F = 5 \\
    1.618 & n_F = 4 \\
    1.661 & n_F = 3 
  \end{array}
\right. ~\label{eq:kCMW}
\end{equation}
for $N_C=3$. The translation of renormalization scale (and scheme) yields then an 
additional term to be added to the definition of $V_3$ in \eqRef{eq:V3},
\begin{equation}
V_{3\mu} ~=~  -\frac{\alpha_s}{2\pi}\frac{11N_C - 2 n_F}{6} \ln \left(
\frac{\mu^2_\mrm{ME}}{\mu^2_\mrm{PS}}\right) ~=~
-\frac{\alpha_s}{2\pi}\frac{\beta_0}{2} \ln \left(
\frac{\mu^2_\mrm{ME}}{\mu^2_\mrm{PS}}\right)~
, \label{eq:V3mu}
\end{equation}
plus any additional flavour-threshold correction terms, cf.~\eqRef{eq:aSthres}. 
By inserting these terms, which enter 
at overall order $\alpha_s^2\ln(\mu_\mrm{ME}^2/\mu_\mrm{PS}^2)$, 
the two calculations can
be compared consistently at one-loop accuracy. 

Note that if several different shower paths populate the same fixed-order
phase-space point, then each path will in general be associated with
a distinct $\mu_{\mrm{PS}}$ value. Thus, one $V_{3\mu}$ 
term arises for each shower path, weighted by the relative contribution of each
path to the total. Since for our case there is only one antenna
contributing to $Z\to 
qg \bar{q}$, this particular complication does not arise here.

We finally alert the reader regarding the use of different flavour number
$\alpha_s$'s in the master equation (\ref{eq:V3}). In that equation cancellation
of $1/\epsilon$ divergences take place, already in the first line of
the right hand side. For this cancellation it is important that the
subtraction terms, originating from the shower expansion and 
listed in \eqRef{eq:AGGG}, use $\alpha_s^{(5)}$ renormalized as in the fixed order
calculation.  All subsequent terms in the master equation
are finite, and constitute differences of unordered and strongly ordered
shower based terms, which are also finite, and beyond NLO.

\subsection{Leading-Colour One-Loop Correction for Z $\mathbf{\to}$ 3 Jets}
Combining the results above, in particular \eqsRef{eq:V3},
\eqref{eq:AGGG}, and \eqref{eq:V3mu}, we obtain the complete
expression for the leading-colour\footnote{We use the usual MC
  definition  of leading colour and include terms $\propto C_A$ and
$\propto n_F$ but neglect ones $\propto 1/C_A$.}  
one-loop correction for Z $\mathbf{\to}$ 3 Jets,
\begin{eqnarray}
V_{3Z}(q,g,\bar{q}) & = & \left[\frac{2\Re[M_3^0 M_3^{1*}]}{|M_3^0|^2}\right]^{\mrm{LC}} \ - \
 \frac{\alpha_s}{\pi} \ - \ 
\frac{\alpha_s}{2\pi}\left(\frac{11N_C - 2 n_F}{6}\right) \ln \left(
\frac{\mu^2_\mrm{ME}}{\mu^2_\mrm{PS}}\right)
 \nonumber\\[1mm]
& & + \ \frac{\alpha_sC_A}{2\pi} \ \Bigg[ 
-2 I_{qg}^{(1)}(\epsilon,\mu^2/s_{qg}) -
2I_{qg}^{(1)}(\epsilon,\mu^2/s_{g\bar{q}}) 
+\frac{34}{3}\Bigg] \nonumber\\[1mm]
& & + \ \frac{\alpha_s n_F}{2\pi}\Bigg[
  -2I_{qg,F}^{(1)}(\epsilon,\mu^2/s_{qg}) -
  2I_{g\bar{q},F}^{(1)}(\epsilon,\mu^2/s_{g\bar{q}}) - 1 \Bigg] \nonumber \\[1mm]
& & + \ \frac{\alpha_s C_A}{2\pi} \Bigg[ 
  \ 8\pi^2\! \int_{Q_3^2}^{m_Z^2} \dPS{\mrm{ant}}
    \ A^\mrm{std}_{g/q\bar{q}} \ + \ 8\pi^2\!\int_{Q_3^2}^{m_Z^2}
    \dPS{\mrm{ant}} \ \dA_{g/q\bar{q}} 
   \nonumber\\[1mm]
  & & \qquad\quad - \  
    \sum_{j=1}^2 8\pi^2\! \int_{0}^{s_{j}}
        \dPS{\mrm{ant}} \ (1-O_{Ej}) \ A_{g/qg}^\mrm{std}
      \ + \ \sum_{j=1}^2 8\pi^2\!\int_{0}^{s_{j}} 
          \dPS{\mrm{ant}} \ \dA_{g/qg}\Bigg] \nonumber \\[1mm]
& & + \ \frac{\alpha_s n_F}{2\pi}\Bigg[
 -\sum_{j=1}^2 8\pi^2\ P_{Aj}\!  \int_{0}^{s_{j}} \dPS{\mrm{ant}} (1-O_{Sj})
 \  A^\mrm{std}_{\bar{q}/qg} 
\ + \ \sum_{j=1}^2 8\pi^2\! \int_{0}^{s_{j}} \dPS{\mrm{ant}} \ 
  \dA_{\bar{q}/qg} \nonumber\\[1mm] 
  & &  \qquad\quad - \frac{1}{6}
  \ \frac{s_{qg} -
    s_{g\bar{q}}}{s_{qg}+s_{g\bar{q}}}\ln\left(\frac{s_{qg}}{s_{g\bar{q}}}\right)
  \Bigg] ~, \label{eq:V3Z}
\end{eqnarray}
where:
\begin{itemize}
\item the first line contains the full (leading-colour) one-loop
  matrix element, the $V_{2Z}$ correction from one-loop matching at
  the preceding order, and the $V_{3\mu}$ term from the choice of
  shower renormalization scale;
\item the second line contains the standardized subtraction term
  arising from the $qg\to qgg$ and $g\bar{q}\to gg\bar{q}$ antennae;
\item the third line contains the standardized subtraction term
  arising from the $qg \to q\bar{q}'q'$ and $g\bar{q}\to
  \bar{q'}q'\bar{q}$ antennae;
\item the fourth to last lines contain the terms arising from the
  difference between the (matched) shower evolution and the
  standardized subtraction terms, including the consequences of
  ordering choices and modification factors such as those arising from
  the Ariadne factor and from matching to the LO matrix elements. 
\end{itemize}
We denote the singular subtracted 1-loop matrix element by \SVirtual

\begin{align}
\SVirtual = & \left[\frac{2\Re[M_3^0 M_3^{1*}]}{|M_3^0|^2}\right]^{\mrm{LC}} + \ \frac{\alpha_sC_A}{2\pi} \ \Bigg[ 
-2 I_{qg}^{(1)}(\epsilon,\mu^2/s_{qg}) -
2I_{qg}^{(1)}(\epsilon,\mu^2/s_{g\bar{q}}) 
+\frac{34}{3}\Bigg] \nonumber \\
& + \ \frac{\alpha_s n_F}{2\pi}\Bigg[
  -2I_{qg,F}^{(1)}(\epsilon,\mu^2/s_{qg}) -
  2I_{g\bar{q},F}^{(1)}(\epsilon,\mu^2/s_{qg}) - 1 \Bigg] \label{eq:SVirtual}
\end{align}

In \secRef{sec:integrals}, we compute the analytical integrals
corresponding to each of the shower-generated terms, for different choices of 
evolution variable, ordering criterion, and antenna functions. 

With the one-loop matrix element expressed as in
\appRef{sec:leading-colour}, it is easy to see that  the infrared singularity
operators in \eqRef{eq:SVirtual} cancel, leaving only explicitly finite 
remainders (which may still contain logarithms of resolved scales). 
This then constitutes the description of the one-loop matching for $Z\rightarrow 3$
jets, having already discussed the case for two jets. In the context
of \eqRef{eq:5} we have now corrected the first two terms on the rhs
to NLO accuracy. 

We round off with a few remarks on the normalizations of the various
 $Z\to n$-parton rates that are obtained by
our procedure, since this is a point on which the various 
approaches to multileg NLO corrections differ. 
We make the following
observations: 
\begin{enumerate}
\item \emph{The total inclusive $Z$ decay rate:}
the matrix-element correction scheme derived in this paper
  maintains a strict unitarity between the real and virtual
  corrections that are applied beyond Born level. An important
  consequence is that the total 
inclusive $Z$ decay rate 
is not changed by switching on the $V_{3Z}$
correction\footnote{The theoretically most sensible choice would be to
normalize the inclusive $Z\to2$-parton rate to the 
full NNLO result, but at the level we work at, one could equally
well normalize to NLO or to data. In either case, the total normalization of
the generated sample is left unchanged by the $V_{3Z}$ correction.}. 
\item \emph{The inclusive $Z\to 3$ jets rate:} both the virtual
  (one-loop) correction to the 3-jet rate and the real (tree-level)
  correction to the 4-jet rate are included here. Hence the inclusive
  3-jet rate is NLO correct. Without these corrections, it would only
  be LO correct. Thus, the 3-jet inclusive rate \emph{does}
  change when switching on the $V_{3Z}$ term\footnote{In
    \secRef{sec:LEP} below (comparisons to LEP measurements), 
this is seen most easily in
  \figRef{fig:LEP1}, where the ``(NLO off)'' curves undershoot the
  ``(NLO on)'' ones, for observable regions dominated by 3 or more jets.}.
\item \emph{The exclusive $Z\to 2$ jets rate:} The strict unitarity
  imposed by our correction method implies that every 3-jet event
  begins life as a 2-jet one. Since the the $V_{3Z}$ 
term modifies the probability for a
  2-jet event to evolve to become a 3-jet one, at the ${\cal
    O}(\alpha_s^2)$ level, the 2-jet exclusive
  rate receives an equal and opposite correction. This
  represents an ${\cal O}(\alpha_s^2)$ ambiguity on the exclusive
  2-jet rate, which is not adressed in our paper (though it could be
  removed by normalizing to the full NNLO result for $Z\to 2$, cf.~
  the inclusive 2-jet rate above).
\item \emph{The exclusive $Z\to 3$ jets rate:} for a given 3-parton
  configuration, the evolution to 4 partons and beyond is not
  changed by $V_{3Z}$ (though it \emph{is} changed by the application of the
  4-parton tree-level corrections, which we take to be included
  throughout this paper). Thus, while switching $V_{3Z}$ on
  does change the total amount of 3-jet events (cf.~the inclusive
  3-jet rate above), it does not directly change the fraction of those
  events which will develop a fourth or more jets. 
\end{enumerate}

\subsection{One-Loop Correction for Born + 2 Partons}

To illustrate how the formalism presented here generalizes to higher
multiplicities, we take the case of the
NLO correction to $Z\to 4$ partons. For simplicity, however, we
continue to restrict our 
 analysis of 
the correction factor to the leading-colour level.
At NLO, the exclusive $Z\to4$ partons rate at ``infinite''
perturbative resolution (similarly to above) is
\begin{equation}
\mbox{Exact} \ \to \ |M_{4}^0|^2 + 2\mrm{Re}[M_4^0M_4^{1*}]~.
\end{equation}

Labeling the 4 partons by $Z\to i,j,k,\ell$, there are two possible
antenna-shower histories leading to each 4-parton configuration, with 
$j$ and $k$ the last emitted parton, respectively. Those two
contributions both enter in the definition of the tree-level 4-parton
matching factor,
\begin{equation}
R_4 = \frac{|M_4^0(i,j,k,\ell)|^2}{A_{j/IK}|M_3^0(I,K,\ell)|^2 +
  A_{k/JL}|M_3^0(i,J,L)|^2} ~, 
\end{equation}
such that their sum reproduces the full 4-parton matrix element. Note
that a separate such factor is applied to $Z\to qgg\bar{q}$ and $Z\to
q\bar{q}'q'\bar{q}$, and that we have suppressed colour and coupling
factors here, for compactness (we ignore the small, non-singular
extra interference terms for the special case where all four quarks
have the same flavour). The
antenna functions, $A$, are understood to include all such factors, as
well as any $P_\mrm{imp}$ and
$P_\mrm{ari}$ factors appropriate to the branchings at hand. For a
general $n$-parton matrix element, the denominator contains one
term for each possible clustering. 

Labeling the $IK\to ijk$ history by $A$ and the $JL\to jk\ell$ one by
$B$, the sum over the two histories yields
\begin{equation}
R_4\Delta_4(Q_4,0)\sum_{\alpha \in A,B}A^\alpha_{3\to 4}|M_{3}^\alpha|^2 
  \Delta_2(m_Z^2,Q_3^\alpha)\Delta_3(Q_3^\alpha,Q_4^\alpha)
  \prod_{m=2}^{3}(1+V_m^\alpha) ~,
\end{equation}
where it is understood that $\alpha$ is an index, not a power, and 
the last product factor takes into account the NLO matching at
the preceding multiplicities. Expanding the Sudakov factors to first
order and using the definition of the NLO correction factor at the
preceding multiplicity, \eqRef{eq:V3formal}, this becomes
\begin{equation}
R_4 \left(1 - \sum_k\int_0^{s_k} \!\!\dPS{\mrm{ant}} R_5\, A_{4\to
    5}\right) \!\sum_{\alpha \in A,B}\!\!\!A^\alpha_{3\to 4}|M_{3}^\alpha|^2 
 \left[1 + \frac{2\mrm{Re}[M_3^0M_3^{1*}]^\alpha}{|M_3^\alpha|^2} + \sum_j
   \int_0^{Q_4^\alpha} \!\!\!\dPS{\mrm{ant}} A_{3\to 4}^\alpha\right],
\end{equation}
which we can rewrite as
\begin{eqnarray}
 & &   |M_4|^2 \left(1 - \sum_k\int_0^{s_k} \!\!\dPS{\mrm{ant}} R_5\, A_{4\to
    5}\right) \nonumber\\ 
 & & + \ R_4 \! \sum_{\alpha \in A,B}\!A^\alpha_{3\to 4}|M_{3}^\alpha|^2 \left(
  \frac{2\mrm{Re}[M_3^0M_3^{1*}]^\alpha}{|M_3^\alpha|^2} + \sum_j
   \int_0^{Q_4^\alpha} \!\dPS{\mrm{ant}} A_{3\to 4}^\alpha
\right)
~,
\end{eqnarray}
where we again emphasize that the antenna functions are understood to
include all relevant coupling, $P_\mrm{imp}$, and $P_\mrm{ari}$ factors. The
first term represents the new subtraction that the shower 
generates at 4 partons, while the second represents part of the NLO
correction to the preceding multiplicity. For one of the histories
(the one followed by the ``current'' event),
this correction has already been evaluated and can be reused. The
contribution from the other history will have to be recomputed,
however. In general, there will be one subtraction to perform 
at the $n$-parton level, and there will be $m\sim n-n_\mrm{Born}-1$ 
new subtractions that
have to be done at the $(n-1)$-parton level, in addition to the one
that was already done during the evolution of the current
event. 

Clearly, there is an undesirable scaling behavior built into
this, which will make NLO matching at many partons quite computing
intensive. An alternative, which eliminates the sum over histories, is
that of sector showers, see e.g.,
\cite{Larkoski:2009ah,LopezVillarejo:2011ap}. Though this 
is not the main avenue pursued in this paper, we nevertheless give some 
comments below on how a sector-based NLO matching scheme could be
constructed. 

\subsection{One-Loop Matching for Sector Showers}

The matching conditions derived above may also be applied to so-called
sector showers~\cite{Larkoski:2009ah,LopezVillarejo:2011ap},
with a few relatively minor modifications. The expansion of the
Sudakov factors generated by the LO matched shower will now contain
integrals over ratios of matrix elements (which are the LO matched
sector antenna functions), multiplied by sector vetos. The presence of
the sector vetos makes analytical phase-space integration more
difficult.

However, since the sector approach merely represents a different way of
decomposing the same singularities as the global one, we may
effectively recycle the integrals carried out for the global case by 
adding and subtracting the terms produced by a smoothly ordered
``standard'' shower (i.e., using the GGG functions). The first four
lines of \eqRef{eq:V3} then remain 
unchanged. The definition of the $\dA$ terms in the last line,
however, changes to
\begin{equation}
\sum_{j=1}^2\int_0^{s_j} \dPS{\mrm{ant}} \ g_s^2 C_A
\left(\Theta_j^{\mrm{sct}} A_{g/qg}^\mrm{sct} - A_{g/qg}^\mrm{std}\right) ~,\label{eq:intSct}
\end{equation}
for the terms arising from the $qg\bar{q}\to qgg\bar{q}$ Sudakov
factor, with analogous ones arising from the gluon-splitting
contributions. The step function, $\Theta_j$, represents the sector
veto applied to the sector antenna functions. The sector antenna
function, up to the tree-level matched orders, is just
\begin{equation}
g_s^2 {\cal C} \ A^{\mrm{sct}} \ = \ \frac{|M^0_n|^2}{|M_{n-1}^0|^2}~.
\end{equation}

Since the individual sector and global antenna functions have
different singularity structures (they are only guaranteed to have the
same singularities at the summed level), the integrals in
\eqRef{eq:intSct} are divergent, and cannot be carried out
numerically. In order to obtain numerically convergent integrals, we
must divide up the contributions of the global terms onto each sector,
and perform a set of correlated integrals in which the singularities
explicitly cancel in the divergent limits,
\begin{eqnarray}
\to &  & \int_0^{s_{qg}} \dPS{\mrm{ant}} \ g_s^2 C_A \ 
\Theta_1^{\mrm{sct}} \left( A_{g/qg}^\mrm{sct} -
A_{g/qg}^\mrm{std}\right) - \int_0^{s_{g\bar{q}}} \dPS{\mrm{ant}} \ g_s^2 C_A \ 
\Theta_1^{\mrm{sct}} A_{g/g\bar{q}}^\mrm{std} \nonumber\\
& + & 
 \int_0^{s_{g\bar{q}}} \dPS{\mrm{ant}} \ g_s^2 C_A \ 
\Theta_2^{\mrm{sct}} \left( A_{g/g\bar{q}}^\mrm{sct} -
A_{g/g\bar{q}}^\mrm{std}\right) - \int_0^{s_{qg}} \dPS{\mrm{ant}} \ g_s^2 C_A \ 
\Theta_2^{\mrm{sct}} A_{g/qg}^\mrm{std}~,\label{eq:intSct2}
\end{eqnarray} 
where each line now corresponds to the sum of contributions to a
single sector, for which the difference between sector and global
antennae is finite. The individual integrals are of course still
divergent, but they can now be treated numerically by collecting the
terms on each line under a single integral sign. Analytically, this is
complicated since the two integrals on each line are not associated with
the same kinematics map\footnote{In the example here, for $Z\to 3$
  partons, one term is associated with branchings in the $qg$ antenna,
  while the other is 
  associated with branchings in the $g\bar{q}$ one.}.
Numerically, however, we may still ensure a
point-by-point cancellation in the singular limits by keeping the two
integrals formally separate, but carrying them out simultaneously, in
a correlated way, as follows.

For each antenna, generate a random uniformly distributed phase-space
point, 
\begin{equation}
[y_{ij},\ y_{jk},\ \phi]~,\label{eq:sample1}
\end{equation}
and evaluate the first term in \eqRef{eq:intSct2}. (If the point is
outside the relevant sector, this term is zero for the time being). 
If the pair $(jk)$, say, corresponds to a sector
shared with a neighboring antenna (as in our example), check whether
the correlated phase-space point defined by 
\begin{equation}
[y_{jk},\ 1-y_{jk}-y_{ij},\ \phi]~, \label{eq:sample2}
\end{equation}
in the neighboring antenna passes the same sector veto as before, 
and if so, subtract the global term corresponding to the second term on the
first line of \eqRef{eq:intSct2}. The sampling represented by
\eqRef{eq:sample2} is uniform, as long as that used to generate the
original point, \eqRef{eq:sample1}, is uniform. The
replacement of $y_{ij}$ by $1-y_{jk}-y_{ij}$ corresponds precisely to
the mapping $z\to 1-z$ in the collinear limit, which is what is
required to reestablish a point-by-point cancellation of the sector
and global singularities. 


\section{Sudakov Integrals}
\label{sec:integrals}

In this section, we work out the standardized Sudakov integrals
appearing in the second and third line of \eqRef{eq:V3}, for each choice of
evolution variable. 
We also study the soft and collinear limits of the Sudakov integrals
and compare them to those of the one-loop matrix
element. This provides an explicit check of whether the first-order
expansion of the Sudakov factors generates the correct logarithms
present in the fixed-order calculation.

Given our choice of the GGG antenna functions as our standard ones, 
the relevant terms are
\begin{equation}
g_s^2 \left[ C_A\int_{Q_3^2}^s \limits a_3^0 \:\dPS{\mrm{ant}}-\sum_{j=1}^2 C_A\int_0^{s_j} (1-O_{E_j})\,d_3^0\: \dPS{\mrm{ant}} -\sum_{j=1}^2 2\,T_R\,n_F P_{A_j} \int_0^{s_j} (1-O_{S_j})\,e_3^0\: \dPS{\mrm{ant}} \right] \\
\label{eq:SudakovInt1}
\end{equation}
The general form of the first term, which originates from 
the $2\rightarrow 3$ branching step, is 
\begin{equation}
  \label{eq:13}
g_s^2 C_A\int_{Q_3^2}^s \limits a_3^0 \:\dPS{\mrm{ant}}=\frac{\alpha_s C_A}{2\pi}\left(\sum_{i=1}^{5} K_i I_i(s,Q_3^2)\right)
\end{equation}
where the definitions for the $K_i$ and the $I_i$ functions are given 
in \appRef{sec:antenna-integrals}, for each type of antenna function
and ordering variable. Their derivation and soft/collinear structure 
will be discussed more closely below, for each choice of ordering and
evolution variable.
The form of the $3\to 4$ integrals depends on whether we work in the
context of strong or smooth ordering. We shall now 
consider each of those cases in turn, beginning with strong ordering.

\subsection{Strong Ordering}

For strong ordering, the inverted ordering conditions 
in \eqRef{eq:V3}, $(1-O_{E_j/S_j})$, reduce to
step functions expressing integration over the unordered region. 
The integration surface is thus limited from below by the phase-space
contour defined by the evolution scale of the first branching, $Q^2$,
and from above by the edge defined by the invariant mass of the antenna.

The expression generated by the $3\rightarrow
4$ splitting case for gluon emission is
\begin{equation}
\label{eq:17}
-g_s^2\sum_{j=1}^2 C_A\int_0^{s_j} (1-O_{E_j})\,d_3^0\: \dPS{\mrm{ant}} =  -  \frac{\alpha_s C_A}{2\pi}\left(\sum_{i=1}^{5} K_i I_i(\sqg,Q_3^2)\right)\,-\frac{\alpha_s C_A}{2\pi}\left(\sum_{i=1}^{5} K_i I_i(\sgq,Q_3^2)\right).
\end{equation}
where $K_i$ and $I_i$ are the same as those for the $2\to 3$ term
above, though they here appear with different arguments. 
The remaining case is the $3\to 4$ gluon splitting defined by   
\begin{equation}
\label{eq:39}
 -g_s^2\sum_{j=1}^2 \,n_F P_{A_j} \int_0^{s_j} (1-O_{S_j})\,e_3^0\: \dPS{\mrm{ant}} 
=  -\frac{\alpha_s n_F}{2\pi}P_{A_{\qg}}H(\sqg,Q_3^2)\,-\frac{\alpha_s n_F}{2\pi}P_{A_{\gq}}H(\sgq,Q_3^2).
\end{equation}
with $H$ defined in  \appRef{sec:antenna-integrals} and $P_{A_j}$ as
defined in \eqRef{eq:Ariadne}. 
We will discuss the derivation
of these terms in more detail in the following three subsections, for
strong $m_D$-, $p_\perp$, and energy-ordering, respectively. 

\subsubsection{Dipole Virtuality}
\label{sec:dipole-virtuality}

We begin with dipole virtuality as evolution variable, which is
perhaps the simplest case.  We start by repeating the integrals of
\eqRef{eq:V3} with the one-particle phase space defined as in
\eqRef{eq:PhiAnt}.  In the case of dipole virtuality the contours are
triangular (\figRef{fig:mDlin}). We recall that, 
for the $g\to q\bar{q}$ terms, it is the $q\bar{q}$ invariant mass
that is used as evolution variable, regardless of what choice is made
for gluon emissions. The $m_D$ scale of the previous emission still
enters, however, since that defines the ordering scale applied to both
emissions and splittings. The explicit forms of the 
terms in \eqRef{eq:SudakovInt1} are:
\begin{align}
&=\frac{\alpha_s }{4\pi} \left[ \frac{C_A}{s} \hspace{-2ex}\int\limits_{\min(s_{qg},s_{g\bar{q}})}^{s-\min(s_{qg},s_{g\bar{q}})} \, \hspace{-5ex}\d{s_{1}}\hspace{1ex}\int\limits_{\min(s_{qg},s_{g\bar{q}})}^{s-s_1}  \hspace{-2ex}\d{s_2} \ 
 a_3^0(s_1,s_2) \right.  \nonumber \\
&- \left\{ \frac{C_A}{\sgq} \Theta\left( \sgq-2\sqg \right)\hspace{-2ex} \int\limits_{\sqg}^{\sgq-\sqg} \hspace*{-2ex}\d{s_{1}}\hspace{-1ex}\int\limits_{\sqg }^{\sgq-s_1}  \hspace*{-1ex}\d{s_2} 
    + \frac{C_A}{\sqg}\Theta\left( \sqg-2\sgq \right) \hspace{-2ex}\int\limits_{\sgq}^{\sqg-\sgq} \hspace*{-2ex}\d{s_{1}} \hspace{-1ex}\int\limits_{\sgq}^{\sqg-s_1}  \hspace*{-1ex}\d{s_2} \
   \right\} d_3^0(s_1,s_2) \nonumber \\ 
 &-  \left. \left\{ \frac{n_F}{\sqg} \Theta\left( \sqg-\sgq \right)P_{A_1} \int\limits_{\sgq}^{\sqg} \d{s_{1}}\int\limits_{0}^{\sqg-s_1}  \hspace{-1ex}\d{s_2}  
+ \frac{n_F}{\sgq}\Theta\left( \sgq-\sqg \right) P_{A_2}  \int\limits_{\sqg}^{\sgq} \d{s_{1}}\hspace{-1ex} \int\limits_{0 }^{\sgq-s_1}  \hspace{-1ex}  \d{s_2}   \right\} e_3^0(s_1,s_2) 
 \right]   \label{eq:md_tot}~,
\end{align}
with $P_{A_j}=\frac{2\sqg}{\sqg+\sgq}$ and
$P_{A_2}=\frac{2\sgq}{\sqg+\sgq}$ as defined in \eqRef{eq:Ariadne} and
the gluon-splitting antenna $e_3^0$ has its singularities in
$s_1$. 

For compactness, we only show the integration for the double-pole 
(soft-collinear eikonal) terms present in both $a_3^0$ and $d_3^0$
here, which are the only sources of transcendentality-2 terms. The full 
antenna integrals, including also the lower-transcendentality terms
originating from single poles and finite terms,
are given in \appRef{sec:antenna-integrals}. The $T=2$ part of the
$a_3^0$  integral is 
\begin{equation}
  \label{eq:4}
\frac{\alpha_s C_A}{4\pi}\left[\int_{\min(s_{qg},s_{g\bar{q}})}^{s-\min(s_{qg},s_{g\bar{q}})} \hspace*{-2mm}
\d{s_{1}}\int_{\min(s_{qg},s_{g\bar{q}})}^{s-s_1}  \hspace*{-3mm}\d{s_2} \ 
 \frac{2}{s_{1}s_{2}} \right]~\,.
\end{equation}
To evaluate this expression, we first rewrite it in a dimensionless
form in terms of the rescaled integration variables $y_i =
s_i/(s-\frac{1}{2}Q_3^2)$, with upper boundary $1$ and lower boundary
\begin{equation}
  \label{eq:7}
  \xi_{\min}  =   \frac{\min(s_{qg},s_{g\bar{q}})}{s - \min(s_{qg},s_{g\bar{q}})} ~.
\end{equation}
The integration is over a triangular surface. The
lower integration boundary cuts off the evolution variable at
the value of the $3$-parton evolution scale. The other boundary is
determined by the total energy of the dipole before branching which
here is $\sqrt{s}$. We use the integral 
\begin{equation}
\int_{x}^{1} 
\frac{\d{y}}{y} \ln\left(\frac{1-y+x}{x}\right) 
=
\ln^2(x)-\ln (x)\ln\left(1+x\right)-\Li\left(\frac{1}{1+x}\right)
+ \Li\left(\frac{x}{1+x}\right)~.
\end{equation}
to obtain
\begin{align}
\label{eq:8}
 \frac{\alpha_s C_A}{2\pi} \Bigg[ 
& \ln\left(\frac{s}{\min(s_{qg},s_{g\bar{q}})}\right)\ln\left(\frac{s-\min(s_{qg},s_{g\bar{q}})}{\min(s_{qg},s_{g\bar{q}})}\right) \nonumber \\
 &-\Li\left(\frac{s-\min(s_{qg},s_{g\bar{q}})}{s}\right)
 +\Li\left(\frac{\min(s_{qg},s_{g\bar{q}})}{s}\right)\Bigg]~.
\end{align}

To discuss the $3\rightarrow 4$ Sudakov terms, let us for definiteness
assume that we are in a 3-parton phase-space point with 
$s_{qg} > s_{g\bar{q}}$. The opposite case is symmetric.
Again we only include the $T=2$ terms explicitly here, with the
details of the full antenna integrals relegated to
\appRef{sec:antenna-integrals}. 
\begin{equation}
\label{eq:6}
\frac{\alpha_s C_A}{4\pi}\left[\ \int_{s_{g\bar{q}}}^{s_{qg}-s_{g\bar{q}}} \hspace*{-2mm}
\d{s_{1}}\int_{s_{g\bar{q}}}^{s_{qg}-s_1}  \hspace*{-3mm}\d{s_2} \
 \frac{2}{s_{1}s_{2}} \right]~\,.
\end{equation}
The integration is again over a triangular surface. The total
energy of the dipole before branching is now $s_{qg}$.
The integral in~\eqRef{eq:6} corresponding 
to the sum over antenna integrals only contains one
$d_3^0$ integral because the other has equal upper and lower
integration boundaries. Note that this integral actually
vanishes for $s_{qg}\le Q_3^2$, which amounts to the
dipole-virtuality ordering allowing the $3\to 4$ branchings to
populate their full respective phase spaces (i.e. no correction term is
necessary).

Focusing on the case $s_{qg} > 2\sgq$ for which the second
integral is nonvanishing (which now amounts to the ordering condition
imposing a nontrivial restriction on the $3\to 4$ phase space, see
\figRef{fig:mdlinOrd}), we obtain, including the $2\to 3$ term
\begin{eqnarray}
\label{eq:18}
& & \frac{\alpha_s
    C_A}{4\pi}\left[\int_{\xi_{\min}}^{1} \hspace*{-2mm} 
\d{y_{1}}\int_{\xi_{\min}}^{1-y_1+\xi_{\min}}  \hspace*{-2mm}\d{y_2}
\ 
 \frac{2}{y_{1}y_{2}} 
\ - \ \int_{\xi_{\min}'}^{1} \hspace*{-2mm}
\d{y'_{1}}\int_{\xi_{\min}'}^{1-y'_1+\xi_{\min}'}  \hspace*{-3mm}\d{y'_2} \
 \frac{2}{y'_{1}y'_{2}} \right]~,
\end{eqnarray}
 $y_i' = s_i/(\sqg-s_{g\bar{q}})$ and boundaries 
\begin{eqnarray}
\xi_{\mrm{min}}' & = &\frac{\sgq}{\sqg-\sgq}~.
\end{eqnarray}
with lower-transcendentality terms again available in appendix \ref{sec:antenna-integrals}.
For the mirror case $\sgq>2\sqg$ essentially symmetric expressions
are obtained, while for the intermediate cases in which the two invariants
are within a factor 2 of each other, the second integral in
eq.~\eqRef{eq:18} simply vanishes. 

The full double-logarithmic term from the expanded Sudakov terms in
\eqRef{eq:md_tot}, for strong ordering in
dipole virtuality, is then
\begin{eqnarray}
& & \frac{\alpha_s C_A}{2\pi} \Bigg[ 
 \ln\left(\frac{s}{\frac{1}{2}Q_3^2}\right)\ln\left(\frac{s-\frac{1}{2}Q_3^2}{\frac{1}{2}Q_3^2}\right)
 -\Li\left(\frac{s-\frac{1}{2}Q_3^2}{s}\right)
 + \Li\left(\frac{\frac{1}{2}Q_3^2}{s}\right)
  \label{eq:sudakovDilogsType2}\\[2mm]
& & \quad +\Theta\left(\smax-Q_3^2\right)\left(
-\ln\left(\frac{\smax}{\frac{1}{2}Q_3^2}\right)\ln\left(\frac{\smax-\frac{1}{2}Q_3^2}{\frac{1}{2}Q_3^2}\right)
+\Li\left(\frac{\smax-\frac{1}{2}Q_3^2}{\smax}\right) 
- \Li\left(\frac{\frac{1}{2}Q_3^2}{\smax}\right)
 \right) \Bigg]~,\nonumber
\end{eqnarray}
where the $\Theta$ function ensures that the second term is only
active if 
\begin{equation}
\smax = \max(\sqg,\sgq) > 2\min(\sqg,\sgq) = Q^2_3~,
\end{equation}
so that the expression applies over all of phase space. 

We shall now consider the infrared limits of this result, and compare them to
those of the one-loop matrix element. For this comparison
we keep only terms that involve logarithms of the invariants. 
The soft limit corresponds to vanishing $Q_3^2$  ($\ximin\rightarrow
0$). The first 
line of \eqRef{eq:sudakovDilogsType2} represents the contribution of
the $2\to 3$ expanded Sudakov. To find the contribution in the soft
limit, we choose to approach the limit along the diagonal of the
phase space triangle. Parametrizing this by 
$\sqg/s=\sgq/s\rightarrow y$ we find for this term 
\begin{align*} 
\ln^2(y)-\frac{\pi^2}{6} \,.
\end{align*}
The contributions of the $3 \to 4$ Sudakovs in the soft limit are
examined in two separate cases corresponding to the two regions in
\figRef{fig:mdlinOrd}. In the first case given by $s_{\max}< 2
s_{\min}$, corresponding to the light grey area in the figure, the
step function in \eqRef{eq:sudakovDilogsType2} yields zero. In the
second case given by $s_{\max}> 2 s_{\min}$, corresponding to the dark
grey area in the figure, the step function is equal to one. The double
logs and dilogarithms now yield a finite contribution that does not
diverge in the soft limit. We can understand this by parametrizing the
soft limit by $\lambda$ 
\begin{align}
\sqg = \lambda s \hspace{5ex} \sgq =p \lambda s \hspace{5ex} s_1'=\lambda \kappa s \hspace{5ex} s_2'=\lambda \mu s \hspace{5ex} p>2~,
\end{align} 
so that the integral becomes
\begin{align}
\int_{s_{\min}}^{s_{\max}-s_{\min}} \d s_1 \int_{s_{\min}}^{s_{\max}-s_1} \d s_2 \frac{1}{s_1s_2} \rightarrow \int_1^{p-1} \d \kappa \int_1^{p-\kappa} \d \mu \frac{1}{\kappa \mu} ~.
\end{align}
This implies that the integration variable scales with the integration
limits and is independent of the soft limit. We can also expect this
behaviour from examing \figRef{fig:mDlin}. The shape of the different
regions does not change for different values of $Q_3^2$, in
contrast with the case of transverse momentum, as we will see below. 

After the poles cancel in \eqRef{eq:V3Z}, 
the pole-subtracted version of the one-loop matrix element, \SVirtual,
defined in \eqRef{eq:SVirtual}, contains all the relevant terms on the
matrix-element side.    
The transcendentality-2 terms of \SVirtual\ are given by
\begin{equation} 
-R(y_1,y_2) = \mrm{Li}_2\left(y_1\right) + \mrm{Li}_2\left(y_2\right)
- \frac{\pi^2}{6} - \ln y_1 \ln y_2 + \ln y_1 \ln(1-y_1) + \ln y_2
\ln(1-y_2) ~.\label{eq:Rfun}
\end{equation}
Including the transcendentality-1 terms (see \appRef{app:FO}),
taking the soft limit by sending $\sqg/s=\sgq/s=y\rightarrow
0$, and keeping only logarithmic terms, the pole-subtracted matrix
element (ME) reduces to  
\begin{align}
\mbox{ME:} \hspace{1cm} \sqg/s=\sgq/s=y \rightarrow 0 \hspace{1cm} &\frac{\alpha_s C_A}{2\pi}\left[ -\ln^2(y)-\frac{10}{3}\ln(y) \right] +  \frac{\alpha_s n_F}{6\pi}\ln(y),  \label{eq:loopSoftLimit} 
\end{align}
The single log proportional to $C_A$
originates from the renormalization term and the single log of the
closed quark loops (proportional to $n_F$) arises due to the
definition of the infrared singularity operator, defined in the
appendix in \eqRef{eq:IRnf}. 

Taking the same limit of the Sudakov
integrals for dipole virtuality \eqRef{eq:md_tot}, but omitting for
the time being the renormalization term, $V_{3\mu}$, we find for the
parton shower (PS),  
\begin{align}
\label{eq:sudSoftLimit}
-\mbox{PS:}\hspace{1cm}\sqg/s=\sgq/s=y\to 0 \hspace{1cm} &\frac{\alpha_s C_A }{2\pi}\left[ \ln^2y + \frac{3}{2}\ln(y) \right]~.
\end{align}

\begin{figure}[t!]
\centering
\newlength{\skipper}
\setlength{\skipper}{2mm}
\begin{tabular}{ccc}
\includegraphics*[scale=0.47]{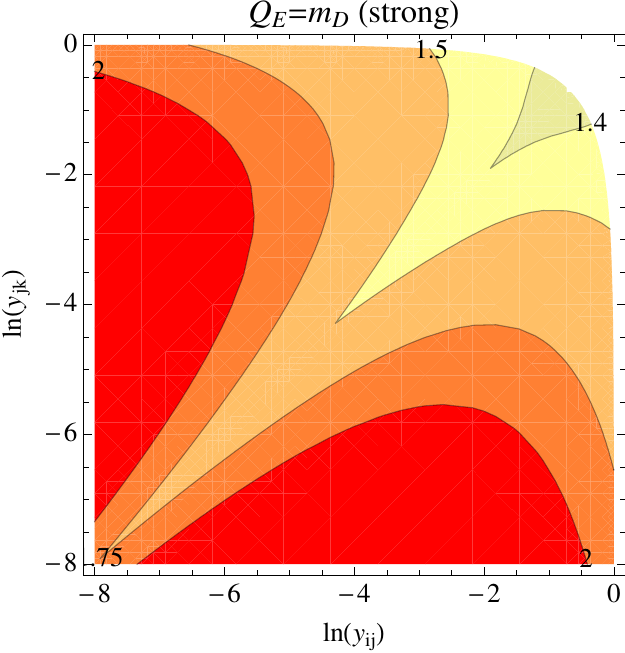}
&\includegraphics*[scale=0.47]{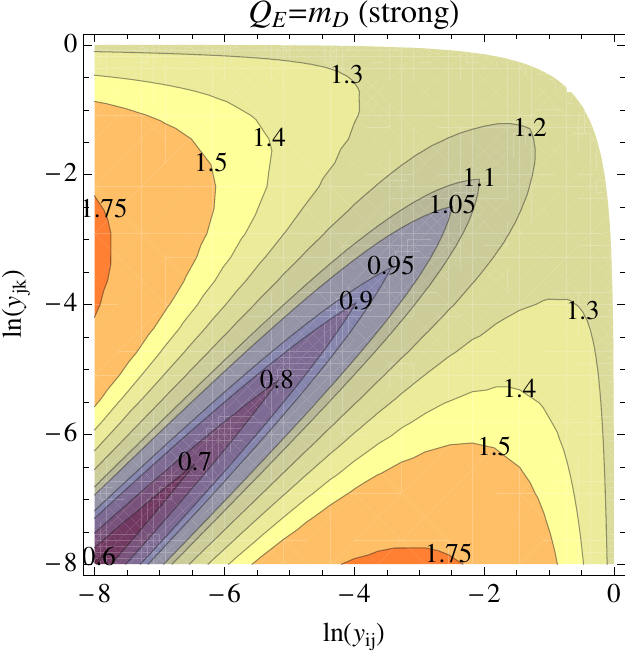}
& 
\includegraphics*[scale=0.47]{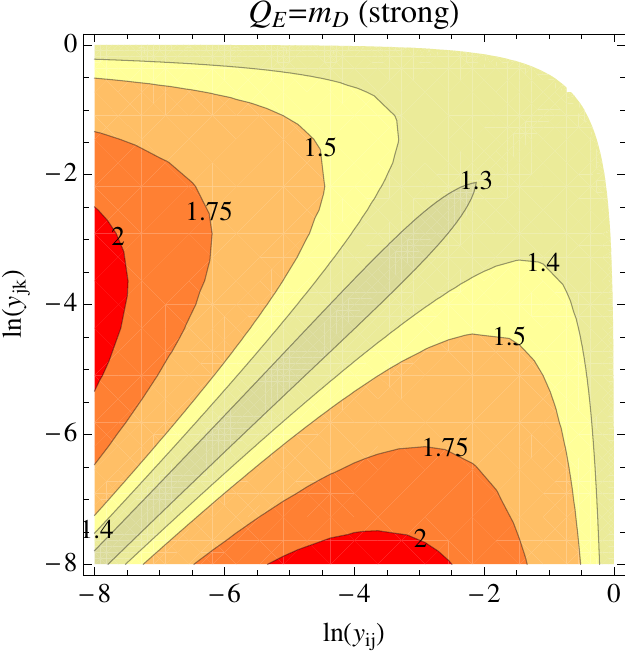} \\[-1mm]
(a) $\mu_\mrm{PS}=\sqrt{s}$
& (b) $\mu_\mrm{PS}=p_\perp$
& (c) $\mu_\mrm{PS}=m_D$
\end{tabular}\\[5mm]
\begin{tabular}{cc}
\includegraphics*[scale=0.47]{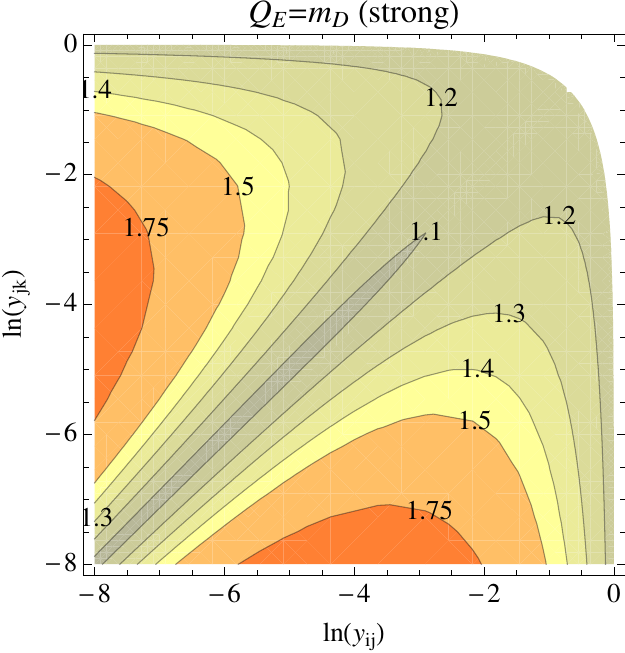}
&
\includegraphics*[scale=0.47]{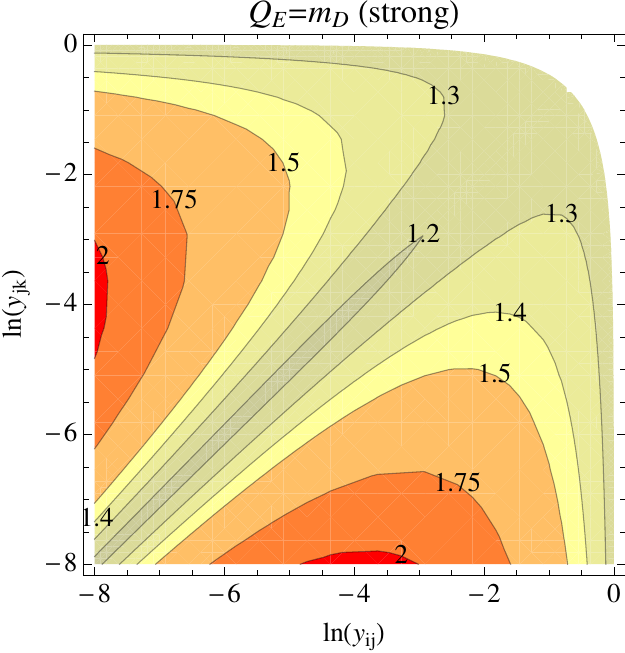} \\[-1mm]
$\mu_\mrm{PS}=\frac12m_D$, with CMW
& $\mu_\mrm{PS}= m_D$, with CMW
\end{tabular}
\caption{NLO correction factor for strong $m_D$-ordering, with GGG
  antennae. {\sl Top row:} $\mu_R=\sqrt{s}$ (left), $\mu_R=p_\perp$ (middle), and
  $\mu_R=m_D$ (right). {\sl Bottom row:} using the CMW
  $\Lambda_\mrm{MC}$, with $\mu_R = \frac12 m_D$ (left) and $\mu_R =
  m_D$ (right). For all plots, $\alpha_s=0.12$, $n_F=5$, and gluon
  splittings were evolved in $m_{qq}$.  
\label{fig:mD_strong_GGG} }
\end{figure}

We see that the soft limit almost cancels against 
\eqRef{eq:loopSoftLimit}. For an NLL-accurate shower, however, all divergent 
terms should match precisely, leaving at most a finite remainder in
the final matching correction, \eqRef{eq:V3Z}. In the expressions
above, this holds for the $\ln^2(y)$
term but not for the single logarithms (different
coefficient). Interestingly, the
remainder is proportional to the QCD $\beta$ function, specifically 
\begin{equation}
\label{eq:19}
\mrm{ME}-\mrm{PS} \ \to \ -\frac{\alpha_s}{2\pi} \frac12 \beta_0 \ln(y)~.
\end{equation}
It can therefore be absorbed in the choice of renormalization
scale by solving for $\mu_\mrm{PS}$ in $V_{3\mu}$, which yields:
\begin{equation}
\mu^2_{\mrm{PS}} ~\propto~y\, s~.
\end{equation}
This tells us that, in the soft limit, 
the specific choice of a renormalization scale
that is linear in the branching invariants will absorb all 
logarithms up to and including $\alpha_s^2\ln(y)$. Interestingly, 
this reasoning would rule out $\mu_R^2 \propto p_\perp^2$, since our
$p_\perp$-definition 
is quadratic in the invariants, $p_\perp^2 = s_{ij}s_{jk}/s$.
A better choice of renormalization scale 
would appear to be $\mu_R\propto m_D$, specifically
\begin{equation}
\mu^2_\mrm{PS} ~=~\min(s_{ij},s_{jk}) ~=~ \frac12 m_D^2~.\label{eq:mDmepsmu}
\end{equation}
Taken at face value, this seems to contradict the standard
literature~\cite{Amati:1980ch} on $p_\perp$ as the
optimal universal renormalization-scale choice. However, 
as we shall see below in \figRef{fig:mD_strong_GGG}, there is in fact no
choice of renormalization scale that absorbs \emph{all} logarithms for
this particular evolution variable;  
the choice $\mu_R\propto m_D$ merely manages to reabsorb 
the additional logarithms that are generated by the ordering
condition as $y\to 0$, but leftover logs in other parts of phase space
will remain uncanceled, ruining the NLL precision. In that sense,
choosing $\mu_R\propto p_\perp$ would simply leave a different set of
uncanceled logs, nonvanishing as $y\to 0$. 

Before we show the results over all of phase space however, we first
investigate a complementary interesting limit, the hard-collinear one, 
which is characterized by one of the
invariants becoming maximal while the other vanishes. In this limit,
the pole-subtracted one-loop matrix element, \SVirtual, becomes
\begin{align}
\mbox{ME:}\hspace{1cm}\sqg/s \rightarrow1,\, \sgq/s=y \rightarrow
0 \hspace{1cm} &  \frac{\alpha_s}{2\pi}\left[-\frac{5}{3}C_A  +
  \frac{1}{6}n_F\right]\, \ln(y) \label{eq:loopHardColLimit}
\end{align}
There are no log-squared terms in this limit, and both of the
single-log terms are half as large here as they were in the soft
limit.

The Sudakov integrals for $m_D$-ordering yield one divergent term,
$-\frac{1}{6}C_A\ln(y)$, in the hard-collinear region, modulo a factor
$\alpha_s/(2\pi)$. The Sudakov integral for 
gluon splitting in the neighbouring antenna, represented by the first
term on the second-to-last line of \eqRef{eq:V3Z} is specified in the last line
of \eqRef{eq:md_tot}. The step function is only non-zero for the first term
in the hard-collinear limit $\sqg\rightarrow s, \sgq  \rightarrow 0$ and produces a term
$\frac{1}{6}P_{A_j}n_F \ln(y)$. The numerator of the corresponding Ariadne factor
contains the invariant of the neighboring dipole $\sgq$ which vanishes
in this limit and causes the dipole splitting contribution to 
reduce to zero. The $n_F$-dependent contribution is instead shifted to 
the last term of \eqRef{eq:V3Z}, which has the same limit but without
the Ariadne pre-factor. The hard-collinear limit of the shower terms,
including only terms involving logarithms of the invariants and not
including the $V_{3\mu}$ term, is therefore
\begin{align}
-\mbox{PS:}\hspace{1cm}\sqg/s \rightarrow1,\, \sgq/s=y \rightarrow
0 \hspace{1cm} &  \frac{\alpha_s}{2\pi}\left[-\frac16 C_A + \frac16
  n_F \right]\, \ln(y)~.\label{eq:mDstrongColl}
\end{align}
Again, 
the combination $(\mrm{ME}-\mrm{PS})$ relevant for computing
the correction factor is proportional to the QCD $\beta$ function, and
in fact has exactly the same form as \eqRef{eq:19}. The
conclusion is therefore that, also in this limit, all logarithms
through $\alpha_s^2\ln(y)$ can be absorbed by choosing a
renormalization scale which is linear in the vanishing invariant. The
particular choice which is linear in both the soft and collinear
limits is $\mu_\mrm{PS} \propto m_D$.
To illustrate this, we show the full NLO $Z\to 3\,$jets correction factors,
$(1+V_{3Z})$, for $m_D$-ordering with a few different choices of
renormalization scale and scheme,
in \figRef{fig:mD_strong_GGG}.
Note that the axes are
logarithmic, in $\ln(y_{ij}) = \ln(s_{ij}/s)$, to make the infrared
limits clearly visible. 

Without the $V_{3\mu}$ term, the correction factor looks as depicted
in the top left-hand plot in \figRef{fig:mD_strong_GGG}. The
increasing contours towards the axes indicate uncanceled logarithms 
in the correction factor. The middle pane
shows the correction factor derived for $\mu_\mrm{PS}=p_\perp$. As
discussed above, there is an uncanceled logarithm in the 
soft limit (lower left-hand corner of the plot), since $p_\perp$
is quadratic in the vanishing invariants there. However, in the
hard-collinear limits (upper left-hand and lower right-hand corners),
$p_\perp$ is linear in the vanishing invariant, and hence the
contours remain bounded there. In the right-hand pane, we show the choice
$\mu_\mrm{PS} = m_D$, which can be seen to lead to bounded 
correction factors (below $\sim 1.3$) in all three phase-space
corners. Nonetheless, there is still an 
uncanceled divergence \emph{between} the soft and hard collinear
limits. We shall see in the section on $p_\perp$-ordering below that
the cure for this is basically to choose a better evolution variable. 

In the bottom row of \figRef{fig:mD_strong_GGG}, we show a few
variations on $\mu_\mrm{PS} = m_D$, specifically we include the CMW rescaling
of $\Lambda_\mrm{QCD}$, as defined by \eqRef{eq:kCMW}, 
and show how a variation of the renormalization
scale by a factor of 2 affects the correction factor. In the left-hand
pane, we show $\mu_\mrm{PS} = \frac12 m_D$ and in the right-hand one
$\mu_\mrm{PS} = m_D$. Of these, the choice $\mu_\mrm{PS} = \frac12
m_D$, with CMW rescaling, leads to the smallest correction factors
(best LO behaviour), and this could therefore be taken as a useful
default for $m_D$-ordering, e.g.\ for uncertainty
estimates.   

\subsubsection{Transverse Momentum}
\label{sec:transverse-momentum-1}

For a shower ordered in $p_\perp$, the antenna phase-space integrals
in \eqRef{eq:V3} are performed over contours such as those 
depicted for $p_T$ squared in \figRef{fig:ev}. 
The curved contours motivate a coordinate transformation from 
$(s_1,s_2)$ to a basis defined as the dimensionless evolution variable
$y=\frac{Q^2}{s}=\frac{4s_1s_2}{s^2}$, complemented by an
energy-sharing variable, which we define as $z=\frac{s_1}{s}$. Note
that the coordinate transformation depends explicitly on the total
invariant mass $s$ of 
the $2 \to 3$ dipole. For the $3 \to 4$ integrations, the invariant
mass $s$ is replaced by the invariant mass of the appropriate dipole
(either $\sqg$ or $\sgq$). 
The integration boundaries in $z$ are determined by the intersection
of the invariant mass of the dipole with the evolution parameter
$Q^2$. The choice of $y$ and its integration boundaries make the
effect of strong ordering especially clear since we see integration
from $Q^2$ to the total invariant mass of the dipole (the unordered
region). As before, the integration over the gluon-splitting antenna
$\left(e_3^0\right)$ makes use of a different phase space integration,
in $m_{q\bar{q}}$, and only uses the evolution parameter as a cut-off
in the singularity of the corresponding dipole. 

The contributing terms are:
\begin{align}
&g_s^2\left[  C_A\int_{Q_3^2}^s a_3^0\, \dPS{\mrm{ant}}-\sum_{j=1}^2 C_A\int_0^{s_j} (1-O_{E_j})\,d_3^0\, \dPS{\mrm{ant}}-\sum_{j=1}^2 2\,T_R\,n_F P_{A_j} \int_0^{s_j} (1-O_{S_j})\,e_3^0\, \dPS{\mrm{ant}} \right] \nonumber \\
&=\frac{\alpha_s}{4\pi }\left[  C_A \,s\, \mathcal{A}_1\left[ \frac{Q_3^2}{s},1 \right]   
-C_A  \sqg  \,\mathcal{A}_2\left[ \frac{4\sgq}{s},\max\left( \frac{4\sgq}{s},1\right) \right]   - \sgq \,C_A\, \mathcal{A}_3\left[ \frac{4\sqg}{s},\max\left( \frac{4\sqg}{s},1\right) \right]   \right. \nonumber \\
&-\left. n_F \left( \frac{P_{A_1}}{\sqg}\int_{Q_3^2}^{\max\left(Q_3^2,\sqg \right)} \d{s_1} \int_{0}^{\sqg-s_1} \d{s_2} +\frac{P_{A_2}}{\sgq} \int_{Q_3^2}^{\max\left({Q_3^2,\sgq}\right)} \d{s_1} \int_{0}^{\sgq-s_1} \d{s_2} \right) e_3^0(s_1,s_2)\right] \label{eq:strongpT}
\end{align}
with 
\begin{align}
\mathcal{A}_n\left[ a,b  \right] = \int_{a}^{b} \d{y_n} 
    \int_{\zmin^n}^{\zmax^n} \d{z_n} \,|\mathcal{J}_n| A_n(y_n,z_n) \hspace{2cm} \mbox{for } n=1,2,3,
\end{align}
and 
\begin{align}
y_n=4 \frac{s_1 s_2}{m^4_{IK}},\; z_n=\frac{s_1}{m^2_{IK}},\; |\mathcal{J}_1|=\frac{m^4_{IK}}{4z_n},\; z^n_{\overset{\scriptsize{\mbox{max}}}{\scriptsize{\mbox{min}}}}=\frac{1}{2}\left(1\pm \sqrt{1- y_n}\right). \label{eq:strongpTvar}
\end{align}
For $n=1$ we set $m^2_{IK}=s$, for $n=2$ $m^2_{IK}=\sqg$ and for $n=3$
$m^2_{IK}=\sgq$. The Ariadne factor $P_{A_j}$ is defined in
\eqRef{eq:Ariadne}. The $\max$ condition on the outer integration
boundary of $\mathcal{A}_2$ and $\mathcal{A}_3$ reflect that the
correction term disappears if the generated $Q_3^2$ is larger than the
invariant mass of the dipole.   
As for $m_D$-ordering, we here work out the most divergent behavior
explicitly, by focussing on the double log terms arising from the
eikonal term in the antenna, and relegate the full form of the antenna
integrals to \appRef{sec:antenna-integrals}. The double poles give
rise to terms 
\begin{align*}
\frac{\alpha_s\,C_A}{2\pi }\int_{\frac{Q_3^2}{s}}^1 \d y_1 \int_{\zmin}^{\zmax} \d z_1 \, \frac{1}{y_1z_1}~,
\end{align*}
which lead to the following generic transcendentality-2 integrals,
\begin{multline}
\label{eq:1}
\int_{x}^1 
 \frac{\d{y_1}}{y_1} \ln\left(\frac{1+\sqrt{1-y_1}}{1-\sqrt{1-y_1}}\right) \ = \ 
  \Li\left(\frac12\left(1-\sqrt{1-x}\right)\right)-\Li\left(\frac12\left(1+\sqrt{1-x}\right)\right)\\
+\frac12 \ln\left(\frac{x}{4}\right)\ln\left[-\left(\frac{-2 + 2 \sqrt{1 - x} + x}{x}\right)\right].
\end{multline}
The double logarithm in the shower expansion is generated by a
combination of the $2\rightarrow 3$ and $3\rightarrow 4$ Sudakov
integrals, with the respective pieces adding up to
\begin{align}
\frac{\alpha_sC_A}{2\pi}\left[  -\frac{\pi^2}{6} + \frac{1}{2}\ln\left( \frac{\sqg \sgq}{s^2} \right)^2 + \frac{\pi^2}{3} -  \frac{1}{2}\ln\left( \frac{\sqg}{s} \right)^2 - \frac{1}{2}\ln\left( \frac{\sgq}{s} \right)^2    \right]~. \label{eq:ptstrongTeq2}
\end{align}
We see that a partial cancellation arises between the first two terms
(which come from the $2\to 3$ Sudakov expansion) and the last three
(which come from the $3\to 4$ expansion). 
What remains is a log squared in both invariants $\ln\left( \sqg/s
\right) \ln \left( \sgq/s \right)$. 

At the single-log level, the $3 \to 4$ terms give a numerically larger
coefficient than the $2\to 3$ ones, leading to a single log remainder.
The gluon-splitting term also reduces to a single log. The overall
result in the soft limit is then
\begin{equation}
-\mbox{PS:}\hspace{1cm}\sqg=\sgq=y \rightarrow 0 \hspace{1cm}
\frac{\alpha_s C_A}{2\pi} \left[ \ln^2(y)  -\frac{1}{3}\ln(y)
  \right]+\frac{\alpha_s n_F}{6\pi}\ln(y)  
\end{equation}

Comparing with the result of the virtual correction in the soft limit,
\eqRef{eq:loopSoftLimit}, we see that the shower generates the double
log terms correctly, and, similarly to the case of $m_D$-ordering,
there is a single-log remainder which is proportional to the QCD
$\beta$ function. However, for $p_\perp$-ordering the
constant of proportionality is $1$, rather than $\frac12$, a
difference which translates to the optimal renormalization-scale choice being
quadratic in the invariants in this case, rather than linear. Before
commenting further on this difference, let us first consider the
complementary, hard-collinear, limit.

In the hard-collinear limit, we find the same as for $m_D$-ordering, 
\begin{align}
-\mbox{PS:}\hspace{1cm} \sqg=y \rightarrow 0, \sgq \rightarrow
s \hspace{1cm} 
\frac{\alpha_s}{2\pi} \left[-\frac16C_A+\frac16n_F\right]\ln(y)~.
\end{align}
Double logs (eikonal parts of the antenna)  also appear at both the
$2\to 3$ and $3\to 4$ levels, but cancel among each other as almost
all other antenna terms do; what remains at the single-log level is
the integrated difference between a quark-antiquark antenna and a
quark-gluon antenna, plus the $n_F$-dependent `Ariadne Log'. The only contributing 
Sudakov gluon splitting contribution is the
second term in the last line of \eqRef{eq:strongpT}. Integration over the $\sgq$ 
dipole, however, is associated with an Ariadne factor carrying $\sqg$ in the numerator
and therefore reduces to zero. 
As before, we can write the remainder as half the QCD $\beta$ function, which
implies that a renormalization scale linear in the vanishing invariants
can absorb the logarithm. 

To summarize, for $p_\perp$-ordering we find that the optimal
renormalization-scale choice must be quadratic in the vanishing
invariants in the soft limit and linear in the hard-collinear
limit. Those conditions are fulfilled by $p_\perp$ itself, thus
\begin{equation}
\mu_\mrm{PS}^2 \ \propto \ p_\perp^2 \ = \ \frac{s_{ij}s_{jk}}{s_{ijk}}
\end{equation}
absorbs all logarithmic terms up to and including $\alpha_s^2\ln(y)$
in the LO couplings. 

\begin{figure}[t!]
\centering
\setlength{\skipper}{2mm}
\begin{tabular}{ccc}
\includegraphics*[scale=0.46]{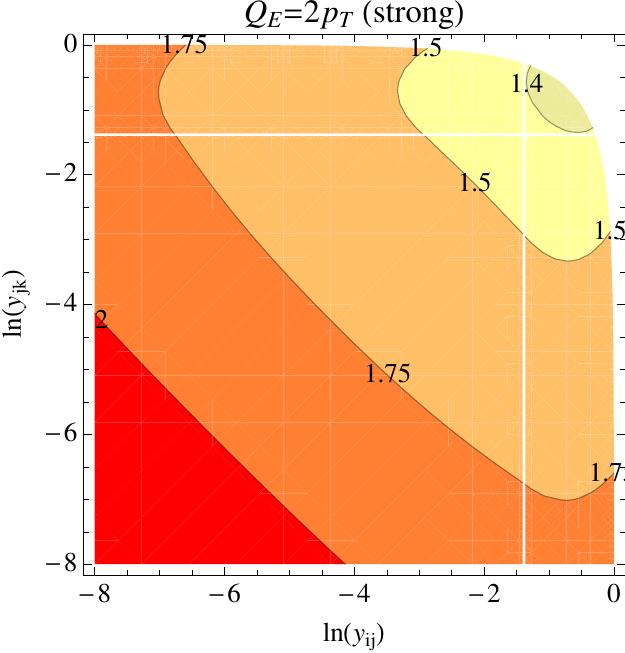}
&\includegraphics*[scale=0.46]{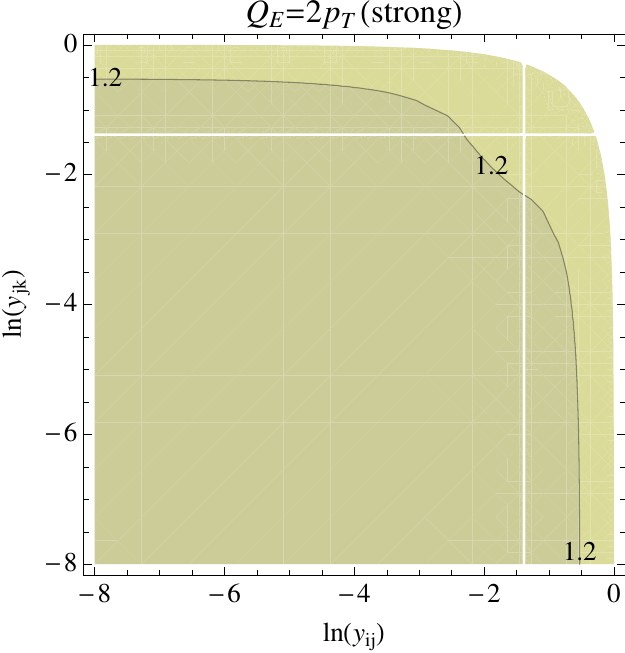}
& \includegraphics*[scale=0.46]{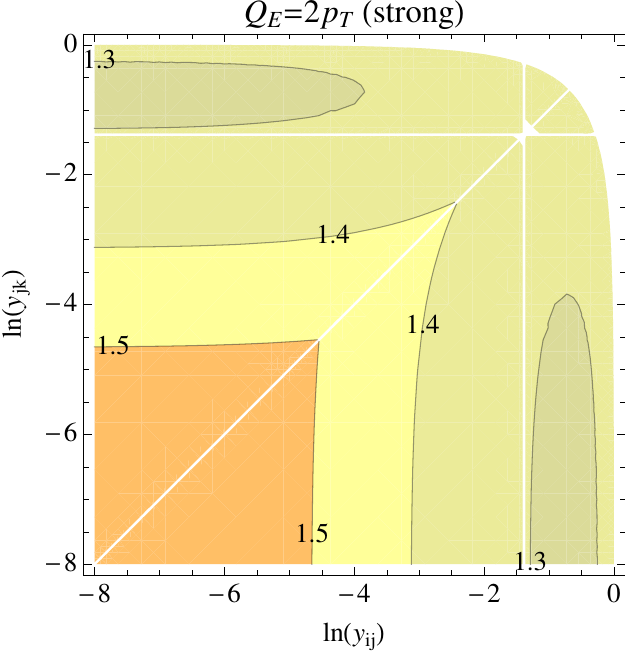} \\[-1mm]
(a) $\mu_\mrm{PS}=\sqrt{s}$
& (b) $\mu_\mrm{PS}=p_\perp$
& (c) $\mu_\mrm{PS}=m_D$
\end{tabular}\\[5mm]
\begin{tabular}{cc}
\includegraphics*[scale=0.46]{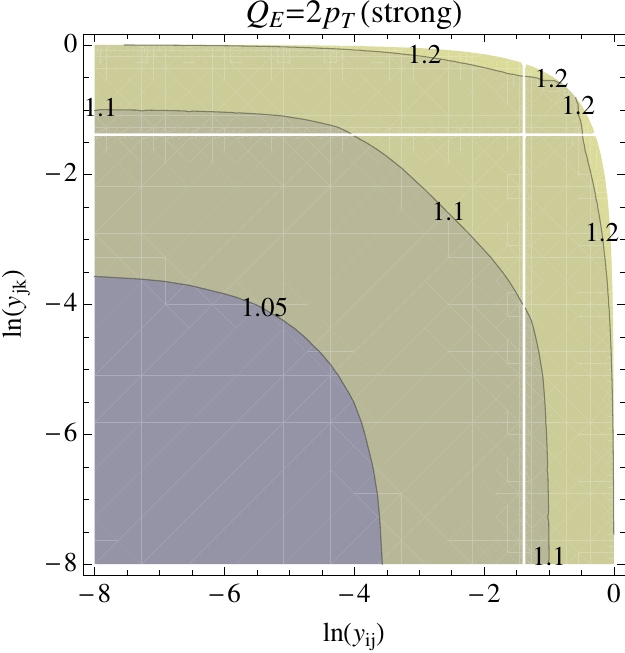}
&
\includegraphics*[scale=0.46]{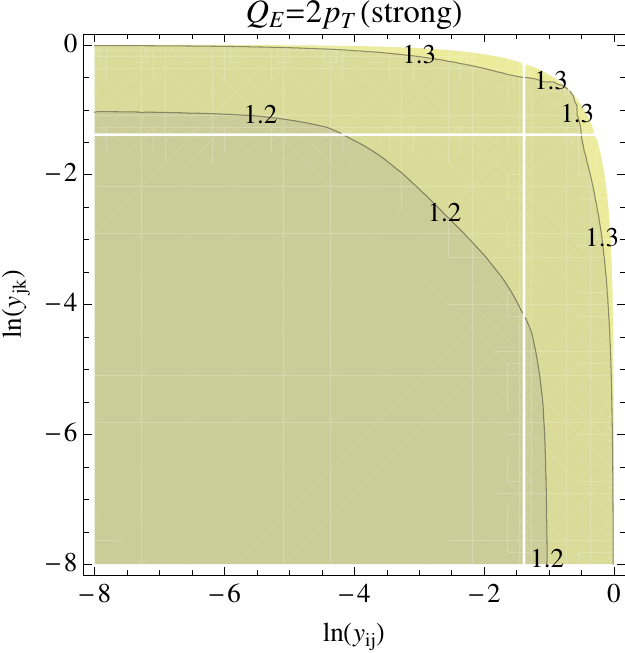} \\[-1mm]
$\mu_\mrm{PS}=p_\perp$, with CMW
& $\mu_\mrm{PS}= 2p_\perp$, with CMW
\end{tabular}
\caption{NLO correction factor for strong $p_\perp$-ordering, with GGG
  antennae. {\sl Top row:} $\mu_\mrm{PS}=\sqrt{s}$ (left), $\mu_\mrm{PS}=p_\perp$ (middle), and
  $\mu_\mrm{PS}=m_D$ (right). {\sl Bottom row:} using the CMW
  $\Lambda_\mrm{MC}$, with $\mu_\mrm{PS} = p_\perp$ (left) and $\mu_\mrm{PS} =
  2p_\perp$ (right). For all plots, $\alpha_s=0.12$, $n_F=5$, and gluon
  splittings were evolved in $m_{qq}$.  
\label{fig:pT_strong_GGG} }
\end{figure}

Illustrations of the full NLO correction factors, $(1+V_{3Z})$, are
given in \figRef{fig:pT_strong_GGG}.
The ordering of the plots in the top row are the same as in
\figRef{fig:pT_strong_GGG}, showing, from left to right, 
$\mu_\mrm{PS} = \sqrt{s}$, $\mu_\mrm{PS}=p_\perp$,
$\mu_\mrm{PS}=m_D$. Similarly to the case of strong $m_D$-ordering,
both of the latter two choices exhibit no logarithmic divergences in
the hard-collinear regions (top left and bottom right corners of the
plots), but in the soft region (bottom left corner) it is here
$\mu_\mrm{PS} = p_\perp$ which leaves the correction factor free of
logarithms. Indeed, we see that the combination of evolution and
renormalization in $p_\perp$ leads to a rather flat correction factor
over all of phase space, showing that this combination is indeed
``better'' than $m_D$-ordering. 

In the bottom row of plots in \figRef{fig:pT_strong_GGG}, we include
the CMW factor and show the correction factors for
$\mu_\mrm{PS}=p_\perp$ (left) and $\mu_\mrm{PS}=2p_\perp$
(right). In particular on the left-hand pane, the 
NLO correction factor is essentially unity in the soft limit, while
the corrections in the hard-collinear regions remain
less than $\sim 20\%$. This gives some additional weight to the
arguments for $p_\perp$-ordered showers with $p_\perp$ as
renormalization scale being the best default choice for strongly
ordered dipole-antenna showers. 
It also provides some rationale why one typically
finds a rather large value of $\alpha_s(m_Z)\sim 0.13$ (with CMW
rescaling, or $\alpha_s(m_Z)\sim 0.14$ without it) when tuning such
models to LEP event shapes; there is still a genuine order 20\% NLO
correction in the hard resolved region (upper right-hand corner). We
return to this in more detail in the context of full LO + NLO matching
in \secRef{sec:results}.

\subsubsection{Energy}
\label{sec:energy}

To put the differences between $m_D$ and $p_\perp$ in context, we now
briefly examine the case of energy ordering, which is known to 
produce the wrong DGLAP evolution in the collinear
limit~\cite{Dokshitzer:2008ia,Nagy:2009re,Skands:2009tb}, 
and hence we should find larger (possibly divergent) NLO corrections.  

Slicing phase space with the energy variable
$Q_3^2=s_{ijk}(y_{ij}+y_{jk})^2$, see~\figRef{fig:ev}, 
requires the use of an explicit
infrared cut-off because the contours otherwise 
allow for the invariants to hit singular regions for
every value of the contour. We will here use a
cut-off in transverse momentum (a cut-off in dipole virtuality is also
possible). 
The cutoff motivates us to switch to a different choice of integration
variables. Therefore integration is transformed from $(s_1,s_2)$ to the
dimensionless evolution parameters
$y_E^2=\frac{Q^2}{s}=\frac{(s_1+s_2)^2}{m^2_{IK}}$ and completed with
the energy sharing variable $\zeta=\frac{s_2}{m^2_{IK}}$.  The
interesting integrals arising from expanding the Sudakov form factor
then are 
\begin{align}
&g_s^2\left[  C_A\int_{Q_3^2}^s a_3^0\, \dPS{\mrm{ant}}-\sum_{j=1}^2 C_A\int_0^{s_j} (1-O_{E_j})\,d_3^0\, \dPS{\mrm{ant}}-\sum_{j=1}^2 2\,T_R\,n_F P_{A_j} \int_0^{s_j} (1-O_{S_j})\,e_3^0\, \dPS{\mrm{ant}} \right] \nonumber \\
&=\frac{\alpha_s}{4 \pi} \left[ C_A \left\{ \mathcal{AE}_1(s,1) - \mathcal{AE}_2(\min[\sqg,1],1) -\mathcal{AE}_3(\min[\sgq,1],1) \right\}  \right. \nonumber \\
&\hspace{1cm} -\left. n_F \left( \frac{P_{A_{\qg}}}{\sqg}\int_{Q_3^2}^{\max\left(Q_3^2,\sqg \right)} \d{s_1} \int_{0}^{\sqg-s_1} \d{s_2} +\frac{P_{A_{\gq}}}{\sgq} \int_{Q_3^2}^{\max\left({Q_3^2,\sgq}\right)} \d{s_1} \int_{0}^{\sgq-s_1} \d{s_2} \right) e_3^0(s_1,s_2)\right] 
\end{align}
with
\begin{align*}
\mathcal{AE}_n(m^2_{IK},1) = \int_{\frac{Q_3^2}{m^2_{IK}}}^1 \d y_E^2 \int_0^1 \d \zeta' \frac{1}{2} \mathcal{AE}^0_n(y_E^2,\zeta').
\end{align*}
With $\mathcal{AE}^0_1 = a_3^0$, $\mathcal{AE}^0_2 = d_3^0$ and $\mathcal{AE}^0_3 = e_3^0$. The inner integral has been rescaled to make it independent of the outer integral with $\zeta = y_E\zeta'$. 
To establish the cut-off, we use the relation $4\frac{s_1s_2}{s^2} = 4p_\perp^2/s$, which we demand to be larger than the cut-off $\Delta$. 
In terms of our variables we then have the condition
\begin{equation}
  \label{eq:23}
  4\zeta'(1-\zeta') > \frac{\Delta}{y_E^2}.
\end{equation}
The upper and lower limits on $\zeta'$ are then
\begin{equation}
  \label{eq:24}
  \zeta_-' < \zeta' < \zeta_+', \quad \zeta_{\pm}' = \frac{1}{2}\left( 1 \pm  \sqrt{1- \frac{\Delta}{y_E^2}}\right)\,.
\end{equation}
Focussing on the eikonal integral
\begin{equation}
  \label{eq:25}
\frac{\alpha_s C_A}{4\pi}  \int_{y_E^2=\frac{Q_3^2}{s}}^1 \frac{dy_E^2}{y_E^2} \int_{\zeta_-'}^{\zeta_+'}  \frac{\d \zeta'}{\zeta'}~,
\end{equation}
the result for this integral is
\begin{align}
&\frac{\alpha_s}{2\pi}\left[\Li\left( \frac{1}{2} \left( 1 - \sqrt{1-\Delta} \right) \right) -  \Li\left( \frac{1}{2} \left( 1 + \sqrt{1-\Delta} \right) \right) + \frac{1}{2} \left[ -2 \;\mbox{atanh}\left( \sqrt{1-\frac{\Delta}{y_E^2}} \right) \ln(4) \right.\right. \nonumber \\
&\left. + \mbox{atanh}\left(\sqrt{1-\Delta}\right) \ln(16) + \ln^2\left( 1 - \sqrt{1-\Delta}\right) - \ln^2\left( 1 + \sqrt{1-\Delta} \right) - \ln^2 \left( 1 - \sqrt{1-\frac{\Delta}{y_E^2}} \right) \right. \nonumber \\
&\left.\left.+ \ln^2 \left( 1 + \sqrt{1-\frac{\Delta}{y_E^2}} \right) \right]-2 \Li\left( \frac{1}{2} \left( 1 - \sqrt{1-\frac{\Delta}{y_E^2}} \right)\right)+ \Li\left( \frac{1}{2} \left( 1 + \sqrt{1-\frac{\Delta}{y_E^2}} \right)\right) \right]~.
\end{align}
In the soft limit $y_E^2=4y^2\to 0$ this reduces to
\begin{align}
-\frac{1}{2}\ln^2(\Delta) - \ln^2\left( \frac{\Delta}{4y^2} \right) - 2 \ln(4y^4) \ln(2) -\Li\left( \frac{\Delta}{64y^2} \right) 
\end{align}
Thus we see that there are explicit non-cancelling double-logarithmic terms that
involve the hadronization cutoff, $\Delta$. Depending on the ratio
between the dipole mass and the cutoff, these would lead to
asymptotically divergent correction factors. 

\begin{figure}[t!]
\centering
\setlength{\skipper}{2mm}
\begin{tabular}{ccc}
\includegraphics*[scale=0.46]{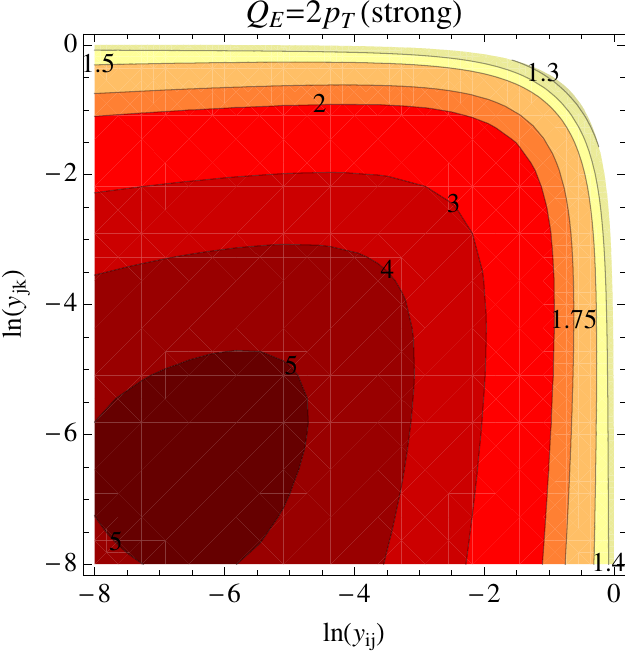}
 & 
\includegraphics*[scale=0.46]{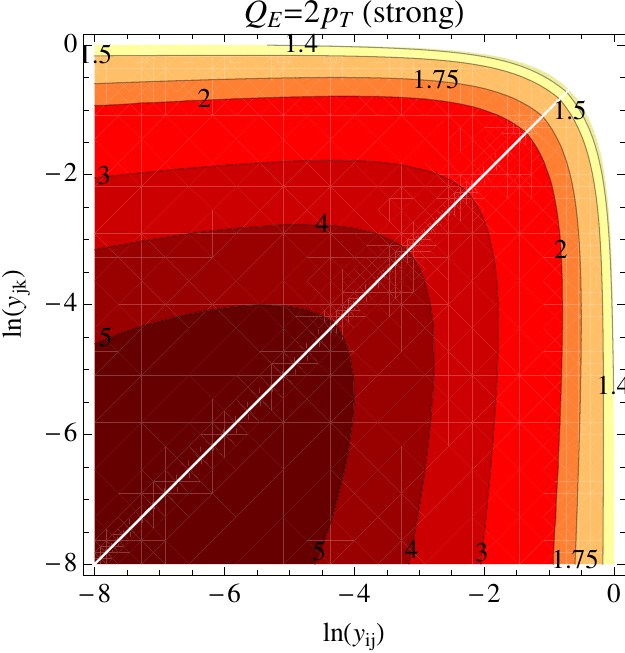}
 & 
\includegraphics*[scale=0.46]{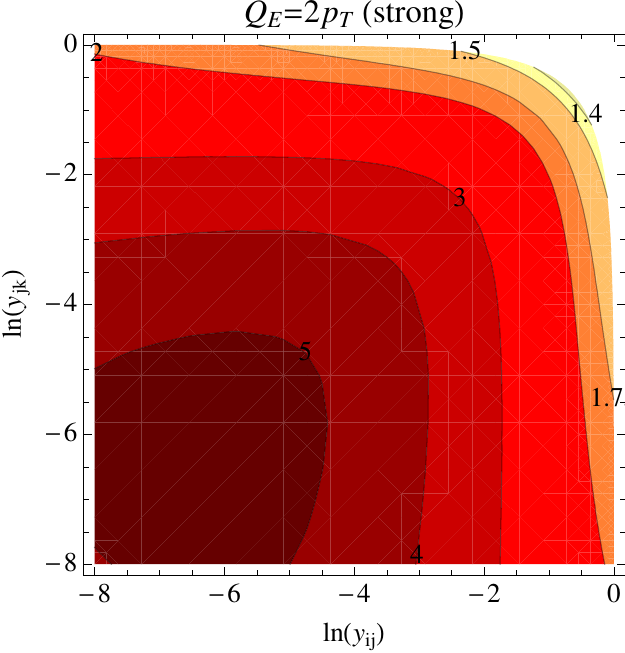} \\[-1mm]
 (a) $\mu_\mrm{PS}=p_\perp$
 & (b) $\mu_\mrm{PS}=m_D$
 & (c) $\mu_\mrm{PS}=E^*$ \\[5mm]
\includegraphics*[scale=0.46]{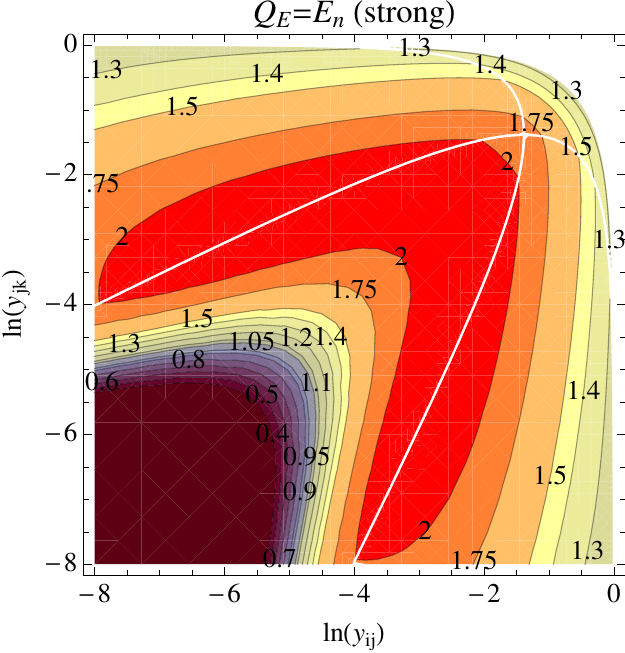}
 & 
\includegraphics*[scale=0.46]{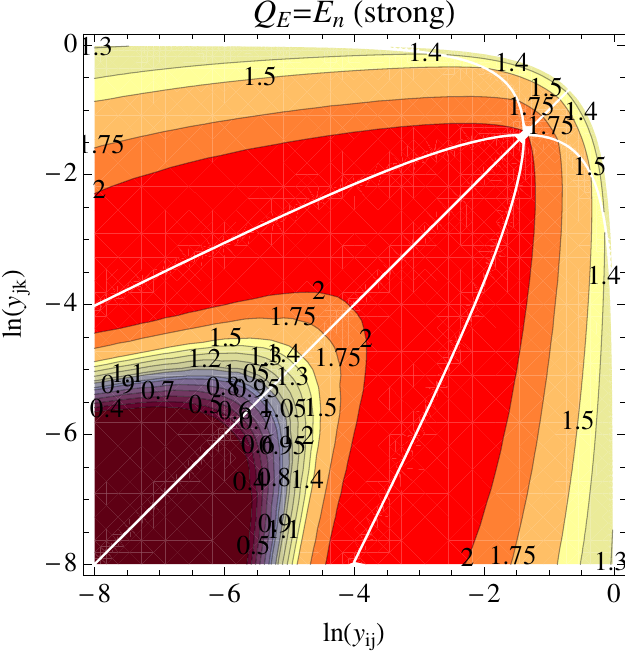}
 & 
\includegraphics*[scale=0.46]{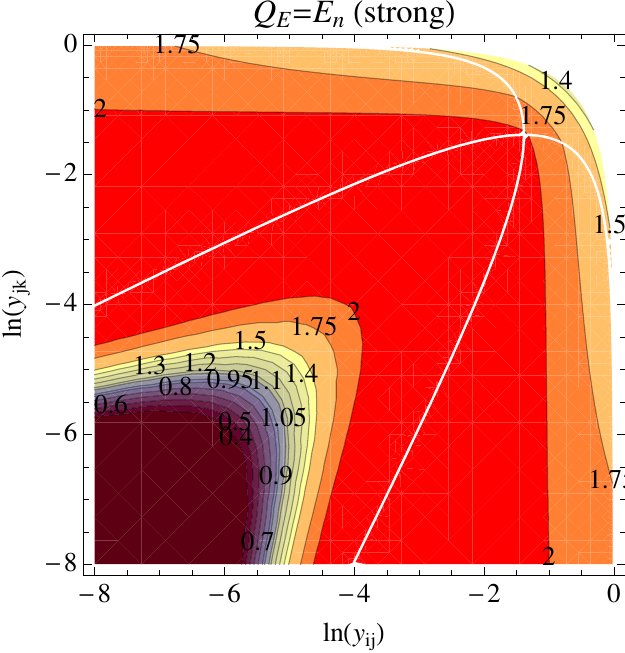} \\[-1mm]
(d) $\mu_\mrm{PS}=p_\perp$
& (e) $\mu_\mrm{PS}=m_D$
& (f) $\mu_\mrm{PS}={E^*}^2/\sqrt{s}$
\end{tabular}
\caption{NLO correction factor for strong energy-ordering, with GGG
  antennae, for various renormalization-scale choices and linear (top
  row) and squared (bottom row) scaling of the evolution variable with
  gluon energy. 
\label{fig:en_strong_GGG}}
\end{figure}

One might wonder whether using a linearized form of
energy ordering would make a difference,
see~\figRef{fig:ev}. Rather than go through the derivations again, 
we merely show the full NLO corrections in \figRef{fig:en_strong_GGG}, 
for both linear (top row) and squared (bottom row) energy ordering, for an
(arbitrary) dimensionless cutoff value of $\Delta = 10^{-7}$. 

From left to right in both rows, we show the three
renormalization-scale choices, $\mu_\mrm{PS} = p_\perp$ (left),
$\mu_\mrm{PS} = m_D$ (middle), and $\mu_\mrm{PS} = Q_E$ (right), with
the latter equal to linear energy in the top row and squared energy in
the bottom one. Interestingly, the contours in the linear case are
increasing towards the soft region, while they ultimately decrease in
the squared case. It is clear, however, 
that no intelligent choice of renormalization
scale can absorb the infrared divergences. Moreover, any other choice
of $\Delta$ would have led to different contours, due to the explicit
$\ln(\Delta)$ terms, hence even if a ``least bad'' choice was found,
it would not be universal.

As mentioned above,
the main point of showing these comparisons is to place the comparison
between $m_D$ and $p_\perp$ in the previous subsections in 
perspective. Thus,
while we saw that $p_\perp$ was generating a better-behaved correction
factor than $m_D$, the one for $m_D$ is still far better behaved than
is the case for energy ordering. From this perspective, we still
believe it could make sense, e.g., to use $m_D$-ordering, with the NLO
correction factor included, as a conservative uncertainty
variation for a central prediction  
based on $p_\perp$-ordering.

\subsection{Smooth Ordering}\label{sec:smooth-ordering}

We will now discuss the same Sudakov integrals as in the previous
subsections but for the case of smooth ordering
(\secRef{sec:SmOrd}). This is especially interesting given that smooth
ordering is the way \Vc is able to fill all of phase space without
significant under- or overcounting at the LO
level~\cite{Giele:2011cb}. As discussed in \secRef{sec:SmOrd},
however, this does involve some ambiguity in what Sudakov factors are
associated with the unordered points, and the NLO 3-jet correction factors
should tell us explicitly whether this ambiguity generates problems
at this level.

The Sudakov integrations are actually more straightforward
for smooth ordering than was the case for strong ordering, 
since the $P_\mrm{imp}$ factor regulates the integrands on the boundaries. 
Therefore the integrations always run over the full phase space of 
the system. The $2\rightarrow 3$ splitting generates the same terms as
in the strong-ordering case, \eqRef{eq:13}. Including also the $3\to
4$ terms, the expanded Sudakov generates the following antenna
integrals,
\begin{align}
&g_s^2 \left[ C_A\int_0^s a_3^0 \:\dPS{\mrm{ant}}-\sum_{j=1}^2
    C_A\int_0^{s_j} \frac{Q_{E_j}^2}{Q_{E_j}^2 + Q_{3}^2}\,d_3^0\:
    \dPS{\mrm{ant}} -\sum_{j=1}^2 2\,T_R\,n_F P_{A_j}\int_0^{s_j}
    \frac{m_{q\bar{q}}^2}{m_{q\bar{q}}^2 + Q_3^2}\,e_3^0\:
    \dPS{\mrm{ant}} \right]~, \label{eq:smoothOrdAnInt} 
\end{align}
where $Q_3$ is the evolution scale evaluated on the 3-parton
configuration and $Q_{E_j}$ ($m_{q\bar{q}}$) 
is the scale of the $3\to 4$ emissions (splittings)
being integrated over. 
The full answer for the $3\rightarrow 4$ case for gluon emission is 
\begin{equation}
\label{eq:40}
-g_s^2\sum_{j=1}^2 C_A\int_0^{s_j} \frac{Q_{E_j}^2}{Q_{E_j}^2+Q_3^2} \,d_3^0\: \dPS{\mrm{ant}} =  -  \frac{\alpha_s C_A}{4\pi}\left(\sum_{i=1}^{5} K_i L_i(\sqg,Q_3^2)\right)\,-\frac{\alpha_s C_A}{4\pi}\left(\sum_{i=1}^{5} K_i L_i(\sgq,Q_3^2)\right).
\end{equation}
where $K_i$ and $L_i$ can be found in \appRef{sec:antenna-integrals}. 
The full answer for the $3\rightarrow 4$ case for gluon splitting is 
\begin{equation}
\label{eq:41}
 -g_s^2\sum_{j=1}^2 \,n_F P_{A_j} \int_0^{s_j}
 \frac{m_{q\bar{q}}^2}{m_{q\bar{q}}^2 + Q_3^2}\,e_3^0\: \dPS{\mrm{ant}} 
=  -\frac{\alpha_s n_F}{4\pi}G(\sqg,Q_3^2)\,-\frac{\alpha_s n_F}{4\pi}G(\sgq,Q_3^2).
\end{equation}
where $G$ can be found in the appendix. 
We will discuss the derivation
of these terms in more detail in the following two subsections, for
smooth $m_D$- and $p_\perp$-ordering, respectively. 
  
\subsubsection{Dipole virtuality}
Since the $2\to 3$ emission terms remain the same as in the case of
strong $m_D$-ordering, we only need to rederive the $3\to 4$
contributions to $V_{3Z}$, which are
\begin{align}
&-g_s^2 \left[ \sum_{j=1}^{2} C_A\int_0^{s_j} \left( 1 - \frac{Q_{3}^2}{Q_{Ej}^2 + Q_{3}^2} \right)\,d_3^0\: \dPS{\mrm{ant}} + \sum_{j=1}^2 2\,T_R\,n_F P_{A_j}\int_0^{s_j} \left( 1-\frac{Q_3^2}{m_{q\bar{q}}^2 + Q_3^2}\right)\,e_3^0\: \dPS{\mrm{ant}} \right] \nonumber \\
&=-\frac{\alpha_s}{4 \pi } \left[ \frac{C_A}{\sqg} \left( \int_0^{\frac{1}{2}\sqg} \d{s_{2}}  \int_{s_2}^{\sqg-s_2}  \d{s_{1}}\frac{2s_2}{Q_3^2 +2 s_2} + \int_0^{\frac{1}{2}\sqg} \d{s_{1}}  \int_{s_1}^{\sqg-s_1}  \d{s_{2}} \frac{2s_1}{Q_3^2 + 2s_1} \right)  \,d_3^0   \right.  \nonumber \\
&\hspace{1.2cm} \left. +\frac{C_A}{\sgq} \left( \int_0^{\frac{1}{2}\sgq} \d{s_{2}}  \int_{s_2}^{\sgq-s_2}  \d{s_{1}} \frac{2s_2}{Q_3^2 + 2s_2} + \int_0^{\frac{1}{2}\sgq} \d{s_{1}}  \int_{s_1}^{\sgq-s_1}  \d{s_{2}} \frac{2s_1}{Q_3^2 + 2s_1} \right)  \,d_3^0   \right.  \nonumber \\
&\left. \hspace{1.2cm}+ 2n_F\left(  \frac{P_{A_1}}{\sqg}\int_0^{\sqg} \d{s_{2}}  \int_{0}^{\sqg-s_2}  \d{s_{1}}   +\frac{P_{A_2}}{\sgq} \int_0^{\sgq} \d{s_{2}}  \int_{0}^{\sgq-s_2}  \d{s_{1}} \right) \frac{s_1}{ s_1 + Q_3^2}\,e_3^0\:  \right] \label{eq:SmOrdDiMa} 
\end{align}
with $Q_3^2=2\min(\sqg,\sgq)$ and $e_3^0$ carrying the singularity in
$s_1$. We will focus again on deriving the transcendality-2 terms explicitly,
with the full expressions given in the appendix. We start by
recalling the expression for the strongly-ordered $2\to 3$ branching, 
\begin{align*}
 \frac{\alpha_s C_A}{2\pi} \Bigg[ 
 \ln\left(\frac{s}{\frac{1}{2}Q_3^2}\right)\ln\left(\frac{s-\frac{1}{2}Q_3^2}{\frac{1}{2}Q_3^2}\right)
 -\Li\left(\frac{s-\frac{1}{2}Q_3^2}{s}\right)
 + \Li\left(\frac{\frac{1}{2}Q_3^2}{s}\right)\Bigg]~.
\end{align*}
To this we add the results from the eikonal term $\frac{2\sqg}{ s_1s_2}$ of
one $3\to 4$ gluon emission, the first line in \eqRef{eq:SmOrdDiMa},
\begin{align}
&-\frac{2\alpha_s C_A}{\pi }  \int_0^{\frac{1}{2}} \d{y_{2}}  \int_{y_2}^{1-y_2}  \d{y_{1}}\frac{1}{y_1(y_3^2 + 2y_2)}  \nonumber \\
&=-\frac{\alpha_s C_A}{2\pi }\left[ -\ln(4)\ln\left( 1- \frac{1}{1+y_3^2}\right) +\ln(4)\ln\left( 1+\frac{1}{1+y_3^2} \right)- 2\,\Li\left(-\frac{1}{y_3^2}\right) +2\, \Li\left(\frac{1}{2+y_3^2}\right) \right. \nonumber \\
&\hspace{1.7cm} \left. - 2\,\Li\left(\frac{2}{2+y_3^2}\right)\right]
\end{align}
where we have transformed $y_i=\frac{s_i}{\sqg}$ for $i=1,2$ and $y_3^2=\frac{Q_3^2}{\sqg}=2\min(1,\frac{\sgq}{\sqg})$.
Taking the limit for the soft region $y_3^2\rightarrow 2$ (since we
take the invariants as vanishing simultaneously), we see that
the remainder is just a finite term that contains no logarithms of the
vanishing invariants, 
\begin{align}
\label{eq:26}
\frac{\alpha_s C_A}{8\pi}\left[2\ln^2(2)+ \Li\left( \frac{1}{4}
  \right)\right]~.
\end{align}
We will receive this contribution twice. Including all divergent logarithmic
contributions and disregarding constant terms such as in
\eqRef{eq:26} , we find the same as in
the strong-ordering case,
\begin{equation}
-\mbox{PS:}\hspace{1cm}\sqg=\sgq=y\to 0 \hspace{1cm} \frac{\alpha_s C_A}{2\pi} \left[ \ln^2(y)+\frac{3}{2} \ln(y) \right]~,
\end{equation}
and hence the preferred choice of scale in the soft limit 
remains one which is linear in the vanishing invariants, such as
$\mu_\mrm{PS} \propto m_D$.  

In the hard collinear limit the Sudakov integrals plus the `Ariadne
Log' reduce to
\begin{equation}
-\mbox{PS:}\hspace{1cm}\sqg=y\to 0, \sgq\to s \hspace{1cm}
\frac{\alpha_s C_A}{2\pi} \left[-\frac16 C_A + 
\frac16 n_F\right]\ln(y)~,
\end{equation}
again the same as in the strongly ordered case, cf.~\eqRef{eq:mDstrongColl}. 

\begin{figure}[t!]
\centering
\setlength{\skipper}{2mm}
\begin{tabular}{ccc}
\includegraphics*[scale=0.48]{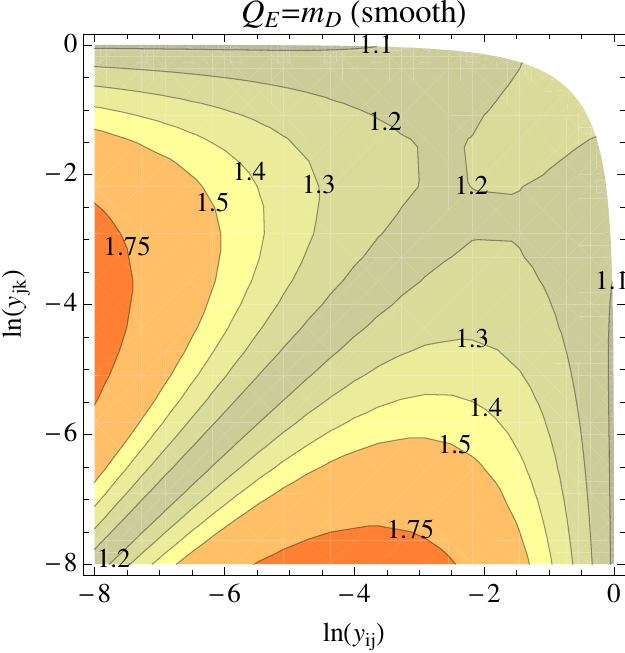}
&\includegraphics*[scale=0.48]{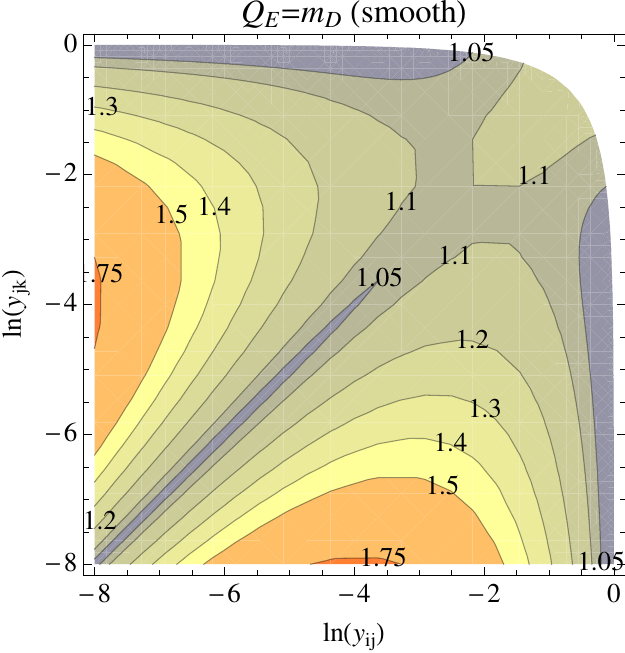}
& \includegraphics*[scale=0.48]{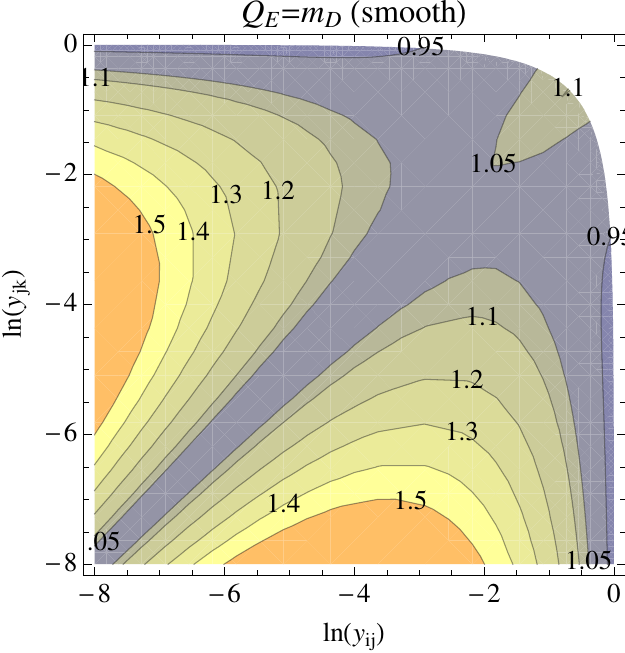} \\
(a) $\mu_\mrm{PS}=m_D$
& (b) $\mu_\mrm{PS}=\frac12m_D$
& (c) $\mu_\mrm{PS}=\frac12m_D$, with CMW
\end{tabular}
\caption{NLO correction factor for smooth $m_D$-ordering, with GGG
  antennae, and $\mu_\mrm{PS}=m_D$ (left), $\mu_\mrm{PS}=\frac12 m_D$
  (middle), and 
  $\mu_\mrm{PS}=\frac12 m_D$ with CMW rescaling (right). For all
  plots, $\alpha_s=0.12$, $n_F=5$, and 
  the evolution scale for gluon splittings was $m_{qq}$.  
\label{fig:mD_smooth_GGG} }
\end{figure}

To summarize, we do not expect
any qualitatively different limiting behaviour in the smoothly ordered
case with respect to the strongly ordered one, though details may of
course still vary. To illustrate this, we include the plots in
\figRef{fig:mD_smooth_GGG}. 
In all cases, we use a renormalization scale $\propto m_D$, but with
different prefactors, from left to right: $\mu_\mrm{PS} = m_D$,
$\mu_\mrm{PS} = m_D/2$, and finally $\mu_\mrm{PS} = m_D/2$ with CMW
rescaling. In particular the latter gives correction factors very
close to unity in both the soft and hard collinear limits, while we
still see the leftover divergence inbetween those limits that was also
present in the case of strong $m_D$-ordering,
cf.~\figRef{fig:mD_strong_GGG}. Nonetheless, it is worth noting that 
for a large region of phase space, say with $m_{ij} > 0.1\, m$
(corresponding to $\ln(y_{ij})>-4.6$),  
the corrections are still quite well behaved and relatively
small, less than $\sim 20\%$. 

\subsubsection{Transverse momentum}

Again we only need to recompute the contributions from the $3\to 4$
Sudakov terms, as the $2\to 3$ ones are the same as 
in the strongly ordered case. The $3\to 4$ Sudakov integrals are
\begin{align}
&-g_s^2 \left[ \sum_{j=1}^{2} C_A\int_0^{s_j} \left( 1 - \frac{Q_{3}^2}{Q_{Ej}^2 + Q_{3}^2} \right)\,d_3^0\: \dPS{\mrm{ant}} + \sum_{j=1}^2 2\,T_R\,n_F P_{A_j}\int_0^{s_j} \left( 1-\frac{Q_3^2}{m_{q\bar{q}}^2 + Q_3^2}\right)\,e_3^0\: \dPS{\mrm{ant}} \right] \nonumber \\
&=-\frac{\alpha_s}{4 \pi } \left[ \left( \frac{C_A}{\sqg}  \int_0^{\sqg} \d{s_{2}}  \int_0^{\sqg-s_2}  \hspace{-0.3cm}\d{s_{1}}\frac{4s_1s_2}{Q_3^2\sqg + 4s_1s_2}+\frac{C_A}{\sgq}  \int_0^{\sgq} \d{s_{2}}  \int_{0}^{\sgq-s_2} \hspace{-0.3cm}  \d{s_{1}} \frac{4s_1s_2}{Q_3^2\sgq + 4s_1s_2} \right) \right.    d_3^0\nonumber \\
&\left. \hspace{1.2cm}+ 2n_F\left(  \frac{P_{A_1}}{\sqg}\int_0^{\sqg} \d{s_{2}}  \int_{0}^{\sqg-s_2}  \d{s_{1}}   +\frac{P_{A_2}}{\sgq} \int_0^{\sgq} \d{s_{2}}  \int_{0}^{\sgq-s_2}  \d{s_{1}} \right) \frac{s_1}{ s_1 + Q_3^2}\,e_3^0\:  \right] \label{eq:SmOrdPT} 
\end{align}
As before we focus on explicitly calculating the transcendentality-2 
contribution arising from the eikonal part of the antenna in
 the first term in the first line of
\eqRef{eq:SmOrdPT},
\begin{align}
&-\frac{\alpha_sC_A}{2 \pi } \int_0^{1} \d{y_{2}}  \int_0^{1-y_2}  \d{y_{1}}\frac{4y_1y_2}{y_3^2 + 4y_1y_2}  \frac{1}{y_1y_2}\nonumber \\
&=-\frac{\alpha_sC_A}{2 \pi }\left[ -\Li\left(-\frac{2}{-1+\sqrt{1+y_3^2}}\right)-\Li\left( \frac{2}{1+\sqrt{1+y_3^2}}\right) \right]
\end{align}
where we have transformed $y_i = \frac{s_i}{\sqg}$ and $y_3^2=\frac{Q_3^2}{\sqg}$. In the limit $s_{\min}/s,s_{\max}/s=y\to 0$ so that $y_3^2\to 0,$ this yields
\begin{align}
&\frac{\alpha_s C_A}{2\pi} \left[  - \frac{1}{2}\ln^2(y)  \right]. 
\end{align}
Adding the contributions from the $2\to 3$ splitting and
transcendentality-1 terms, we find the following result for the soft
limit 
\begin{equation}
-\mbox{PS:}\hspace{1cm}\sqg=\sgq=y\to 0 \hspace{1cm} \frac{\alpha_s C_A}{2\pi} \left[  \ln^2(y) - \frac{1}{3}\ln(y)   \right]+ \frac{\alpha_s }{6 \pi}n_F\ln(y)~,
\end{equation} 
as in the strongly ordered case.
The double logarithm matches with \SVirtual and the single logarithm can be
absorbed by choosing a renormalization scale that is quadratic in the
vanishing invariants, such as $\mu_\mrm{PS}\propto p_\perp$. 

In the hard collinear limit, the shower integrals behave as
\begin{align}
-\mbox{PS:}\hspace{1cm}\sqg=y\to 0, \sgq\to s \hspace{1cm} 
\frac{\alpha_s}{2\pi} \left[ -\frac{1}{6}C_A + \frac16n_F \right] \ln(y)~,
\end{align} 
the same as in all the other cases. This completes the argument that
indeed $\mu_\mrm{PS}\propto p_\perp$ is the appropriate choice also
for smooth $p_\perp$-ordering. 

\begin{figure}[t!p]
\centering
\setlength{\skipper}{0mm}
\begin{tabular}{cc}
\includegraphics*[scale=0.48]{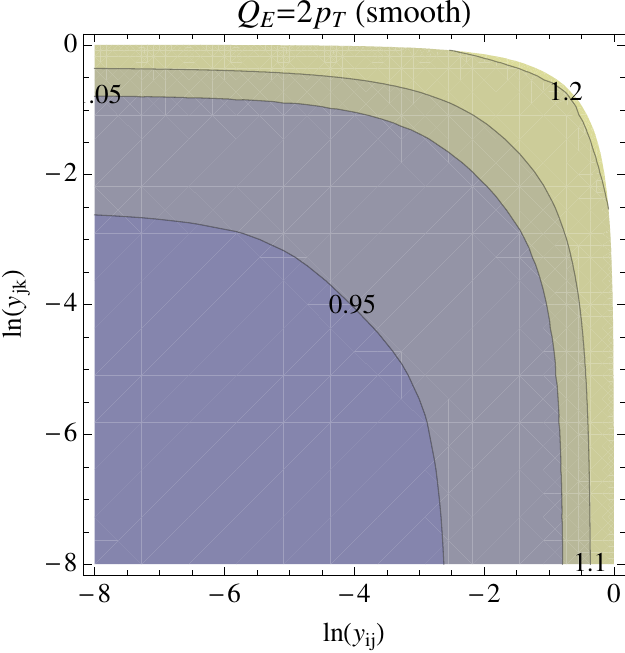}
&\includegraphics*[scale=0.48]{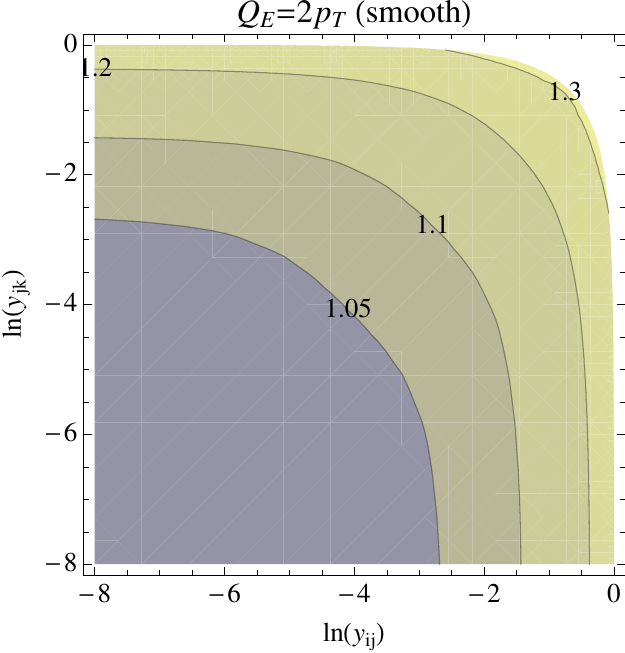}\\
(a) $\mu_\mrm{PS}=p_\perp$
& (b) $\mu_\mrm{PS}=2p_\perp$ \\
\includegraphics*[scale=0.48]{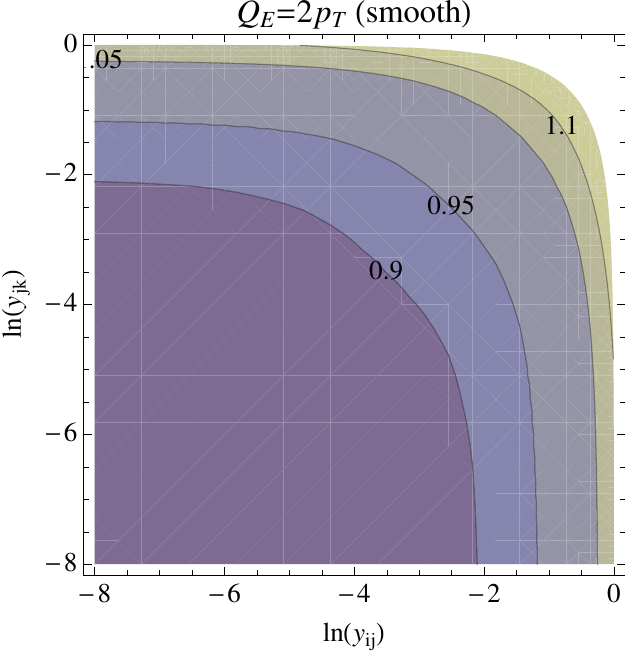}
&\includegraphics*[scale=0.48]{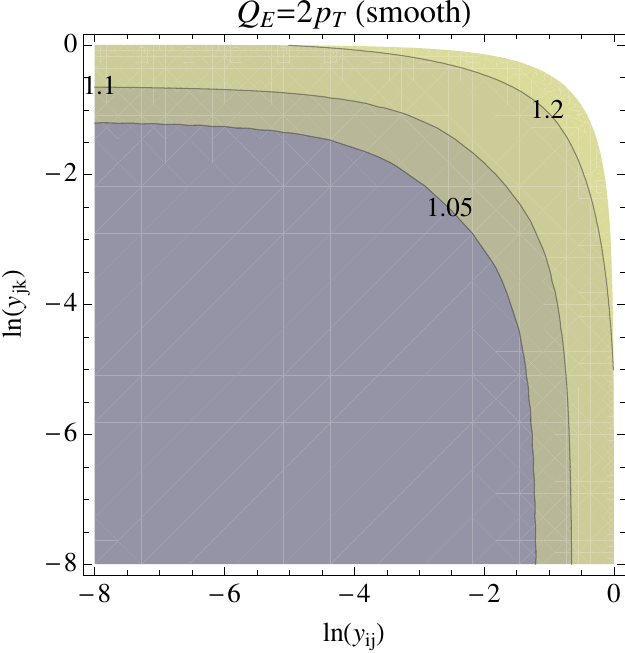} \\
(c) $\mu_\mrm{PS}=p_\perp$, with CMW
& (d) $\mu_\mrm{PS}=2p_\perp$, with CMW
\end{tabular}
\caption{NLO correction factor for smooth $p_\perp$-ordering, with GGG
  antennae, without (top row) and with (bottom row) the CMW rescaling
  of $\Lambda_\mrm{QCD}$. The left-hand panes use $\mu_\mrm{PS}=p_\perp$ 
and the right-hand ones $\mu_\mrm{PS} = 2p_\perp$. 
For all plots, $\alpha_s=0.12$, $n_F=5$, and
  the evolution scale for gluon splittings was $m_{qq}$.  
\label{fig:pT_smooth_GGG} }
\end{figure}

\begin{figure}[t!p]
\centering
\setlength{\skipper}{0mm}
\begin{tabular}{ccc}
\includegraphics*[scale=0.46]{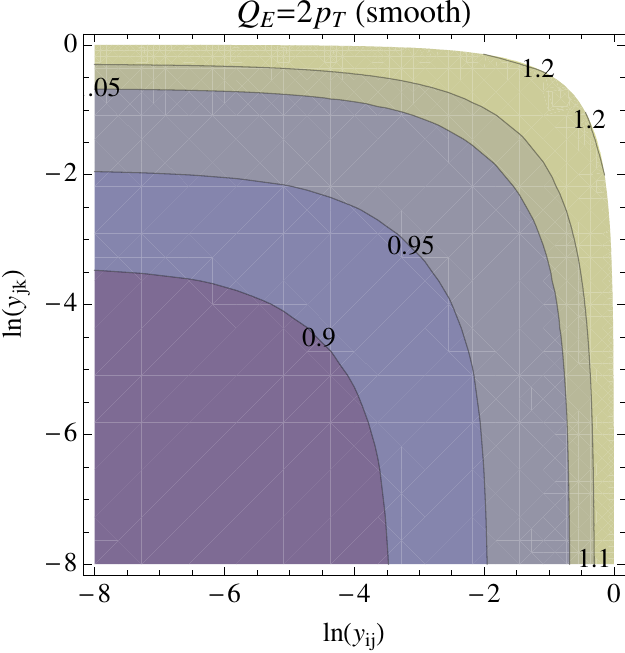}
&
\includegraphics*[scale=0.46]{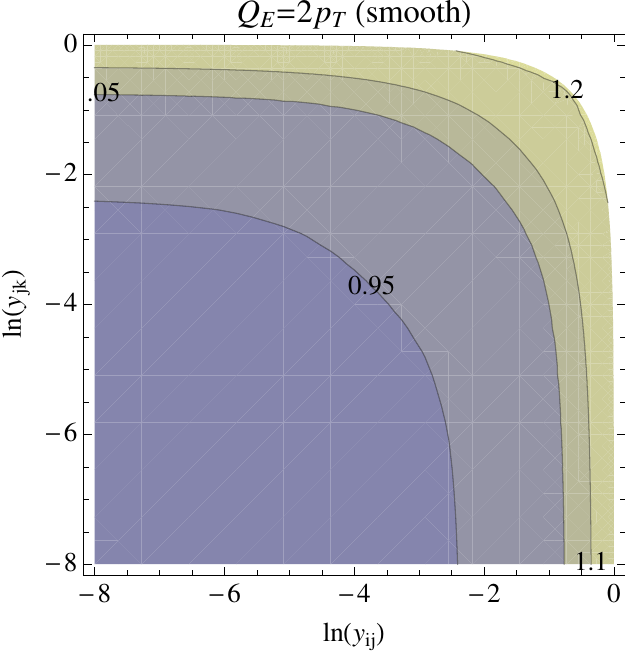}
&
\includegraphics*[scale=0.46]{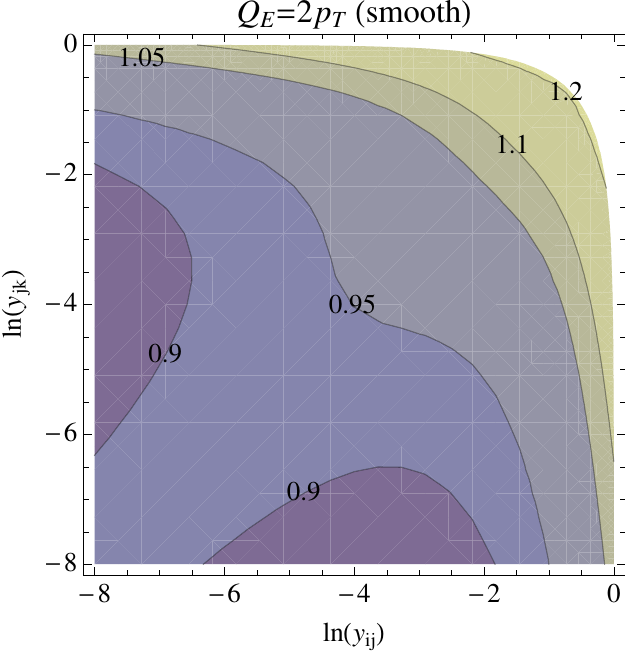} \\
(a) $\mu_\mrm{PS}=p_\perp$, $N_\perp = 1$
& (b) $\mu_\mrm{PS}=p_\perp$, $N_\perp=4$, 
& (c) $\mu_\mrm{PS}=p_\perp$, $N_\perp=4$,  \\
& $g\to q\bar{q}$ in $p_\perp$ & $P_{Aj}=1$
\end{tabular}
\caption{NLO correction factor for smooth $p_\perp$-ordering, with GGG
  antennae: variations of how gluon
  splittings are interleaved with gluon emissions, see text. We used
 $\alpha_s=0.12$, $n_F=5$, and $\mu_\mrm{PS}=p_\perp$.  
\label{fig:pT_smooth_2} }
\end{figure}

In \figRef{fig:pT_smooth_GGG}, we show the NLO correction factors,
$(1+V_{3Z})$, for smooth $p_\perp$-ordering.
The top row shows the correction factors without using the CMW
rescaling of $\Lambda_\mrm{QCD}$, and the plots in the bottom row
include it. For the left-hand panes, we used a shower renormalization scale of
$\mu_\mrm{PS}=p_\perp$, and for the right-hand ones we used $\mu_\mrm{PS} =
2p_\perp$. 

We see that all correction factors are essentially well-behaved, with
no runaway logs, similarly to the case of strong $p_\perp$-ordering.
However, for the case of smooth $p_\perp$-ordering, it looks as if
the CMW rescaling (bottom row) is almost doing ``too much'' in the
soft region. Given that the CMW arguments~\cite{Catani:1990rr} were
derived explicitly by considering the subleading behaviour of strongly
ordered (coherent) parton showers, we do not perceive of this as any major
drawback. Instead, one should merely be aware of the slight shifts in
the NLO corrections that result from applying it or not, recalling
that a rescaling of $\Lambda$ by the CMW factor $\sim$ 1.5 is within
the ordinary factor 2 variation of the renormalization scale that
is often taken as a standard for uncertainty estimates. 

The shifts caused by CMW rescaling and/or by
renormalization-scale prefactors are of
course fully taken into account in our implementation in the \Vc
code, and are thus reabsorbed into the one-loop matching coefficient
at the matched order, stabilizing the prediction. 
Differences at higher orders will result from
the fact that the 
CMW rescaling, if applied, is used for all shower branchings, while
the NLO correction derived here is only applied at the $Z\to 3$
stage of the calculation. 

Because smooth $p_\perp$-ordering is the default in \Vc  we
wish to understand this case as best as we can, and therefore we include some further
comparisons with non-default settings in \figRef{fig:pT_smooth_2}.

In the left figure of \figRef{fig:pT_smooth_2}, we modify the normalization of the
evolution variable from the \Vc default $Q^2_E=4p_\perp^2$ to the
\Ar choice $Q^2_E = p_\perp^2$. Though the normalization factor
cancels in the $P_\mrm{imp}$ factor for sequential gluon emissions, it is
relevant for deciding the relative ordering between gluon emissions
and gluon splittings. As this plot shows, however, the modification
only produces quite small differences in the NLO correction factor,
and with the ``wrong'' sign. Thus,
we retain $N_\perp=4$ as the default in \Vc. 
In middle figure of \figRef{fig:pT_smooth_2}, we change the evolution variable
for gluon splittings to be the same as that for gluon emissions,
i.e., $p_\perp$, with similar conclusions as for the previous
variation. 
In the right figure of \figRef{fig:pT_smooth_2}, we switch off the Ariadne factor
for gluon splittings. We notice that the NLO corrections get slightly
larger. There is no change along the diagonal $y_{ij} = y_{jk}$ since
the Ariadne factor is unity there, but along the edges of the plots, 
the NLO corrections become larger, which further motivates the choice
of keeping the Ariadne factor switched on by default in \Vc. 

The overall result is that the infrared limits are generally
well-behaved for $p_\perp$ evolution with $\mu_\mrm{PS}\propto
p_\perp$. Remaining differences amount to small finite shifts of order
10\%-20\% away from unity. At that level, the effective finite terms of the
antenna functions also play a role, hence it is too early to draw
definite conclusions just based on the plots presented here. 
The impact of finite terms will be studied in
\secRef{sec:results} in the context of matching to the LO matrix
elements for $Z\to 4$ partons, which effectively fixes the 
finite terms with respect to the pure-shower answers studied here.  

\subsection{Tables of Infrared Limits}
The results of the preceding subsections on the infrared limits of the
pole-subtracted matrix elements and of the 
Sudakov integrals generated by the various evolution-scale choices are
collected here, in parametric form, for easy reference. The
renormalization terms, $V_{3\mu}$, are not
included. \TabRef{tab:SVirtual} expresses the limits of \SVirtual,
while \tabRef{tab:Sudakovs} contains the Sudakov-integral
limits.
\begin{table}[h!]
\centering 
\begin{tabular}{| l | l | l |} 
\hline                        
\multirow{2}{*}{\SVirtual} & soft                     & $\left(-L^2-\frac{10}{3} L - \frac{\pi^2}{6}\right) C_A + \frac{1}{3} n_F L$  \\ \cline{2-3}
                                             & hard collinear &  $-\frac{5}{3} L C_A + \frac{1}{6} n_FL$ \\  \hline
\end{tabular}
\caption{Limits of \SVirtual, with $L$ denoting $\ln(y)$, where $y$ parametrizes the limit in the soft region taken along the diagonal of the phase space triangle $y=\sqg/s=\sgq/s\rightarrow 0$. The hard collinear limit only takes one invariant $\sqg/s$ or $\sgq/s$ to the soft limit while the other is set to $1$. We have omitted an overall factor of $\alpha_s/2\pi$.} 
\label{tab:SVirtual}
\end{table}

\begin{table}[h!]
\centering 
\begin{tabular}{| l | l | l | l | l |} 
\hline
\multicolumn{2}{|c|}{} & strong & smooth &  $V_{3Z} $ \\ \hline 
\multirow{2}{*}{$p_\perp$} & soft & $\left(L^2-\frac{1}{3}L+\frac{\pi^2}{6}\right)C_A + \frac{1}{3}n_FL$ & 
                                                        $\left(L^2-\frac{1}{3}L-\frac{\pi^2}{6}\right)C_A+ \frac{1}{3}n_FL$  & $-\beta_0L$ \\ \cline{2-5} 
                                         & hard collinear & $-\frac{1}{6}LC_A+\frac{1}{6}n_F L$ & $\left( -\frac{1}{6}L - \frac{\pi^2}{6} \right) C_A+\frac{1}{6}n_F L$ & $-\frac{1}{2}\beta_0 L$ \\ \hline                        
\multirow{2}{*}{$m_D$} & soft & $\left( L^2+\frac{3}{2}L-\frac{\pi^2}{6}  \right) C_A$ & $\left( L^2+\frac{3}{2}L-\frac{\pi^2}{6}\right) C_A$ & $-\frac{1}{2}\beta_0 L$ \\ \cline{2-5} 
                                         & hard collinear & $-\frac{1}{6}LC_A+\frac{1}{6}n_F L$ & $\left( -\frac{1}{6}L -\frac{\pi^2}{3}\right) C_A+\frac{1}{6}n_F L$ &$-\frac{1}{2}\beta_0 L$ \\ \hline                        
\end{tabular}
\caption{Limits of strong and smooth $p_\perp$ and $m_D$ ordering, with naming conventions as defined in \tabRef{tab:SVirtual}.
Non divergent terms, such as $\pi^2$ have been omitted in
  the calculation of $V_{3Z}$, and the renormalization term in
  $V_{3Z}$ is set to zero. An overall factor of $\alpha_s/2\pi$ is suppressed.
 } 
\label{tab:Sudakovs}
\end{table}

\section{Results including both LO and NLO corrections}
\label{sec:results}

In the preceding section, we focussed on deriving the analytic forms
of the shower integrals and comparing their infrared limits to the
matrix-element expressions. It is now time to include also the finite
terms arising from matching to the 4-parton tree-level matrix element,
expressed by the $\delta A$ terms in \eqRef{eq:V3Z}. Our ultimate aim
in this section is to include the full leading-colour one-loop
corrections through second order in $\alpha_s$ (i.e., up to and
including $Z\to 3$ partons) and combine these with the full-colour
tree-level corrections through third order in $\alpha_s$ (i.e., up to
and including $Z\to 5$ partons, the default in \Vc). However, since
we shall perform the $\delta A$ integrals numerically, adding those
terms to the analytic ones derived in the previous section, we first wish
to examine the numerical size and precision required on the $\delta A$
terms themselves. 

\subsection{Finite antenna terms and LO matching corrections}
Finite-term variations of the antenna functions (and in particular
fixing the finite terms via unitary LO matching corrections, such as
is done in \Vc~\cite{Giele:2011cb}) will
affect the terms 
generated by the $3\to 4$ Sudakov expansions in the following
way. Larger finite terms will cause an increased amount of $3\to 4$
branchings, which in turn will \emph{decrease} the associated 
Sudakov factor (in the sense of driving it closer to zero). This will
feed into the NLO correction factor, which compensates and 
drives the final answer back towards its NLO-correct value. (Note that 
similar variations will not occur
for the $2\to 3$ branching step, since we treat that as fixed
to the LO $3$-parton matrix element throughout.) This feedback
mechanism is encoded in the $\delta A$ terms in \eqRef{eq:V3Z}.

\begin{figure}[t]
\setlength{\skipper}{2mm}
\begin{tabular}{ccc}
\includegraphics*[scale=0.46]{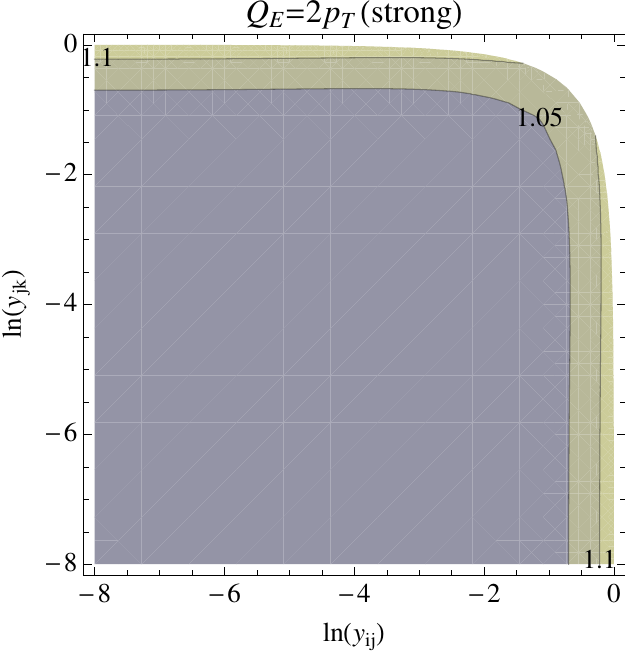}
&
\includegraphics*[scale=0.46]{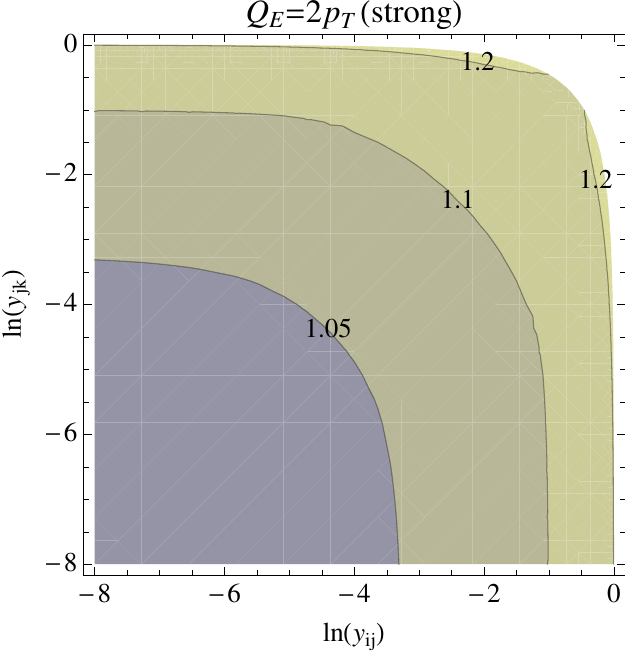}
&
\includegraphics*[scale=0.46]{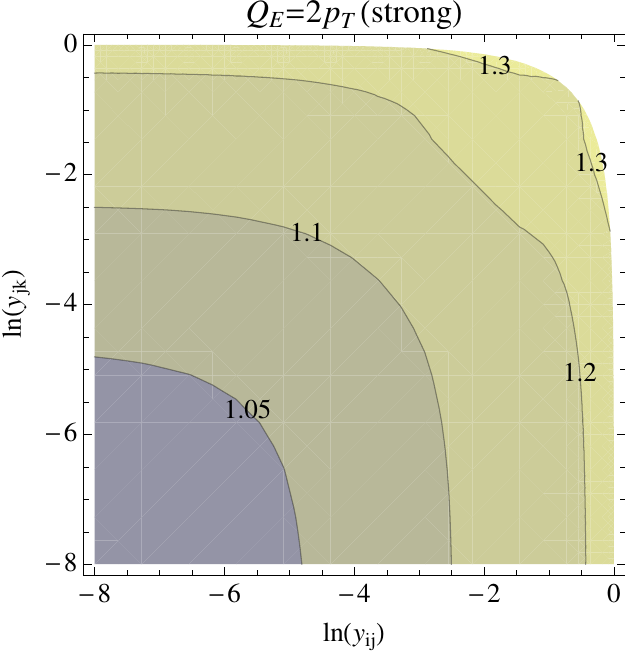} \\[-1mm]
MIN Antennae,
& VINCIA Antennae, 
& MAX Antennae,   \\[-1mm]
Strong & Strong & Strong \\[5mm]
\includegraphics*[scale=0.46]{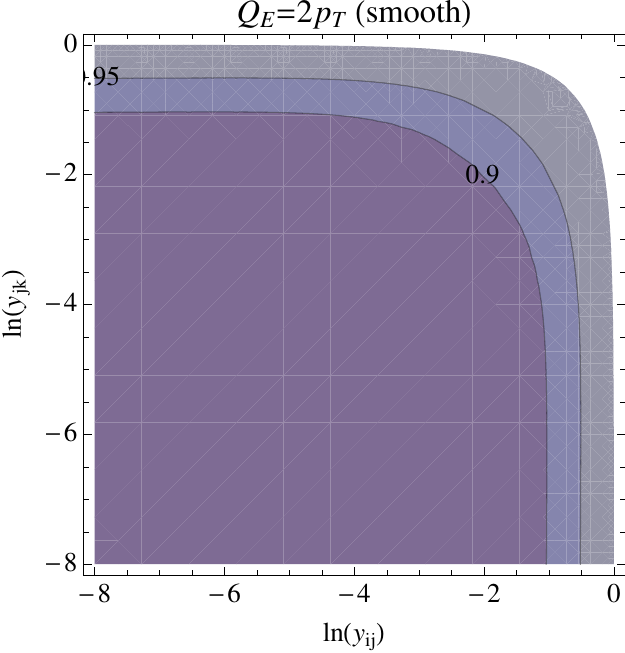}
&
\includegraphics*[scale=0.46]{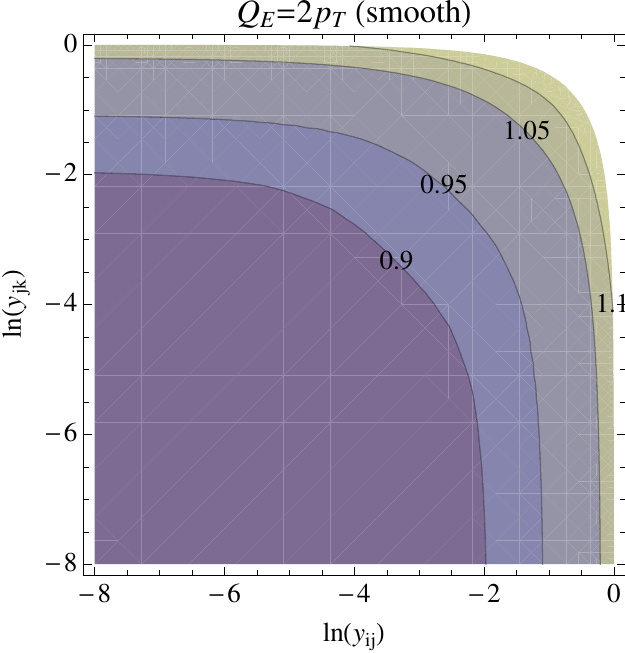}
&
\includegraphics*[scale=0.46]{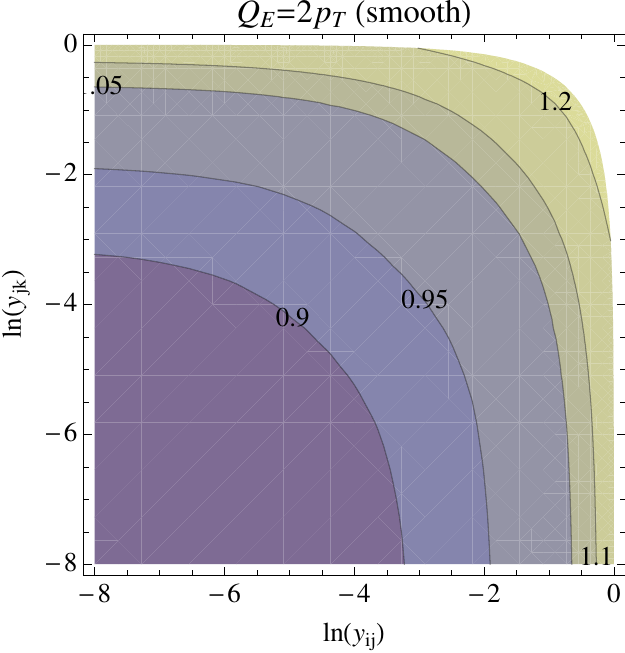} \\[-1mm]
MIN Antennae, 
& VINCIA Antennae, 
& MAX Antennae, \\[-1mm]
Smooth & Smooth & Smooth
\end{tabular}
\caption{NLO correction factor for strong (top row) and smooth (bottom
  row) $p_\perp$-ordering, for MIN (left), \Vc default (middle),
  and MAX (right) antenna functions. We use $\mu_R=p_\perp$ combined with
  CMW rescaling, $\alpha_s=0.12$, and gluon splitting in $m_{qq}$.  
\label{fig:finite} }
\end{figure}

Following the reasoning above, we should expect larger antenna 
finite terms to \emph{increase} the NLO correction factor (since, to
stabilize the 3-parton exclusive rate, it 
must compensate for losing more 3-parton phase-space points to
4-parton ones), and vice versa: 
smaller finite terms should result in a \emph{decrease} of the NLO
correction. At the pure-shower level (i.e., without LO matrix-element
corrections to fix the finite terms), this is illustrated by
\figRef{fig:finite}.
For ease of comparison, all plots use the CMW rescaling of
$\Lambda_\mrm{QCD}$, $\mu_\mrm{PS}=p_\perp$, $n_F=5$, and $\alpha_s(m_Z)=0.12$. 
The default antenna functions in \Vc\footnote{Note that
  \Vc was recently updated with a set of 
helicity-dependent antenna functions~\cite{Larkoski:2013yi}, so the
defaults shown here are  
not identical to the GGG ones, but are instead helicity sums/averages over
the functions defined in \cite{Larkoski:2013yi}.}
 are shown in the middle 
panes, for strong (top row) and smooth (bottom row) ordering,
respectively. A variation with smaller finite terms for the $3\to 4$
antenna functions is shown to the left, and one with larger finite
terms on the right. As expected, the NLO correction factors react by
becoming lower for smaller finite terms and higher for larger finite
terms, for both strong and smooth ordering.

We emphasise that the plots in \figRef{fig:finite} are 
shown purely for illustration, to give a feeling for the changes
produced by finite-term variations. In the actual matched shower
evolution, the constraint imposed by matching to the LO
4-parton matrix elements fixes the finite terms, via the unitary 
procedure derived in \cite{Giele:2011cb}, which was briefly recapped in
\secRef{sec:LO}. The effective finite terms then depend on 
the full LO 4-parton matrix elements, and have a more complicated 
structure than the simple antenna functions we have so far been
playing with. We shall therefore not attempt to integrate them analytically,
but prefer instead to let \Vc compute a numerical MC estimate for
us. 

Each point in that MC integration will involve computing at least
one LO 4-parton matrix element, hence it is crucial to know how many
points will be needed to obtain sufficient accuracy. Since everything
else is handled analytically, this will be the deciding factor in
determining the speed of the NLO-corrected algorithm. We shall perform a
speed test below in \secRef{sec:speed}, but first we need to determine
the accuracy we need on the $\delta A$ integral. 

\begin{figure}[t!]
\centering
\begin{tabular}{ccc}
\includegraphics*[scale=0.48]{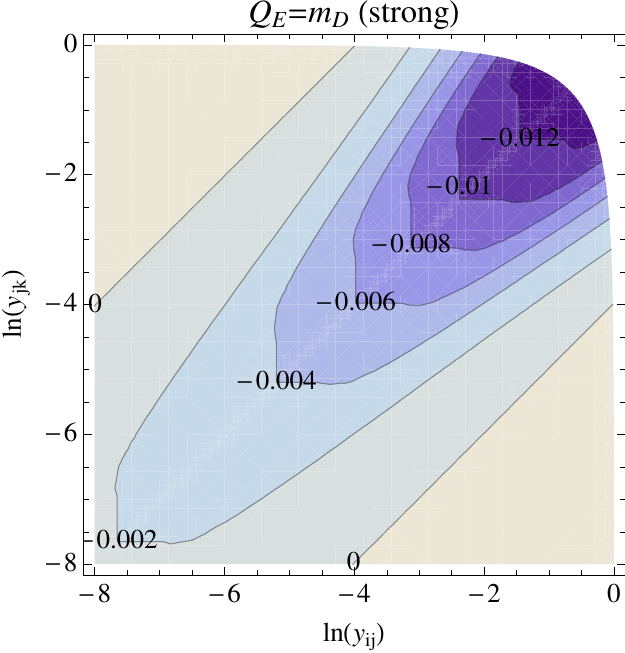}
&
\includegraphics*[scale=0.48]{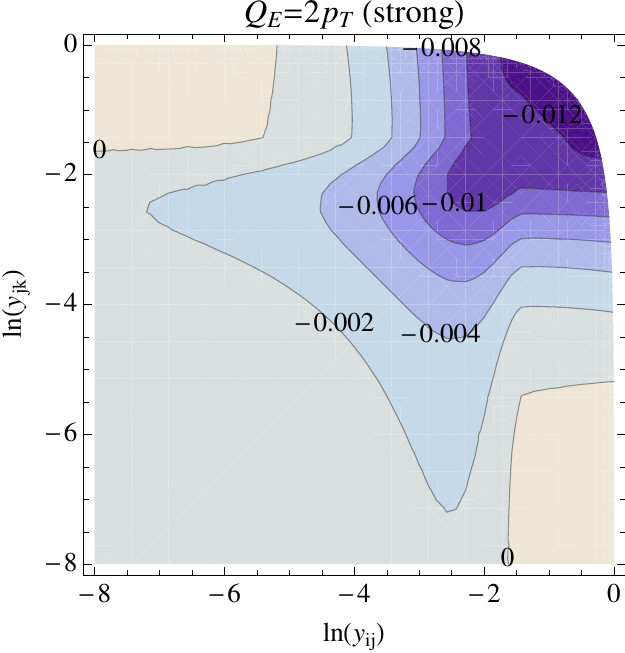}
&
\includegraphics*[scale=0.48]{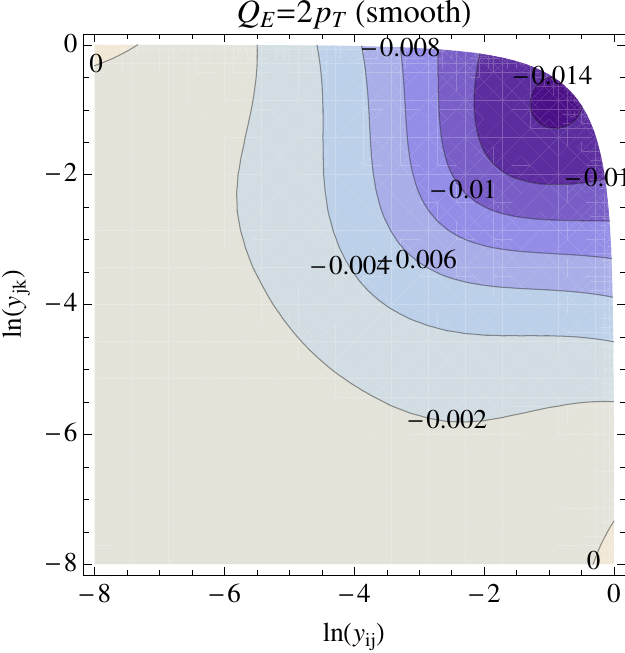}
\end{tabular}
\caption{Size of $\delta A$ terms differences between GGG and \Vc
  default antennae. 
\label{fig:deltaA}}
\end{figure}

A first analytic estimate of the size of the $\delta A$ terms can be
obtained by simply computing the ones produced by switching from GGG
 to the \Vc default antennae (summed and averaged over
 helicities~\cite{Larkoski:2013yi}),  
with the following ${\cal O}(1)$ finite-term differences:
\begin{align}
qg \to qgg &: & 
  F^{\mbox{\scriptsize\Vc}}_\mrm{Emit} - F^{\mbox{\scriptsize
      GGG}}_\mrm{Emit} & \ = \ 1.5 \ -\ (2.5 - y_{ij} - 0.5 y_{jk}) & & = 
 -1 + y_{yij} + 0.5 y_{jk} ~, \\[2mm]
qg \to q\bar{q}'q' &: &
  F^{\mbox{\scriptsize \Vc}}_\mrm{Split} - F^{\mbox{\scriptsize 
      GGG}}_\mrm{Split} & \ = \ 0.0 \ - \ (-0.5 + y_{ij}) & &=  \ 
0.5 - y_{ij}
~,
\end{align}
with $F_\mrm{Emit}$ and $F_{\mrm{Split}}$ defined in \eqsRef{eq:aEmit}
and \eqref{eq:aSplit}. The $\delta A$ terms produced by these
differences are plotted in \figRef{fig:deltaA}, for strong ordering in
$m_D$ (left) and $p_\perp$ (center), and for smooth
ordering in $p_\perp$ (right), respectively. 
As expected, they do come out to be numerically subleading, roughly of
order $\alpha_s/(2\pi)$, relative to
LO (unity), yielding corrections ranging from a few permille to about
a percent of the LO result. 

\begin{figure}[t!]
\centering
\vspace*{-3.5cm}\hspace*{-0.5cm}%
\includegraphics*[scale=0.38]{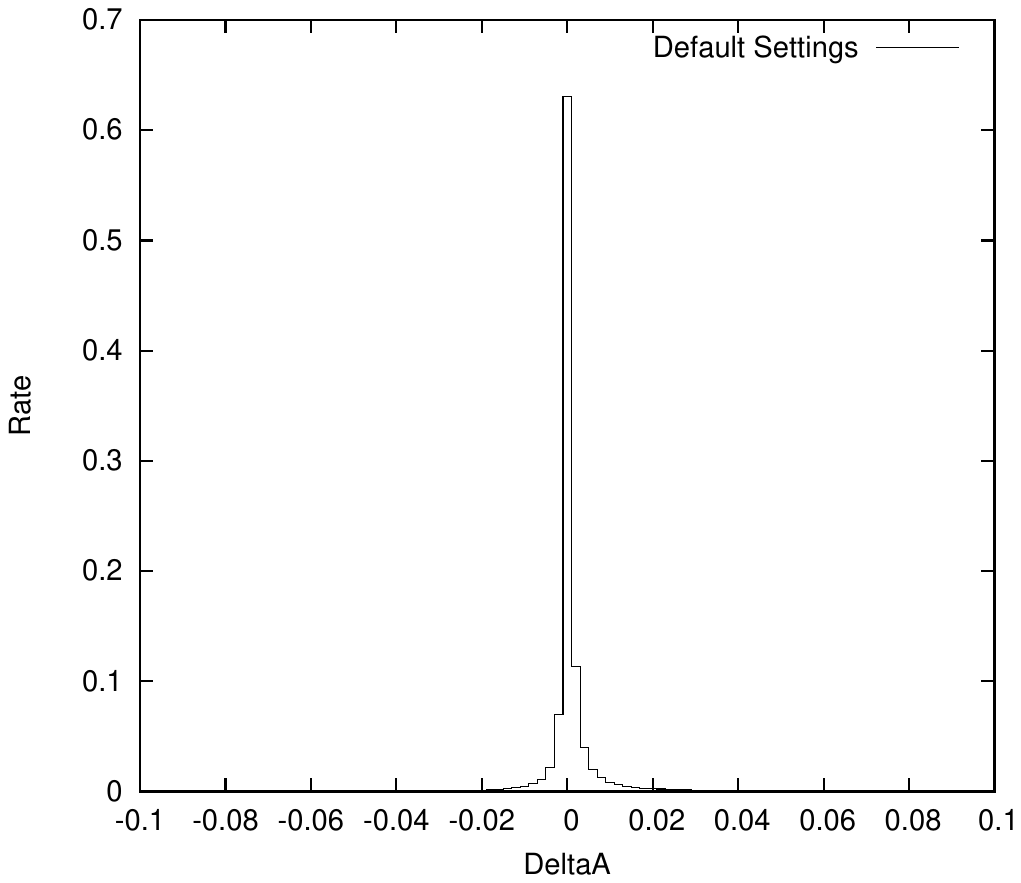}\hspace*{-1cm}%
\includegraphics*[scale=0.38]{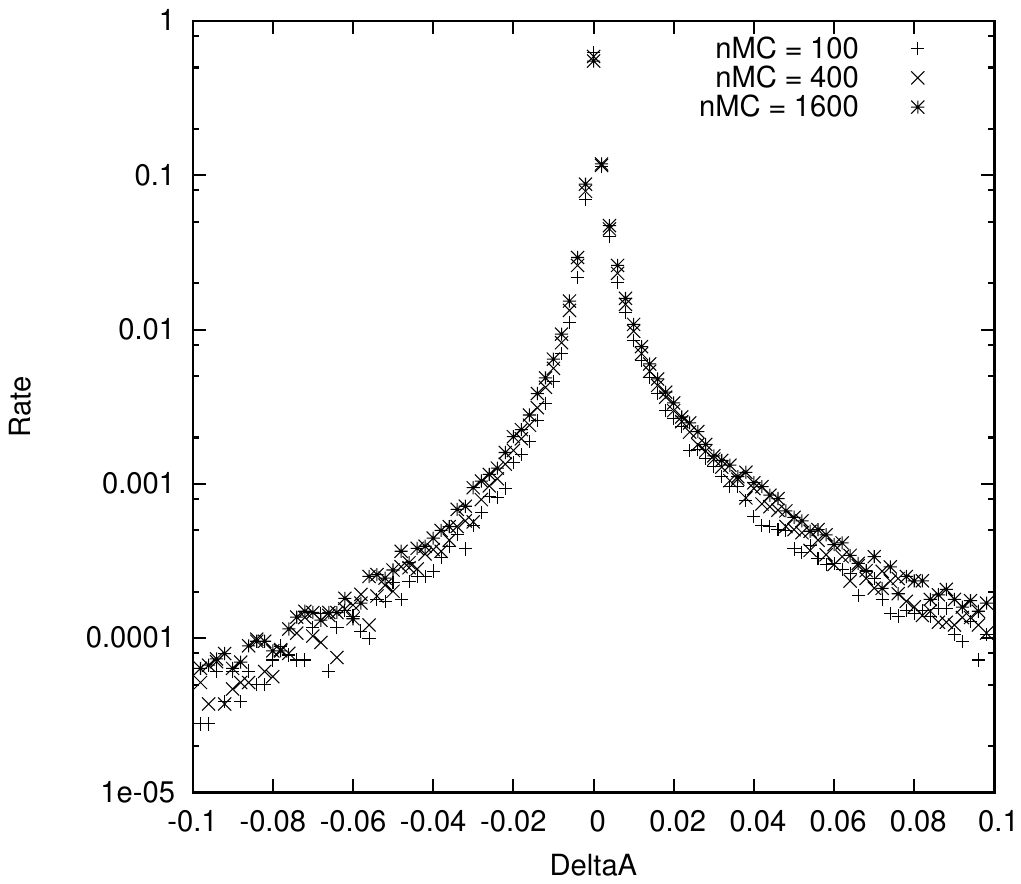}\hspace*{-0.5cm}\vspace*{-0.5cm}%
\caption{Distribution of the size of the $\delta A$ terms (normalized
  so the LO result is unity) in actual \Vc runs (v.1.1.00). {\sl Left:} linear
  scale, default settings. {\sl Right:} logarithmic scale, with
  variations on the minimum number of MC points used for the
  integrations (default is 100).
\label{fig:deltaArun}}
\end{figure}

Finally, in \figRef{fig:deltaArun}, we include the full LO 4-parton
matrix elements and plot the distribution of numerically computed
$\delta A$ terms during actual \Vc runs, for 100,000 events. 
The result is now represented
by a one-dimensional histogram, with $\delta A$ on the $x$-axis and
relative rate on the $y$-axis. On the left-hand pane, the $\delta A$
distribution with default settings is shown on a linear scale, while
the right-hand pane shows the same result on a logarithmic scale, 
including variations with higher numerical accuracy.

As mentioned above, the integration is done by a uniform Monte Carlo
sampling of the $\delta A$ integrands. We require a numerical
precision better than $1\%$ on 
the estimated size of the term (relative to LO) 
and, by default, always sample at least 100 MC
points for each antenna integral. In the left-hand pane of
\figRef{fig:deltaArun}, we see that, even with the full 4-parton LO
matrix-element corrections included, the size of the $\delta A$
terms remains below one percent for the vast majority of 3-parton
phase-space points. 

On the logarithmic scale in the right-hand pane of \figRef{fig:deltaArun},
however, it is evident that there is also a tail of quite rare
phase-space points which are associated with larger $\delta A$
corrections. Numerical investigations reveal that this tail is mainly 
generated by the integrals over the $g\to q\bar{q}$ terms, in
particular in phase-space points in which the gluon is collinear to
one of the original quarks. 
This agrees with our expectation that these terms are the ones to
which the pure shower gives the ``worst'' approximation, and hence
they are the ones that receive the largest matrix-element corrections. 
As a test of the numerical stability of the NLO corrections for
these points, we increased the minimum number of MC points used for the
$\delta A$  integration from the default 100 (shown with ``$+$'' symbols) to 400
(``$\times$'' symbols) and 1600 (``$*$'' symbols), 
cutting the expected statistical MC error in half with each
step, at the cost of increased event-generation
time. Though we do observe a slight broadening of the distribution
between the default and the higher-precision settings, the shifts
should be interpreted horizontally and remain well under the required
percent-level precision with respect to LO. The default settings are
therefore kept at a minimum of 100 MC points, though we note that future
investigations, in particular of more complicated partonic topologies, 
may require developing a better understanding of, and perhaps a better
shower approximation to, these integrals, 
especially the $g\to q\bar{q}$ contribution.

For completeness, we note that the runs used to obtain these
distributions were performed using the new default ``Nikhef'' tune of \Vc's
NLO-corrected shower, which will be described in more detail in the
following subsection. Parameters for the tune are given in
\appRef{app:tunes}. 

\subsection{LEP Results \label{sec:LEP}}

Since we have restricted our attention to massless
partons in this work, we shall mainly consider the
light-flavour-tagged event-shape and fragmentation distributions
produced by the L3 experiment at LEP for our validations and tuning,
see~\cite{Achard:2004sv}. We consider three possible \Vc settings:
\begin{itemize}
\item New default (NLO on): uses two-loop running for $\alpha_s$, with
  CMW rescaling of $\Lambda_\mrm{QCD}$. From the comparisons to
  event-shape variables presented in this section, we settled on a
  value of $\alpha_s(M_Z) = 0.122$. A few modifications to the
  string-fragmentation parameters were made, relative to the old
  default, to compensate for differences in the region close to the
  hadronization scale. The revised parameters are 
 listed in \appRef{app:tunes}, under the ``Nikhef'' tune. 
\item New default (NLO off). Identical to the previous
  bullet point, but with the NLO correction factor switched off. 
\item Old default (LO tune): uses one-loop running for $\alpha_s$, without
  CMW rescaling of $\Lambda_\mrm{QCD}$, and $\alpha_s(M_Z) =
  0.139$. The string-fragmentation parameters are those of the
  ``Jeppsson 5'' tune, see \appRef{app:tunes}.
\end{itemize}

\begin{figure}[t!p]
\centering
\begin{tabular}{ccc}
\includegraphics*[scale=0.245]{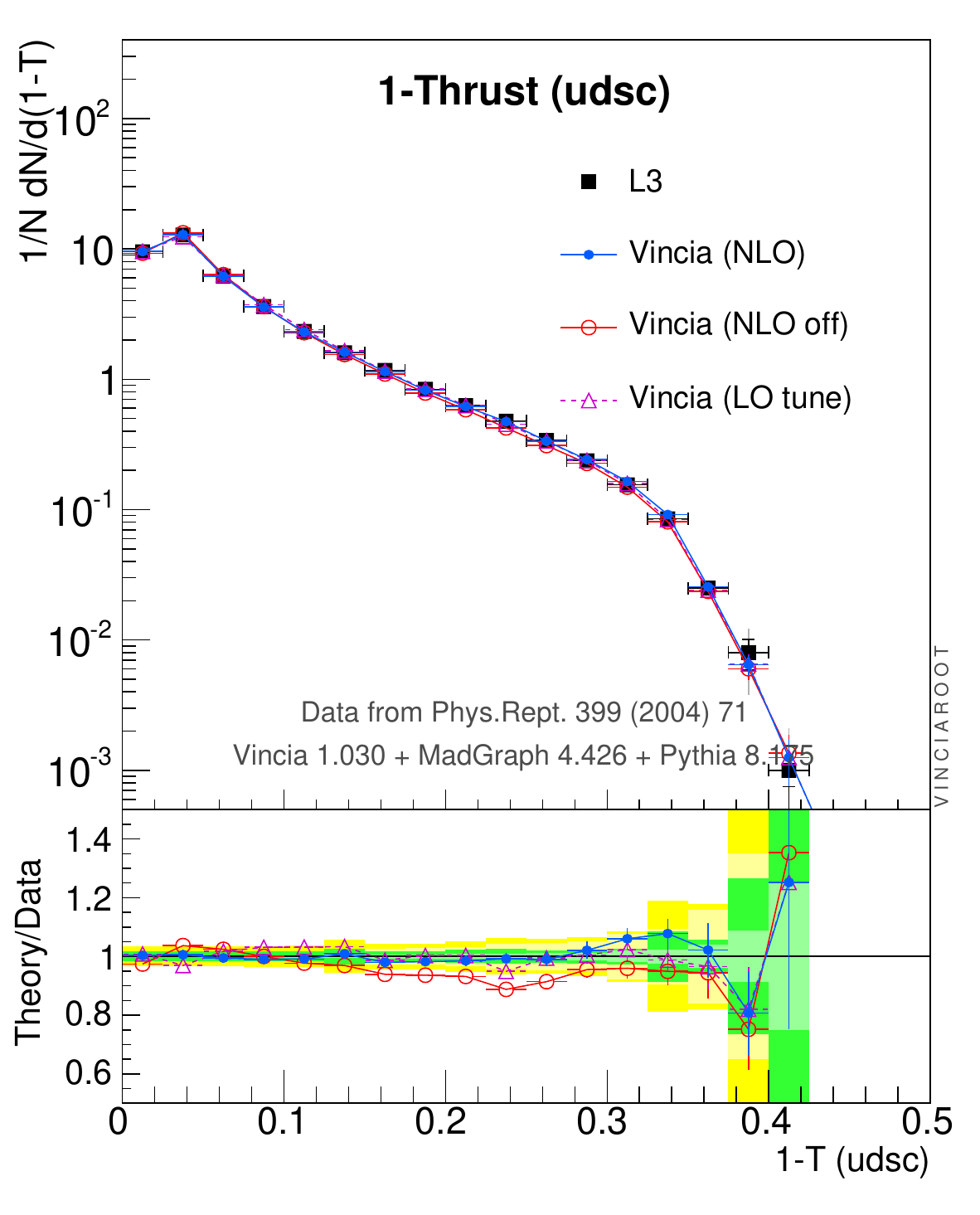}
&
\includegraphics*[scale=0.245]{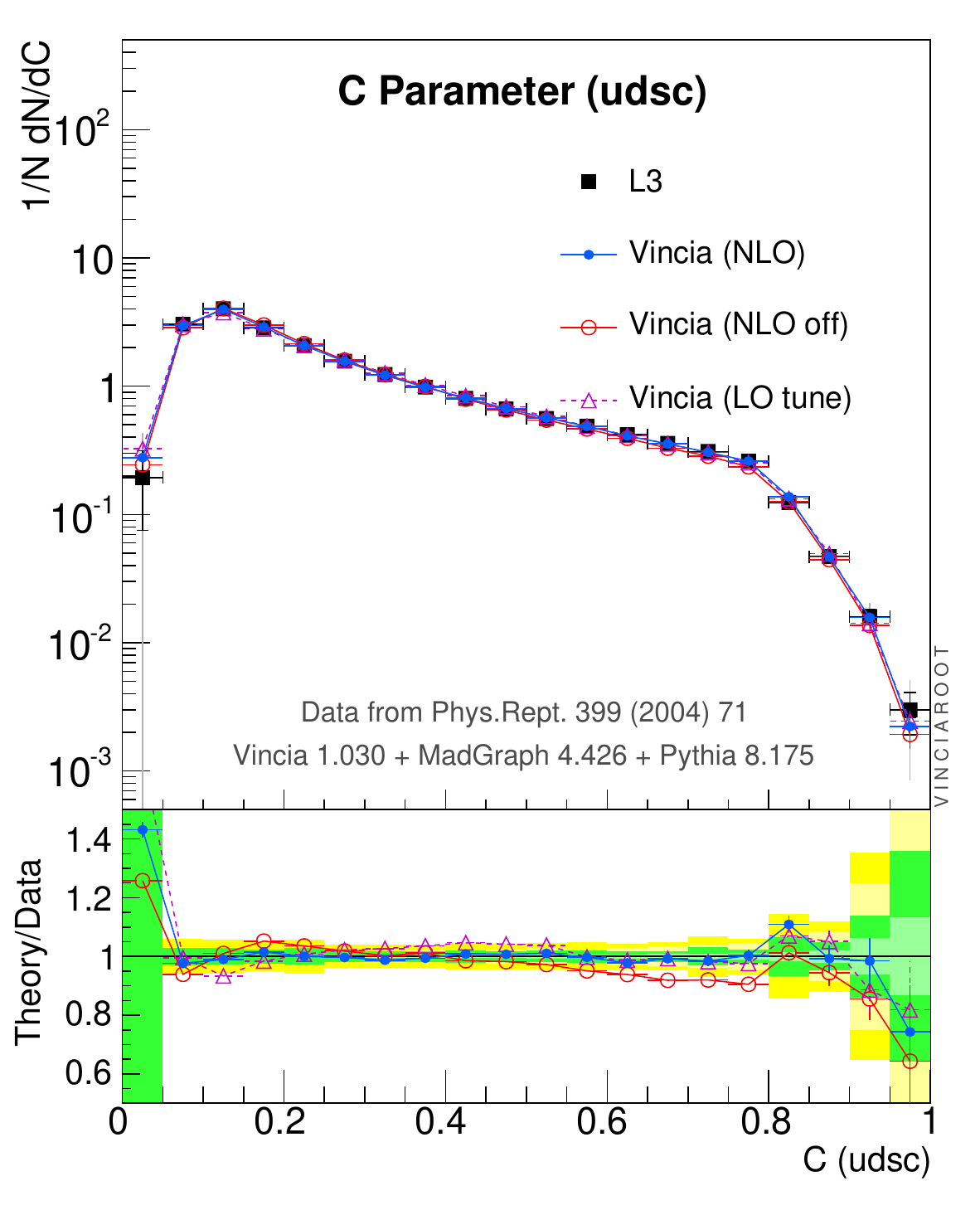}
&
\includegraphics*[scale=0.245]{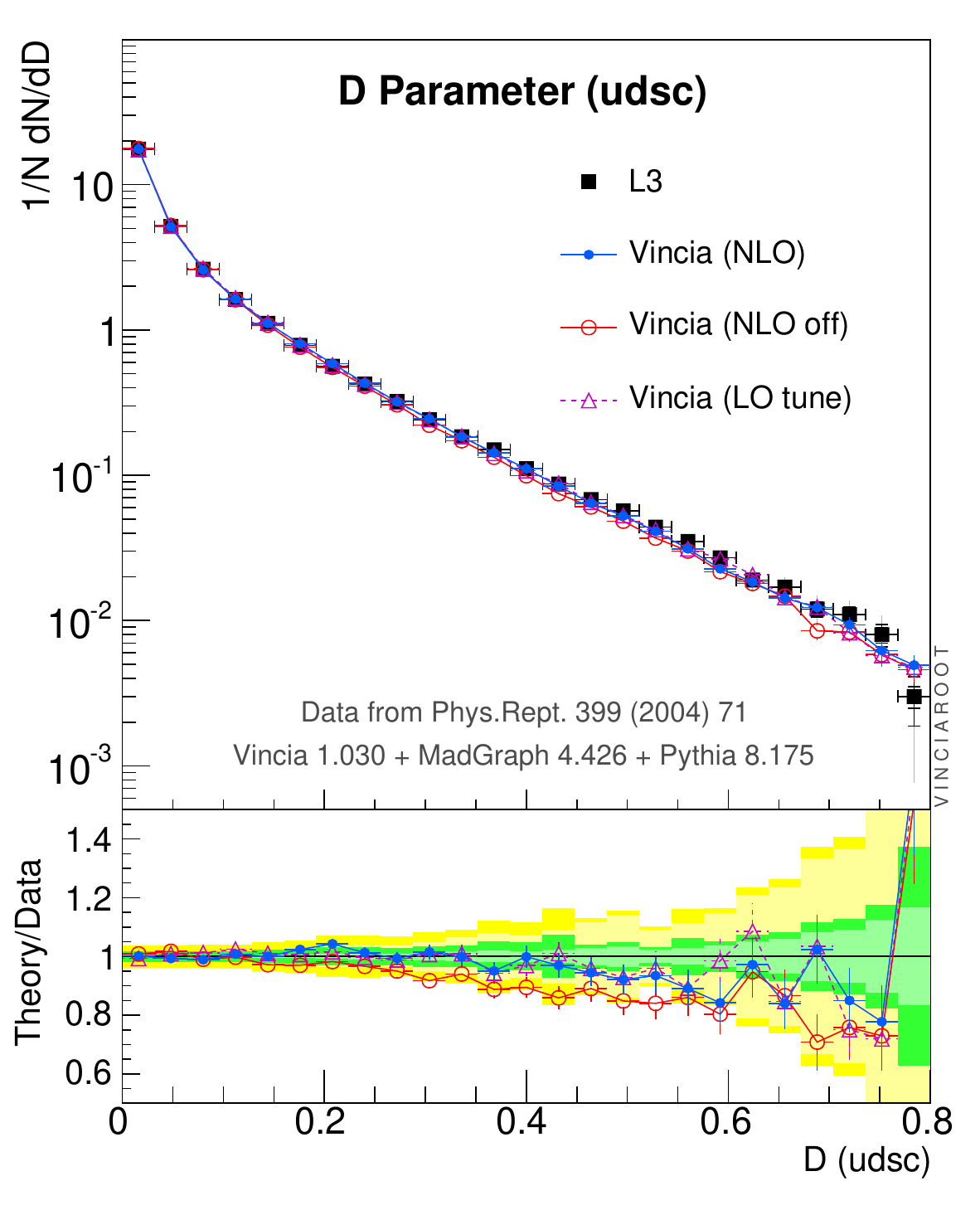} \\
(a) & (b) & (c)
\end{tabular}
\caption{L3 light-flavour event shapes: Thrust, $C$, and $D$. 
\label{fig:LEP1}}
\end{figure}

The three main event-shape variables that were used to determine the
value of $\alpha_s(M_Z)$ are shown in \figRef{fig:LEP1}, with upper
panes showing the distributions themselves (data and MC) and lower panes
showing the ratios of MC/data, with one- and two-sigma uncertainties
on the data shown by darker (green) and lighter (yellow) shaded bands,
respectively. The Thrust ({\sl left}) and $C$-parameter ({\sl middle}) 
distributions both have perturbative expansions that start at ${\cal
  O}(\alpha_s)$ and hence they are both explicitly sensitive to the
corrections considered in this paper. The 
expansion of the $D$ parameter ({\sl right}) begins at ${\cal
  O}(\alpha_s^2)$. It is sensitive to the NLO 3-jet corrections mainly
via unitarity, since all 4-jet events begin their lives as 3-jet
events in our framework. It also represents an important cross-check on the
value extracted from the other two variables. 

For a pedagogical description of the variables, see~\cite{Achard:2004sv}.
Pencil-like 2-jet configurations are to the left (near zero) for all
three observables. This region is particularly sensitive to
non-perturbative hadronization corrections. More spherical events,
with several hard perturbative emissions, are towards the right (near
$0.5$ for Thrust and 1.0 for $C$ and $D$). The maximal $\tau=1-T$ for
a 3-particle 
configuration is $\tau = 1/3$ (corresponding to the Mercedes
configuration), beyond which only 4-particle (and higher) states can
contribute. This causes a noticeable change in slope in the
distribution at that point, see the left pane of \figRef{fig:LEP1}. 
The same thing happens for the $C$ parameter at $C=3/4$, in
the middle pane of \figRef{fig:LEP1}. 
The $D$ parameter is 
sensitive to the smallest of the eigenvalues of the sphericity
tensor, and is therefore zero for any purely planar event, causing it
to be sensitive only to 4- and higher-particle configurations over its
entire range. 

Both the new NLO tune (solid blue line with filled-dot symbols) 
and the old LO one (dashed magenta line with open-triangle symbols)
reproduce all three event shapes very well. With the NLO corrections
switched off (solid red line with open-circle symbols), the new tune
produces a somewhat too soft spectrum, consistent with its low value
of $\alpha_s(M_Z)$ not being able to describe the data without the
benefit of the NLO 3-jet corrections.

\begin{figure}[t!]
\centering
\begin{tabular}{cc}
\includegraphics*[scale=0.265]{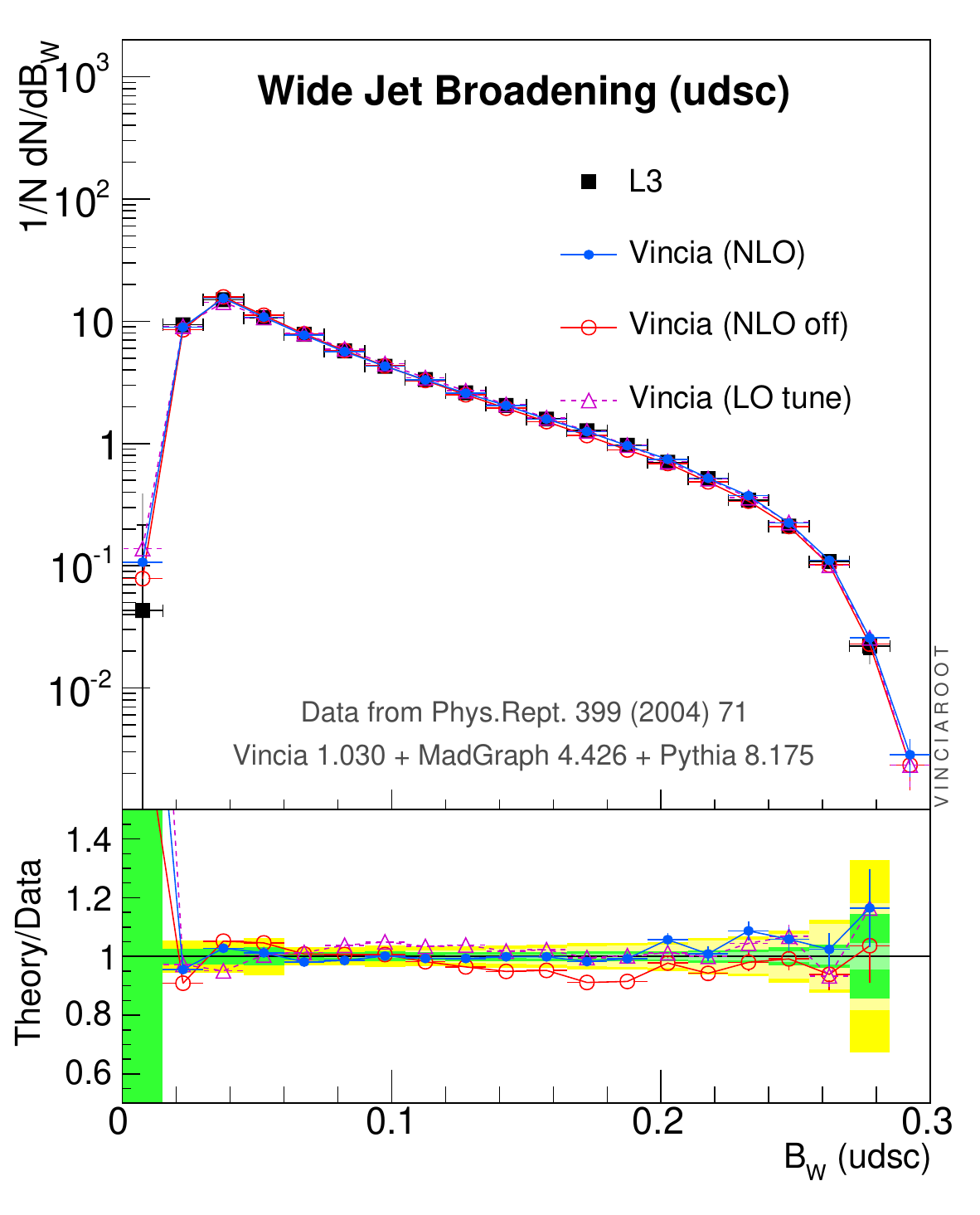}
&
\includegraphics*[scale=0.265]{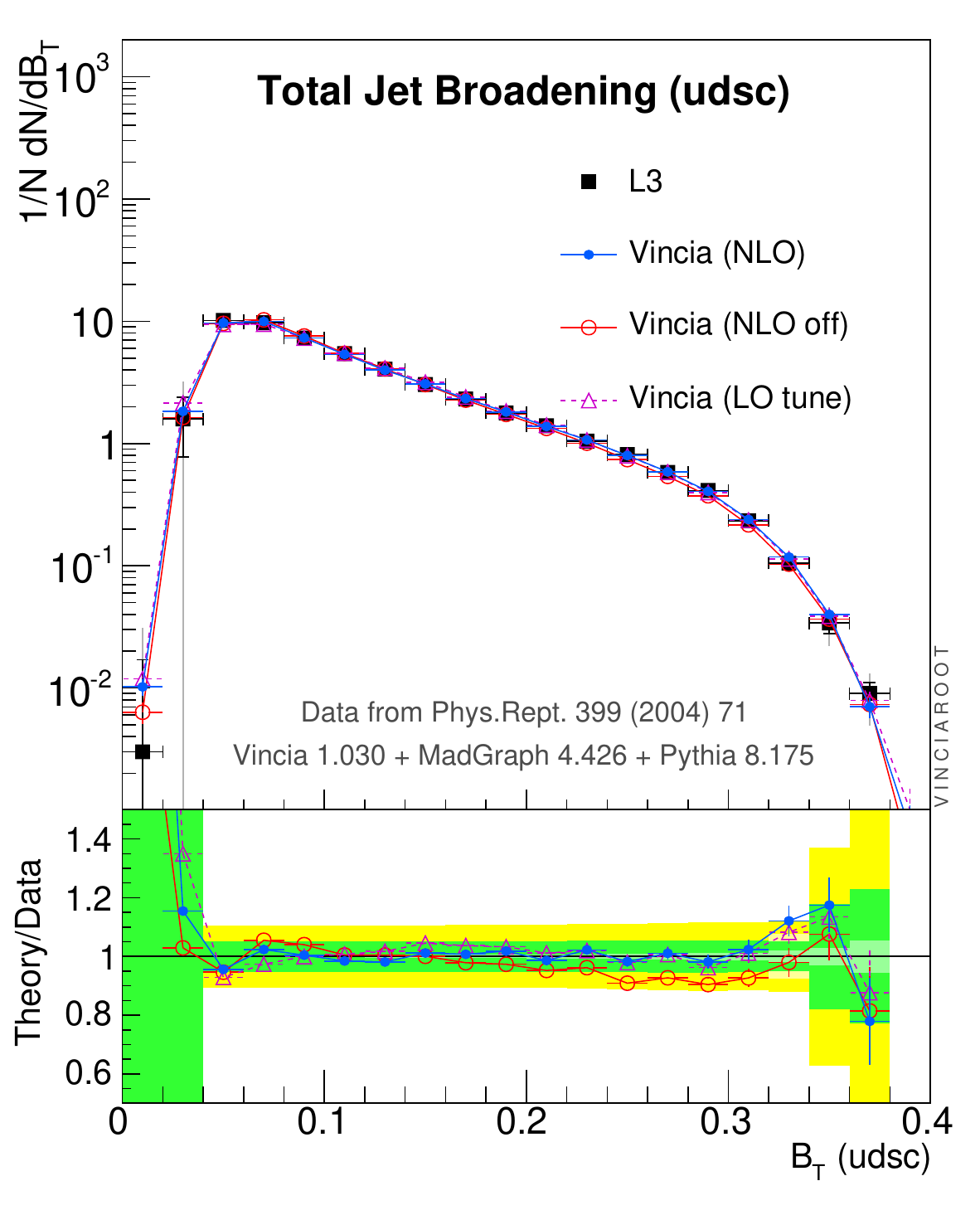}\\
(a) & (b)
\end{tabular}
\caption{L3 light-flavour event shapes: jet broadening
\label{fig:LEP2}}
\end{figure}

As a further cross check, we show two further event-shape variables
that were included in the L3 study in \figRef{fig:LEP2}: the Wide and
Total Jet Broadening parameters, $B_W$ and $B_T$,
respectively. 
These have a somewhat different and complementary
sensitivity to the perturbative corrections, compared to the
variables above, picking out mainly the transverse component of jet
structure. They are equal at ${\cal O}(\alpha_s)$, 
but $B_T$ receives somewhat larger ${\cal O}(\alpha_s^2)$ corrections
than $B_W$. Again, we see that both the old (LO) and new (NLO) 
defaults are able to describe the data, and that the spectrum with the
new default value for $\alpha_s(M_Z)$ is too soft if the NLO
corrections are switched off.

\begin{figure}[t!]
\centering
\begin{tabular}{cc}
\includegraphics*[scale=0.265]{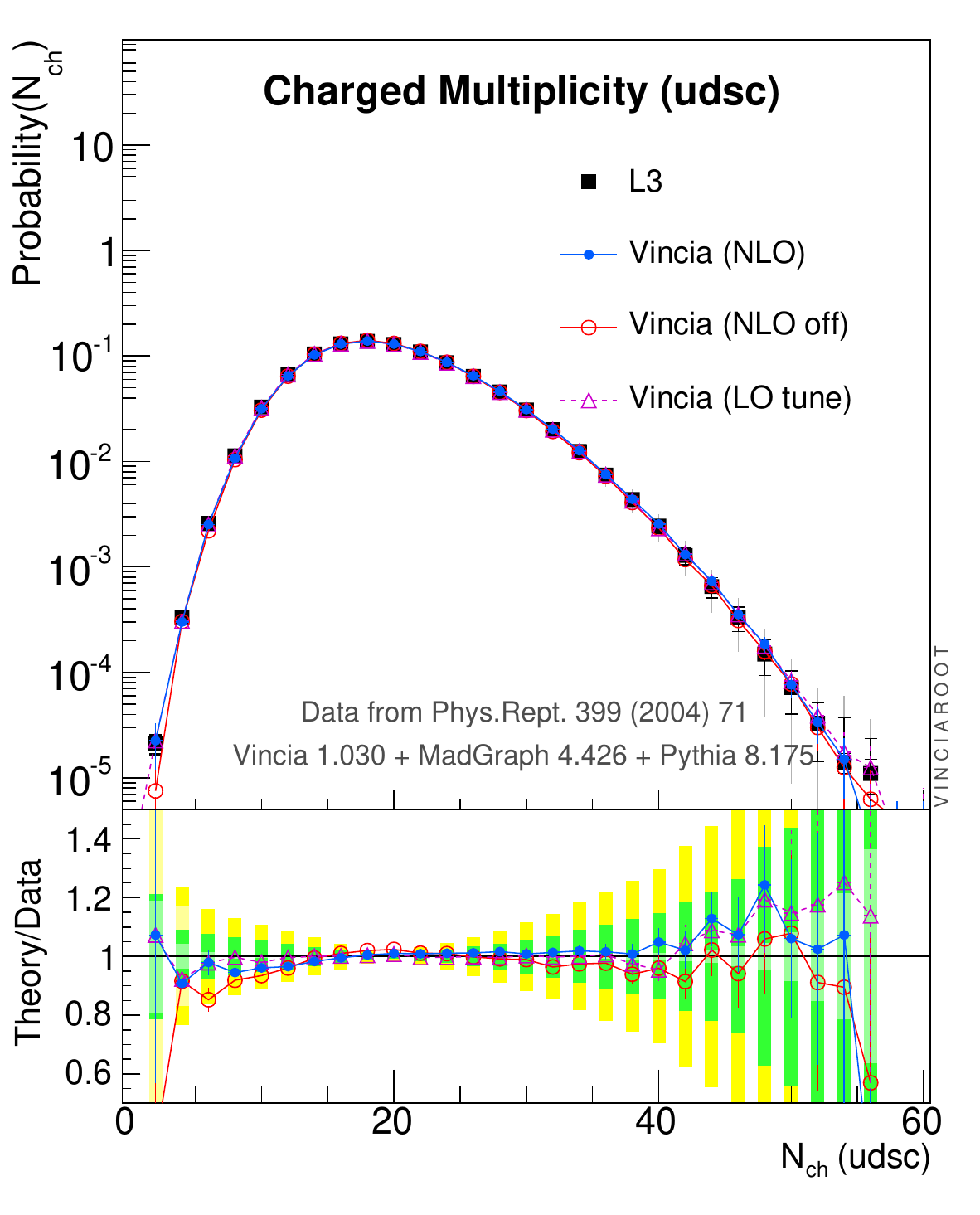}
&
\includegraphics*[scale=0.265]{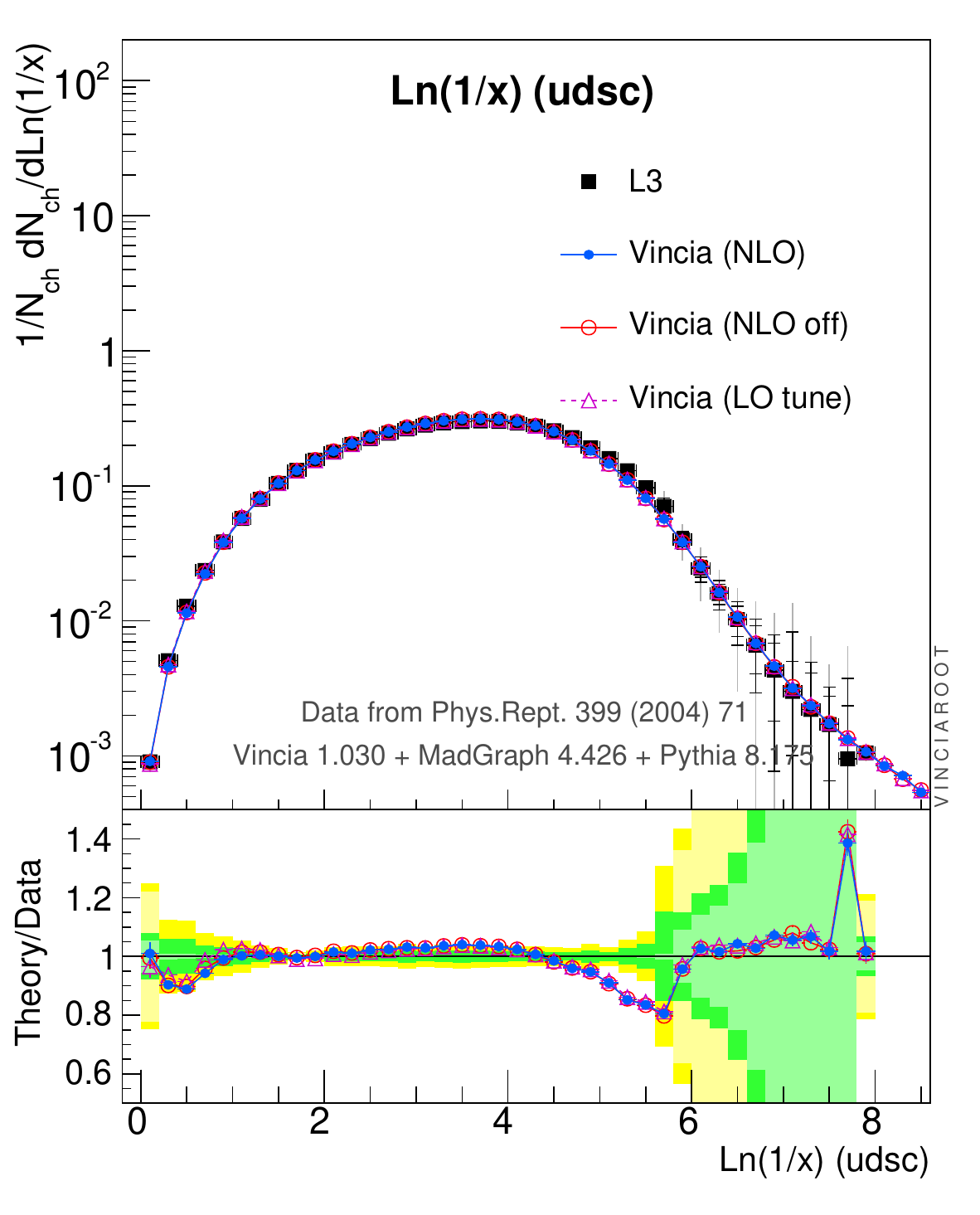} \\
(a) & (b)
\end{tabular}
\caption{L3 light-flavour fragmentation observables: charged-track
  multiplicity and momentum distribution.
\label{fig:LEP3}}
\end{figure}
Finally, as an aid to constraining the Lund fragmentation-function
parameters, the L3 study also included two infrared-sensitive
observables: the charged-particle 
multiplicity and momentum distributions, to which we compare 
in \figRef{fig:LEP3}, with the momentum fraction defined as 
\begin{equation}
x = \frac{2|p|}{\sqrt{s}}~.
\end{equation}
There is again no noteworthy differences between the old and new
default tunes.

Having determined the value of $\alpha_s(M_Z)$ and the parameters of
the non-perturbative fragmentation function, we extended the
validations to include a set of jet-rate and jet-resolution
measurements by the ALEPH experiment~\cite{Heister:2003aj} (now without
the benefit of light-flavour tagging), using the standard Durham $k_T$
algorithm for $e^+e^-$ collisions~\cite{Stirling:1991ds}, 
as implemented in the \Fj
code~\cite{Cacciari:2011ma}. We also compared to
default \Pp and, for completeness, checked that the relative production
fractions of various meson and baryon species were indeed unchanged
relative to the old \Vc default. 

Rather than presenting all of this information in the form of 
many additional plots, \tabRef{tab:LEP} instead provides a condensed
summary of all the validations we have carried out, 
via $\left<\chi^2\right>$ values for
each of the models with respect to each of the LEP distributions,
including a flat 5\% ``theory uncertainty'' on the MC numbers. 
Already from this simple set of $\chi^2$ values, it is clear that the LO
models/tunes are already doing very 
well\footnote{Both \Vc and \Py are known to give quite 
  good fits to LEP
  data~\cite{Buckley:2009bj,Giele:2011cb,AlcarazMaestre:2012vp,Larkoski:2013yi}.   
For comparisons including other generators and
  tunes, see \href{http://mcplots.cern.ch}{mcplots.cern.ch}~\cite{Karneyeu:2013aha}.}. 
This agreement, however, comes at the price of using a very large
(``LO'') value for $\alpha_s$, which is not guaranteed to be
universally applicable. 
\begin{table}[tp]
\centering
\small
\begin{tabular}{lccccc}
{\bf $\mathbf{\left<\chi^2\right>}$ Shapes}
 & $T$ & $C$ & $D$ & $B_W$ & $B_T$ \\ 
\Pp & \gbox{0.4} & \gbox{0.4} & \gybox{0.6} & \gbox{0.3} & \gbox{0.2} \\
\Vc (LO) & \gbox{0.2} & \gbox{0.4} & \gbox{0.4} & \gbox{0.3} & \gbox{0.3} \\
\Vc (NLO) & \gbox{0.2} & \gbox{0.2} & \gybox{0.6} & \gbox{0.3} &
\gbox{0.2} \\ 
\end{tabular} \hfill
\begin{tabular}{lcccc}
{\bf $\mathbf{\left<\chi^2\right>}$ Frag}
 & $N_\mrm{ch}$ & $x$ & Mesons & Baryons \\  
\Pp & \gybox{0.8} & \gbox{0.4} & \gybox{0.9} & \ybox{1.2} \\
\Vc (LO)  & \gbox{0.0} & \gybox{0.5} & \gbox{0.3} & \gybox{0.6} \\
\Vc (NLO) & \gbox{0.1} & \gybox{0.7} & \gbox{0.2} & \gybox{0.6} 
\\ 
\end{tabular}\\[4mm]
\begin{tabular}{lcccccccccccc}
{\bf $\mathbf{\left<\chi^2\right>}$ Jets}
 & $r^\mrm{exc}_{1j}$ & $\ln(y_{12})$ & $r^\mrm{exc}_{2j}$ & $\ln(y_{23})$ 
 & $r^\mrm{exc}_{3j}$ & $\ln(y_{34})$ & $r^\mrm{exc}_{4j}$ & $\ln(y_{45})$ 
 & $r^\mrm{exc}_{5j}$ & $\ln(y_{56})$ & $r^\mrm{inc}_{6j}$ \\ 
\Pp & \gbox{0.1} & \gbox{0.2} & \gbox{0.1} & \gbox{0.2} 
    & \gbox{0.1} & \gbox{0.3} & \gbox{0.2} & \gbox{0.3} 
    & \gbox{0.2} & \gbox{0.4} & \gbox{0.3} \\
\Vc (LO) & \gbox{0.1} & \gbox{0.2} & \gbox{0.1} & \gbox{0.2} 
         & \gbox{0.0} & \gbox{0.2} & \gbox{0.3} & \gbox{0.1}
         & \gbox{0.1} & \gbox{0.0} & \gbox{0.0}
\\
\Vc (NLO) & \gbox{0.2} & \gbox{0.4} & \gbox{0.1} & \gbox{0.3}
          & \gbox{0.1} & \gbox{0.3} & \gbox{0.2} & \gbox{0.2}
          & \gbox{0.1} & \gbox{0.2} & \gbox{0.1}\\  
\end{tabular}
\caption{$\left<\chi^2\right>$ values for: {\sl Top:} L3 light-flavour
  event shapes ({\sl left}) and fragmentation variables~\cite{Achard:2004sv},  
  and LEP average 
  meson and baryon
  fractions~({\sl right})~\cite{Buckley:2010jn,Beringer:1900zz}. {\sl Bottom:} 
  Durham $k_T$ $n$-jet rates, $r_{nj}$, and jet resolutions, 
  $y_{ij}$, measured by the ALEPH
  experiment~\cite{Heister:2003aj}. For the latter, the
  $\left<\chi^2\right>$ calculation 
  was restricted to the perturbative region, $\ln(y) > -8$. 
  A flat 5\% theory uncertainty was included on the MC
  numbers. Both default \Py and the \Vc (LO) tune 
  use $\alpha_s(m_Z)=0.139$ while the \Vc
  (NLO) tune uses $\alpha_s(m_Z)=0.122$. \label{tab:LEP}} 
\end{table}

The main 
point of the overview in \tabRef{tab:LEP} is that an equally good
agreement can be obtained with an $\alpha_s(m_Z)$ value that is
consistent with other NLO determinations~\cite{Pich:2013sqa},
specifically 
\begin{equation}
\alpha_s(m_Z) \ = \ 0.122~,
\end{equation} 
once the NLO 3-jet corrections are included. This should carry over to
other NLO-corrected processes, and hence the fragmentation parameters
we have settled on should be applicable to future NLO-corrected
studies with \Vc, and can also serve as a starting point for NLO-level
matching studies with \Pp. In the latter context, the 2-loop running
in particular could be retained, while the soft  
fragmentation parameters would presumably have to be somewhat
readjusted to absorb differences between \Vc and \Pp near the
hadronization scale\footnote{The differences in soft fragmentation parameters 
between existing LO \Vc and \Py-8 tunes could be used as an initial
guideline for such an effort, see, e.g., \appRef{app:tunes}.}.

\subsection{Uncertainties \label{sec:uncertainties}}
As in previous versions of \Vc, we use the method proposed
in~\cite{Giele:2011cb} to compute a comprehensive set of uncertainty
bands, which are provided in the form of a vector of alternative
weights for each event. Each set is separately unitary, with average
weight one\footnote{\Vc currently does not attempt to give a separate
estimate of the uncertainty on the total inclusive cross section. The
uncertainties it computes only pertain to shapes of distributions and
the effects of cuts on the total inclusive rate.}.
The difference with respect to previous
versions is that each variation now benefits fully from the inclusion
of NLO corrections. 

When setting the parameter \texttt{Vincia:uncertaintyBands = on}, the
uncertainty weights are accessible through the method 
\begin{center}
\ttt{double vincia.weight(int i=0);}
\end{center}
with $i=0$ corresponding to the ordinary event sample, normally
with all weights equal to unity, and the following variations, for $i=$:
\begin{enumerate}
\item Default: since the user may have chosen other settings than the
  default, the default is included as the first variation. 
\item alphaS-Hi: all renormalization scales are decreased to 
$\mu = \mu_\mrm{def} / k_\mu$, where $\mu_\mrm{def} = p_\perp$ for
  gluon emission and $\mu_\mrm{def} = m_{q\bar{q}}$ for gluon
  splitting. The default size of the variation ($k_\mu=2$) can be changed by the
  user, if desired. A second-order compensation for this variation is
  provided by  
  the renormalization-scale sensitive term $V_{3\mu}$. 
\item alphaS-Lo: all renormalizzation scales are increased to 
$\mu = \mu_\mrm{def} * k_\mu$, with similar comments
  as for alphaS-Hi above.  
\item ant-Hi: antenna functions with large finite terms
  (MAX~\cite{Larkoski:2013yi}). 
This
  variation is already compensated for by LO matching (up to the LO
  matched orders) and is not
  explicitly affected by the NLO corrections.
\item ant-Lo: antenna functions with small finite terms
  (MIN~\cite{Larkoski:2013yi}), with
  similar comments as above.
\item NLO-Hi: branching probabilities are multiplied by a factor
  $(1+\alpha_s)$ to represent unknown (but finite) NLO
  corrections. Is canceled by NLO matching (up to the NLO matched order).
\item NLO-Lo: branching probabilities are divided by a factor
  $(1+\alpha_s)$. Is canceled by NLO matching.
\item Ord-pT: smooth $p_\perp$ ordering for all branchings, including $g\to
  q\bar{q}$ ones. Compensated at first order by LO matching, and at 
second order (Sudakov corrections) by NLO matching
via ordering-sensitive terms in $V_{3Z}$. 
\item Ord-mD: smooth $m_D$ ordering for gluon emissions
(with $m_{q\bar{q}}$ used for gluon splittings). Similar comments as for Ord-pT
above.
\item NLC-Hi: $qg$ emission antennae use $C_A$ as color
  factor. Compensated at first order by LO matching. Not affected by
  NLO matching since those are so far only only done at leading color.
\item NLC-Lo: $qg$ emission antennae use $2C_F$ as color factor, with
  similar comments as above.
\end{enumerate}
We emphasize that these variations are not all independent (for
instance the $\alpha_s$ and NLO variations are highly correlated) 
and hence the corresponding uncertainties should not be summed in
quadrature. In the \Vr plotting tool included with
\Vc, the uncertainty band is constructed by taking the max and min of
the variations. See the \Vc HTML manual for more information about the
uncertainty bands and \cite{Giele:2011cb} for details on their algorithmic
construction.

\begin{figure}[t]
\centering
\begin{tabular}{cc}
\includegraphics*[scale=0.32]{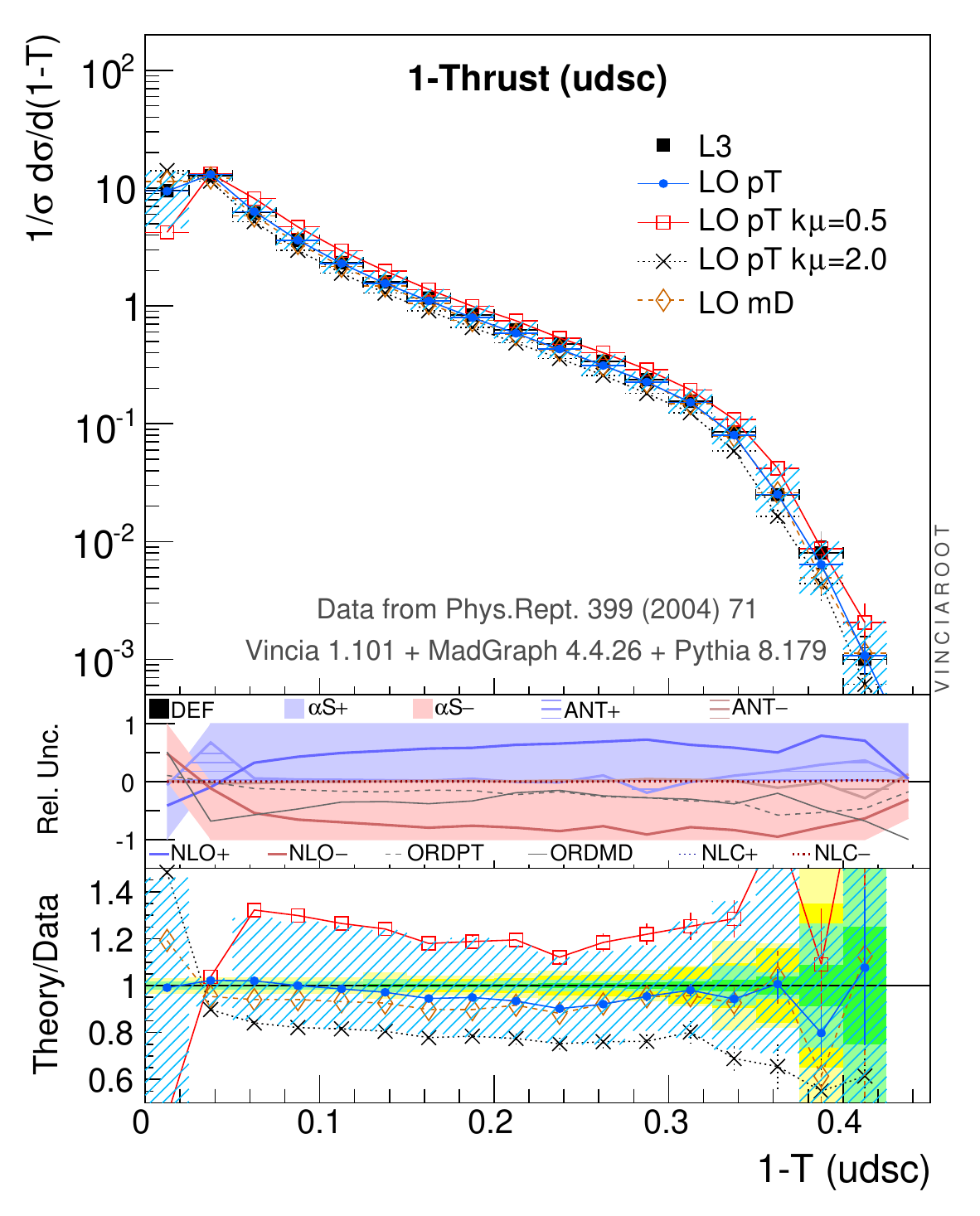}
&\includegraphics*[scale=0.32]{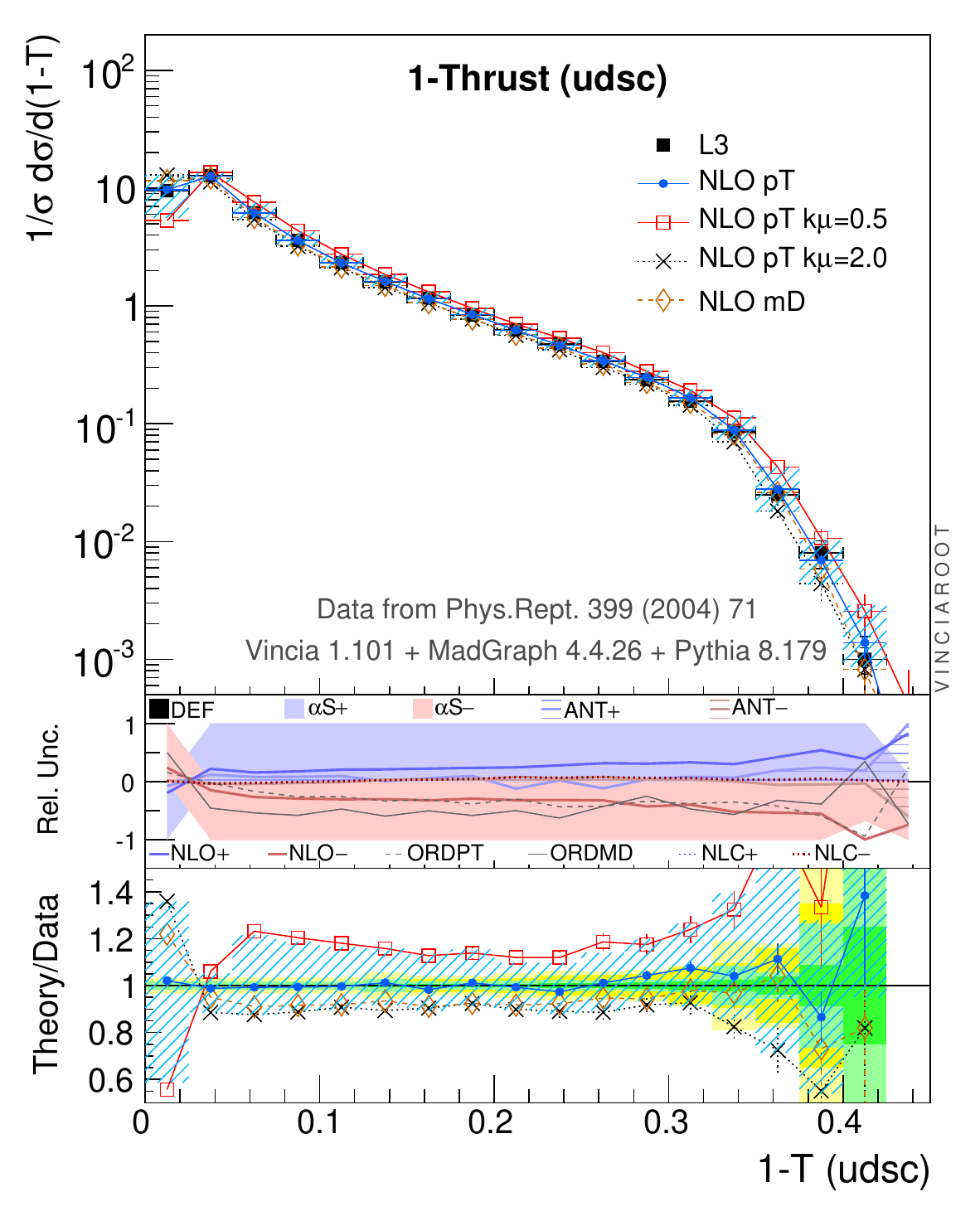}\\
(a) NLO$_{2}$ + LO$_{2,3,4,5}$ + shower & (b) NLO$_{2,3}$ + LO$_{2,3,4,5}$ + shower
\end{tabular}
\caption{Comparison of explicit and automated uncertainty variations
  without ({\sl left}) and with ({\sl right}) NLO 3-jet
  corrections. The individual curves each represents an explicit run, 
  while the shaded blue areas represent the automated uncertainty estimates
  calculated from the central run. 
\label{fig:uncertainties}}
\end{figure}

To illustrate these variations, and the effect of the NLO
3-jet corrections upon them, we include the two plots shown in
\figRef{fig:uncertainties}. We here take the Thrust observable as a
representative example. (More such plots can be generated 
using the \textsc{vinciaroot} interface and the 
\texttt{vincia03-root.cc} example program included with the \Vc code.) 
Similarly to previous plots in this paper, 
the top pane shows the normalized distribution, $1/\sigma\,
\mrm{d}\sigma/\mrm{d}(1-T)$, and the bottom one shows the ratio of
theory to data. Now, however, there is also a middle pane, which gives 
the relative breakdown of the automated uncertainty variations into their
respective components (normalized to unity). In each plot, we compare four
individual runs of \Vc\ to the automated uncertainty variations, with
the latter based on the central run. This provides a useful cross
check of whether the variations are indeed well represented by the
automated estimates, before ({\sl left}) and after ({\sl right}) 
including the NLO 3-jet corrections. For the individual runs, we have
chosen to show the renormalization-scale ($\mu_\mrm{PS} = p_\perp$ for the
central run and factor-2 variations) and evolution-variable ($p_\perp$
for the central run and $m_D$ as the last variation) dependence. (The
antenna-function variations are canceled already at LO for this observable, 
so they are not interesting in the present context.) The
automated uncertainty bands include all 10 variations, with the middle
panes showing the contributions from each. In both plots, the
scale-variation uncertainty dominates over the full range of the
observable, highlighting that this is the main component that would
need to be improved in order to obtain more precise results. 
Note, however, that both the central value at large
$1-T$ and the amount of scale variation, \emph{are} improved by the
introduction of the NLO 3-jet corrections in the right-hand pane. We
also note that the distributions obtained from the explicit variation 
runs are faithfully reproduced by the automated variations, thus
validating our confidence in the automated approach. 

\subsection{Speed \label{sec:speed}}

Although the CPU 
time required by matrix-element and shower/hadronization generators is
still typically small 
in comparison to that of, say, full detector simulations,
their speed and efficiency are still decisive for all
generator-level studies, including tuning and validation,
parameter scans, development work, phenomenology studies, 
comparisons to measurements corrected to the hadron level, and even
studies interfaced to fast 
detector simulations. For this wide range of applications, the
high-energy simulation itself constitutes the main
part of the calculation. An important benchmark relevant to practical
work is for instance whether the calculation can be performed easily on a
single machine or not.  

Higher matched orders are characterized by increasing complexity and
decreasing unweighting efficiencies, resulting in an 
extremely rapid growth in CPU requirements (see e.g.
\cite{Larkoski:2013yi}). At NLO, the additional issues of 
negative weights and/or so-called counter-events can contribute further 
to the demands on computing power. With this in mind, high
efficiencies and fast algorithmic structures were a primary concern in 
the development of the formalism for leading-order matrix-element
corrections in
\Vc~\cite{Giele:2011cb,LopezVillarejo:2011ap,Larkoski:2013yi}, and 
this emphasis carries through to the present work. We 
can make the following remarks.
\begin{itemize}
\item The only fixed-order phase-space generator is the Born-level
  one. All higher-multiplicity phase-space points are generated by
  (trial) showers off the lower-multiplicity ones. This essentially
  produces a very fast 
  importance-sampling of phase-space that automatically  
  reproduces the dominant QCD structures.
\item Likewise, the only cross-section estimate that needs to be
  precomputed at initialization is the total inclusive one. Thus,
  initialization times remain at fractions of a second regardless of
  the matching order. 
\item The matrix-element corrected algorithm works just like an
  ordinary parton shower, with modified (corrected) splitting
  kernels. In particular, all produced events have the same weights, 
  and no additional unweighting step is required.
\item Since the corrections are performed multiplicatively, in the
  form of $(1+\mrm{correction})$, with 1 being the LO answer, there
  are no negative-weight events and no counter-events. The only
  exception would be if the correction becomes larger than the LO
  answer, and negative. This would correspond to a point with
  a divergent fixed-order expansion, in which case the use of NLO
  corrections would be pointless anyway. Moreover, as demonstrated by
  the plots in the previous  sections, our definitions of the
  corrections are analytically stable (and numerically subleading with
  respect to LO) over all of phase space,
  including the soft and collinear regions, 
  for reasonable renormalization- and evolution-variable choices. 
\item The parameter variations described in \secRef{sec:uncertainties}
  can be performed together with the
  matching corrections to provide a set of uncertainty bands in which
  each variation benefits from the full corrections up to the matched
  orders. These 
  are provided in the form of a vector of alternative weights for each
  event~\cite{Giele:2011cb}, at a cost in CPU time which is only a
  fraction of that  
  of a comparable number of independent runs.
\end{itemize}
These attributes, in combination with helicity dependence in the case
of the leading-order formalism~\cite{Larkoski:2013yi}, allow \Vc to
run comfortably on a single machine even with full-fledged matching
and uncertainty variations switched on. 

\begin{table}[t]
\centering
\begin{tabular}{lllcc}
 & LO level & NLO level & {\bf Time / Event} & Speed relative to \Py\\
 &  $Z\to$  &  $Z\to$   & {\bf [milliseconds]} & $\frac{1}{\mrm{Time}}$ / \Pp
\\  
  \Pp &  $2, 3$ & $2$ & {\bf 0.6} & 1
\\\Vc (NLO off) & $2, 3, 4, 5$ & $2$ & {\bf 2.5} & $\sim 1/4$
\\\hfill+ uncertainties & $2, 3, 4, 5$ & $2$ & {\bf 2.9} & $\sim 1/5$
\\\Vc (NLO on) & $2, 3, 4, 5$ & $2, 3$ & {\bf 3.9} & $\sim 1/7$
\\
\hfill+ uncertainties & $2, 3, 4, 5$ & $2, 3$ & {\bf 4.0} & $\sim 1/7$
\\  
\end{tabular}
\caption{Event-generation time in default \Vc 1.1.01 (NIKHEF tune),
  with and without automated uncertainty
  evaluations and NLO 3-jet corrections, 
compared to default \Py~8.179.
\label{tab:speed}}
\end{table}
The inclusion of NLO corrections will necessarily slow down the
calculation. The 
relative increase in running time relative to \Pp, 
is given in \tabRef{tab:speed}, including the
default level of tree-level matching, with and without the NLO 3-jet
correction\footnote{The numbers include both showering and hadronziation and 
were obtained on a single
2.53 GHz CPU, with gcc 4.7 -O2, using default settings for \Pp and
the ``Nikhef'' NLO tune for \Vc.}. Without it (but
still including the default tree-level corrections which go up to
$Z\to 5$ partons), \Vc is 5 times slower than \Py. With the
NLO 3-jet correction switched on, this increases only slightly, to a factor
7. For a fully showered and hadronized
calculation which includes second-order virtual and third-order tree-level
corrections, we consider that to still be acceptably fast. 
Importantly, an event-generation time of a few
milliseconds per event implies that serious studies can still be
performed on an ordinary laptop computer.

\section{Outlook and Conclusions}
\label{sec:conclusions}

In this work, we have investigated the expansion of a Markov-chain
QCD shower algorithm to second order in the strong coupling, for $e^+
e^- \to 3\,$partons, and made systematic comparisons to matrix-element
results obtained at the same order. Using these results, we have 
subjected the subleading properties of shower algorithms with
different evolution/ordering variables and different
renormalization-scale choices to a rigorous examination. At the
analytical level, we have compared the logarithmic structures at the
edge of phase space, and at the numerical level we have illustrated
the difference between the expanded shower algorithm and the one-loop
matrix element. 

We find that the choice of $p_\perp$-ordering, with a renormalization
scale proportional to $p_\perp$ yields the best agreement with the
one-loop matrix element, over all of phase space. This elaborates on,
and is consistent with, earlier
findings~\cite{Amati:1980ch,Catani:1990rr}. 
Using the antenna
invariant mass, $m_D$, for the evolution variable still gives reasonable
results in the hard regions of phase space, but leads to
logarithmically divergent corrections for soft emissions, the exact
form of which depends on the choice of renormalization variable. In
the \Vc code, we retain the option of using $m_D$ mainly as a way of
providing a conservative uncertainty estimate. 

With the NLO 3-jet corrections included as multiplicative corrections to the
shower branching probabilities, we find that we can obtain good agreement
with a large set of LEP event-shape, fragmentation, and jet-rate 
observables with a value of the strong coupling constant of
$\alpha_s(M_Z) = 0.122$. This is in strong contrast with earlier (LO)
tunes of both \Py and \Vc which employed much larger values $\sim
0.14$ to obtain agreement with the LEP measurements. The parameters
for the NLO tune are collected in \appRef{app:tunes} and represent the
first dedicated NLO-corrected tune to LEP data.

This paper is intended as a first step towards a systematic embedding of
one-loop amplitudes within the \Vc shower and matching
formalism. To arrive at a full-fledged prescription, this will need 
to be extended to hadron collisions, ideally in a way that allows for
convenient automation. A first step towards developing the underlying
shower formalism for $pp$ collisions was recently 
taken~\cite{Ritzmann:2012ca}, and more work is in 
progress~\cite{Giele:2013ema}. 

In addition, further studies should be
undertaken of the impact of unordered sequences of radiation that can
occur for the smooth-ordering case (it may be necessary to adopt a strategy
similar to the truncated showers of the \Fw approach), and the
mutually related issues of total normalization and how much of the
(hard) corrections are exponentiated (similar to the differences
between the \Pw and \Fw formalisms, but here occurring at one
additional order, where the relevant total normalization is the NNLO
one). Finally, it would be interesting to develop an extension of this
formalism that would allow second-order-corrected antenna functions to be
used at every stage in the shower, thereby upgrading the precision of
the all-orders resummation, a project that would involve examining 
the second-order corrections to branchings of $qg$ and $gg$ mother
antennae as well. We look forward to following up on these issues in
the near future.  

\acknowledgments

We are grateful to A.~Larkoski, L.~L\"onnblad, J.~Lopez-Villarejo, 
S.~Prestel, G.~Salam, and T.~Sj\"ostrand for discussions and comments
on the manuscript. We are indebted to H.~Mantler for careful
cross-checks of many of the formulae in this paper, several typo corrections, 
and especially for providing us with alternative forms for some of the
integrals listed in the appendices which are more manifestly numerically
stable than those written in the original (arXiv v1 and v2) versions
of our paper. We also thank C.~Duncan for validating the resulting
changes. PS thanks the Galileo Galilei Institute for Theoretical
Physics, LH 
would like to thank Lund University and both of us thank CEA-Saclay for their hospitality. We also thank the INFN for partial support during the completion of
this work.  
EL and LH have been supported by the Netherlands Foundation for Fundamental
Research of Matter (FOM) programme 104, entitled  ``Theoretical 
Particle Physics in the Era of the LHC'', 
and the National Organization for Scientific Research (NWO).
\clearpage
\appendix

\section{Infrared singular operators \label{app:I}}

Here we list the IR singularity operators from
\cite{GehrmannDeRidder:2005cm,GehrmannDeRidder:2004tv,GehrmannDeRidder:2007jk}
as they are used in section \ref{sec:match-antenna-show}.
\begin{align}
{I}_{q\bar{q}}^{(1)}\left(\epsilon,\mu^2/s_{q\bar{q}}\right) &= -\frac{e^{\epsilon \gamma}}{2\Gamma\left(1-\epsilon\right)} \left[ \frac{1}{\epsilon^2}+\frac{3}{2\epsilon} \right] \Re \left(- \frac{\mu^2}{s_{q\bar{q}}} \right)^\epsilon \\
{I}_{qg}^{(1)}\left(\epsilon,\mu^2/s_{qg}\right) &= -\frac{e^{\epsilon \gamma}}{2\Gamma\left(1-\epsilon\right)} \left[ \frac{1}{\epsilon^2}+\frac{5}{3\epsilon} \right] \Re \left( -\frac{\mu^2}{s_{qg}} \right)^\epsilon \\
{I}_{qg,F}^{(1)}\left(\epsilon,\mu^2/s_{qg}\right) &= \frac{e^{\epsilon \gamma}}{2\Gamma\left(1-\epsilon\right)} \frac{1}{6\epsilon} \Re\left( -\frac{\mu^2}{s_{qg}} \right)^\epsilon \label{eq:IRnf}
\end{align}

\section{One-Loop Amplitudes}
\label{app:FO}

\subsection{Renormalization}
\label{sec:renormalization}

Because a detailed derivation of the calculation of $Z\to 3$ jets 
can be found in~\cite{Ellis:1980wv} we restrict ourselves to listing
the result in form that is convenient
for our purpose. Divergences are regulated using
dimensional regularization with $d=4-2 \epsilon$. Our results, before
ultraviolet renormalization, are cross-checked with
\cite{Ellis:1980wv} where one must undo the renormalization in their
case.  In order to cancel the ultraviolet poles we need to
renormalize the coupling according to  (see also section \ref{sec:renormalization-term})
\begin{equation}
\alpha_s^{bare} = \alpha_s(\mu_R^2) \mu^{2\epsilon}\left[ 1 - \frac{\beta_0}{\epsilon}S_\epsilon \left( \frac{ \alpha_s(\mu_R^2)}{4 \pi} \right) \left( \frac{\mu^2}{\mu_R^2} \right)^\epsilon \right]
\end{equation} 
where 
\begin{equation}
\beta_0 = \frac{11 N_c - 2 n_F}{3} \label{eq:betaQCD}
\end{equation} 
and $S_\epsilon=(4\pi)^\epsilon \exp(-\epsilon \,\gamma_E)$ 
contains the factors characterizing the
$\overline{\mrm{MS}}$ scheme. 
Due to the renormalization, the leading
order calculation will generate a term quadratic in
$\alpha_s(\mu_R^2)$,
\begin{equation}
-\frac{\alpha_s(\mu_R)^2}{4\pi} \frac{\beta_0}{\epsilon}\left[ 1 + \epsilon \ln \left( \frac{\mu_R^2}{\mu^2} \right) \right] \mbox{Born}~,
\end{equation}
 which directly cancels the ultraviolet poles of the next-to-leading
 order calculation.

\subsection{One-loop Matrix Element}
\label{sec:leading-colour}

The fixed-order expression relevant to matching in the \Vc context
is the one-loop matrix element normalized by the tree-level one. 
We decompose this into leading- and subleading-colour pieces, as
follows: 
\begin{equation}
\frac{2\Re\left[M_3^{(1)}M_3^{0*} \right]}{|M_3^0|^2} =  \frac{\alpha_s}{2\pi} (LC+QL+SL) ~,
\end{equation}
with the $LC$ piece containing the $C_A$ part of the gluon loops,
the $QL$ one containing the quark loops, proportional to $n_F
T_R$, and the $SL$ piece containing
the subleading gluon-loop corrections, proportional to 
$-1/N_C$. As usual in MC applications, we usually refer to
``Leading Colour'' as including both the $N_C$ and $T_R$ pieces. 
These are both
associated with so-called planar colour flows that are simple to relate 
to the colour-flow representations used in Monte Carlo event generators,
see e.g.~\cite{Boos:2001cv,Buckley:2011ms}. 
The subleading-colour piece is included below for completeness, 
but has so far been left out of the NLO matching
corrections implemented in the \Vc code.    

The notation of the infrared pole structure of these terms has been
written similar to the integrated antenna in
\cite{GehrmannDeRidder:2005cm}, with 
the difference that we have chosen to expand the scale
factor $\mu$ in the integrated antenna terms in
order to obtain explicitly dimensionless logarithms.

The quark has been labelled $1$, the anti-quark $2$ and the gluon $3$.
\begin{align}
\mbox{LC}  = & N_C \left[ 
           \left( 2I_{qg}^{(1)}(\eps,\mu^2/s_{13}) + 2I_{qg}^{(1)}(\eps,\mu^2/s_{23})  \right) \right.\label{eq:VirtualLC}  \\
             &\quad+ \left( -R(y_{13},y_{23}) +\frac{3}{2}\ln \left( \frac{s_{123}}{\mu_R^2} \right) 
                  + \frac{5}{3} \ln \left(\frac{\mu_R^2}{s_{23}} \right) 
                  + \frac{5}{3} \ln \left(\frac{\mu_R^2}{s_{13}} \right) - 4 \right) \nonumber\\
              &\quad+\frac{1}{s_{123}\;a_3^0} \Big[ 2\ln(y_{13}) \left( 1 + \frac{s_{13}}{s_{12}+s_{23}} -\frac{s_{23}}{s_{12}+s_{23}} -\frac{s_{23}s_{13}}{4(s_{12}+s_{23})^2} \right) \nonumber\\       
              & \quad + 2\ln(y_{23}) \left( 1 - \frac{s_{13}}{s_{12}+s_{13}} +\frac{s_{23}}{s_{12}+s_{13}} -\frac{s_{23}s_{13}}{4(s_{12}+s_{13})^2} \right)    \nonumber  \\  
          & \quad + \frac{1}{2} \left( \frac{s_{13}}{s_{23}} -\frac{s_{13}}{s_{12}+s_{13}}+\frac{s_{23}}{s_{13}}-\frac{s_{23}}{s_{12}+s_{23}}+\frac{s_{12}}{s_{23}}+\frac{s_{12}}{s_{13}} +1 \right)
             \Big] \Big{]}  \nonumber\\[3mm] 
\mbox{QL} = & 2n_FT_R \left[ 
             \left( 2I_{qg,F}^{(1)}(\eps,\mu^2/s_{13}) + 2I_{qg,F}^{(1)}(\eps,\mu^2/s_{23})  \right) \right.  \label{eq:VirtualQL}
\\
             &\quad+ \frac{1}{6} \left(
            \ln\left( \frac{s_{23}}{\mu_R^2}
            \right) + \ln \left(
            \frac{s_{13}}{\mu_R^2} \right)
            \right)  \Big{]}  \nonumber   \\[3mm] 
            \mbox{SL}  = &  \frac{1}{N_C} \left[ 
            \left( 2I_{q\bar{q}}^{(1)}(\eps,\mu^2/s_{12})  \right) \right.  \\
             &\quad-  \left( 4 + \frac{3}{2} \ln(y_{12})+ R(y_{12},y_{13}) + R(y_{12},y_{23}) \right) \nonumber \\
             &\quad
+\frac{1}{s_{123}\;a_3^0}\Big[ R(y_{12},y_{13})
  \left(\frac{s_{13}}{s_{23}} + 2 \frac{s_{12}}{s_{23}}\right) 
                + R(y_{12},y_{23}) \left(\frac{s_{23}}{s_{13}} + 2
                \frac{s_{12}}{s_{13}}\right)  \nonumber\\ 
             &\quad +\ln(y_{12}) \left( \frac{4s_{12}}{s_{13}+s_{23}} + \frac{2 s_{12}^2}{(s_{13}+s_{23})^2} \right) \nonumber\\
              &\quad   +\frac{1}{2} \ln(y_{13}) \left( \frac{s_{13}}{s_{12}+s_{23}} 
                 +\frac{4s_{12}}{s_{12}+s_{23}}+ \frac{s_{12}s_{13}}{(s_{12}+s_{23})^2} \right) \nonumber\\
              &\quad   +\frac{1}{2} \ln(y_{23}) \left( \frac{s_{23}}{s_{12}+s_{13}} 
                 +\frac{4s_{12}}{s_{12}+s_{13}}+ \frac{s_{12}s_{23}}{(s_{12}+s_{13})^2} \right) \nonumber\\  
              &\quad  - \frac{1}{2} \left( \frac{s_{13}}{s_{23}} -\frac{s_{23}}{s_{13}}-\frac{s_{12}}{s_{23}}-\frac{s_{12}}{s_{13}}+\frac{s_{12}}{s_{12}+s_{13}}+\frac{s_{12}}{s_{12}+s_{23}} +\frac{4s_{12}}{s_{13}+s_{23}} \right)
             \Big] \Big{]}                   \nonumber
\end{align}
with 
\begin{equation}
R(y,z) = \ln(y) \ln(z)-\ln(y)\ln(1-y)-\ln(z)\ln(1-z)+\frac{\pi^2}{6}-\mbox{Li}_2(y)-\mbox{Li}_2(z) ~,
\end{equation}
\begin{equation}
a_3^0 
\ = \ 
  \frac{|M_3^0|^2}{g_s^2\ C_F |M_2^0|^2} 
\ = \ 
  \frac{1}{s_{123}} \left( \frac{(1-\eps)s_{13}}{s_{23}} + \frac{(1-\eps)s_{23}}{s_{13}} + 2 \frac{ s_{12}s_{123}-\eps s_{13}s_{23}}{s_{13}s_{23}} \right) (1-\eps)~,
\end{equation}
and the infrared singular operators, $I^{(1)}$, given in
\appRef{app:I}. 

With the one-loop matrix element expressed in this form, 
cancellation of the infrared poles against the integrated antennae
(see below) coming from the shower will be 
 particularly simple and will yield an expression purely
dependent on the renormalization scale, $\mu_R$, and on the kinematic
invariants $s_{12}$, $s_{23}$, and 
$s_{12}$, but not on the scale factor $\mu$. 


\section{Antenna integrals}
\label{sec:antenna-integrals}

In this appendix we list the results of antenna integrals over phase space corresponding
to the various evolution variables. 

\subsection{Strong Ordering Gluon Emission}
\label{app:StrongOrderingGR}

The expressions for a gluon emitting antenna is given in \eqRef{eq:aEmit}. With a redefinition the same antenna function reads
\begin{align}
a_{g/IK}(y_1,y_2) &=\frac{1}{m^2_{IK}} \left[ \frac{2(1-y_1-y_2)}{y_1 y_2}+ \frac{y_1}{y_2} + \frac{y_2}{y_1} - \delta_{Ig} \frac{y_2^2}{y_1}-\delta{Kg}\frac{y_1^2}{y_2} + C_{00} + C_{01}y_1 + C_{10} y_2 \right] \label{eq:gluon_radiation}
\end{align}
where $y_1,y_2$ correspond to $y_{ij},y_{jk}$ of \eqRef{eq:yDef}, respectively. Recall that the last three terms serve to give a flexible and explicit way of tracking extra non-singular terms in antennae.
The phase space integral over these antenna, as determined by the evolution variable, can be written as

\begin{equation}
\frac{1}{16\pi^2 m^2_{IK}}\int_{Q_E^2}^{m^2_{IK}} a_{g/IK}\,  |\mathcal{J}(Q^2,\zeta)| \, \d{Q^2} \, \d{\zeta}.\label{eq:AnIntGR}
\end{equation}
All antenna integrals in \eqRef{eq:V3} have been written in such a way that they are integrated over their whole invariant mass plus a correction term running from the evolution variable to the total invariant mass. The integrals running over the whole invariant mass contain singular regions and therefore poles while the correction terms yield finite corrections. These finite corrections are discussed per evolution variable below. We define the integrals\\

\begin{align}
\mathcal{D}_{Q\zeta} =\frac{1}{m^4_{IK}} \int \mbox{d}Q^2\; \mbox{d}\zeta \;|J|\ 
\end{align}
\begin{align}
I_1  &=\mathcal{D}_{Q\zeta} \frac{1}{y_1(Q^2,\zeta)y_2(Q^2,\zeta)}\label{eq:I1}\\[3mm]
I_2 &=\mathcal{D}_{Q\zeta}\frac{1}{y_2(Q^2,\zeta)}=\mathcal{D}_{Q\zeta}\frac{1}{y_1(Q^2,\zeta)}\\[3mm]
I_3  &=\mathcal{D}_{Q\zeta}\frac{y_1(Q^2,\zeta)}{2y_2(Q^2,\zeta)}=\mathcal{D}_{Q\zeta}\frac{{y_2(Q^2,\zeta)}}{2y_1(Q^2,\zeta)}\\[3mm]
I_4  &=\mathcal{D}_{Q\zeta} \frac{y_2^2(Q^2,\zeta)}{2y_1(Q^2,\zeta)}=\mathcal{D}_{Q\zeta}\frac{y_1^2(Q^2,\zeta)}{2y_2(Q^2,\zeta)}\\[3mm]
I_5 &=\frac12\mathcal{D}_{Q\zeta} \left[ C_{00} + C_{01}y_1(Q^2,\zeta)  + C_{10}y_2(Q^2,\zeta) \right]. \label{eq:I5}
\end{align}
So that, in these terms, the results read
\begin{equation}
  \label{eq:11}
\frac{1}{16\pi^2 m^2_{IK}} \int_{Q_E^2}^{m^2_{IK}} \d{Q^2}\d{\zeta} \,|J(Q^2,\zeta)|\, a_{g/IK}   =
 \frac{1}{8\pi^2}\left(\sum_{i=1}^{5} K_i I_i  \right)
\end{equation}
where
\begin{equation}
  \label{eq:14}
  K_1 = 1\, , \;\; K_2 = -2\, \;\; K_3 = 2, \;\; K_4 = -\delta_{Ig}-\delta_{Kg}, \;\; K_5 =1.
\end{equation}
We now turn to specific cases.

\subsubsection{Dipole Virtuality}
\label{sec:dipole-virtuality-1}

The results for the individual contributing parts of the antenna function as defined in  \eqRef{eq:I1} - \eqRef{eq:I5} with $\xi = \frac{\min(\sqg,\sgq)}{m^2_{IK}-\min(\sqg,\sgq)}$ are
\begin{align}
I_1 &=
\frac{\pi^2}{6} + \ln^2(\xi) - \ln^2(1 + \xi) - 2
\operatorname{Li_2}\left[\frac{1}{1 + \xi}\right] \\[3mm]
I_2 &=\frac{\xi-1-\ln(\xi)}{1+\xi} \\[3mm]
I_3&=
\frac{-3+3\xi^2-(2+4\xi)\ln(\xi)}{8(1+\xi)^2} \\[3mm]
I_4&= \frac{\left( \xi -1 \right) \left( 11+\xi \left( 20+11\xi \right)\right) - 6 \left( 1+3\xi \left( 1+\xi\right)\right) \ln \left( \xi \right) }{36 \left( 1 + \xi \right)^3}\\[3mm]
I_5&=
 \frac{(\xi-1)^2((C_{01}+C_{10})(1+2\xi)+3
   C_{00}(1+\xi))}{12(1+\xi)^3}
\end{align}
In the case of integration over the $3 \to 4$ splittings, the definition of the integrals remains the same. Only the definition of $\xi$ changes with 
\begin{align}
\xi_{3\to 4} = \frac{\min\left(\sqg,\sgq\right)}{\max\left(\sqg,\sgq\right)-\min\left(\sqg,\sgq\right)}
\end{align}

\subsubsection{Transverse Momentum}
\label{sec:transverse-momentum}

The results for the individual contributing parts of the antenna function as defined in  \eqRef{eq:I1} - \eqRef{eq:I5} are

\begin{align} 
I_1 &=  
\frac{\pi^2}{6} + \frac{1}{2}\ln^2\left[\frac{y_3^2}{2\left(1 +
    \sqrt{1 - y_3^2}\right) - y_3^2}\right]
 - \ln^2\left[\frac{1}{2}\left(1 +
  \sqrt{1 - y_3^2}\right)\right] - 2
\operatorname{Li_2}\left[\frac{1}{2}\left(1 + \sqrt{1 -
    y_3^2}\right)\right]\\[3mm]
%
I_2 &=  
-\left(\ln\left[\frac{y_3^2}{2\left(1 + \sqrt{1 - y_3^2}\right) -
    y_3^2}\right]
 + 2 \sqrt{1 - y_3^2}\right)\\[3mm]
I_3 &=  -\frac{1}{4} \left(\ln\left[\frac{y_3^2}{2\left(1 + \sqrt{1 - y_3^2}\right) -
    y_3^2}\right] + 2 \sqrt{1 - y_3^2}\right)\\[3mm]
I_4 &= -\frac{13 \sqrt{1 - y_3^2}}{36} + \frac{1}{36}y_3^2\sqrt{1 -
  y_3^2} - \frac{1}{6} 
\ln\left[\frac{y_3^2}{2\left(1 + \sqrt{1 - y_3^2}\right) - y_3^2}\right]\\[3mm]
I_5 &= \frac{1}{24} \left(2\left(3 C_{00} + \left(C_{01} +
C_{10}\right)\left(1 - y_3^2\right)\right) \sqrt{1 - y_3^2} + 3 C_{00}
y_3^2 \ln\left[\frac{y_3^2}{2\left(1 + \sqrt{1 - y_3^2}\right) - y_3^2}\right]
\right)
\end{align}
with $y_3^2=\frac{Q_3^2}{m^2_{IK}}$ and $m^2_{IK}=s$. In the case of the $3\to 4$ splittings the only adaptation takes place in the former definition where $m^2_{IK}$ is set equal to $\sqg$ or $\sgq$ dependent on which dipole is being integrated over.
\clearpage
\subsubsection{Energy ordering}
\label{sec:energy-ordering}

The results for this evolution parameter are
\begin{align}
I_1 &= 2 \operatorname{Li_2}\left[\frac{1}{2}\left(1+\sqrt{1-\frac{\Delta}{y_3}}\right)\right] + \ln^2\left[\frac{1}{2}\left(1+\sqrt{1-\frac{\Delta}{y_3}}\right)\right] - \frac{1}{2} \ln^2\left[\frac{\Delta}{2 y_3\left(1+\sqrt{1-\frac{\Delta}{y_3}}\right)-\Delta}\right] \nonumber\\
&\quad - 2 \operatorname{Li_2}\left[\frac{1}{2}\left(1+\sqrt{1-\Delta}\right)\right] - \ln^2\left[\frac{1}{2}\left(1+\sqrt{1-\Delta}\right)\right] + \frac{1}{2} \ln^2\left[\frac{\Delta}{2\left(1+\sqrt{1-\Delta}\right)-\Delta}\right] \\[3mm]
I_2 &= 2 \left(\sqrt{y_3-\Delta}-\sqrt{1-\Delta}\right) + \sqrt{y_3}\ln\left[\frac{\Delta}{2 y_3\left(1+\sqrt{1-\frac{\Delta}{y_3}}\right)-\Delta}\right] - \ln\left[\frac{\Delta}{2\left(1+\sqrt{1-\Delta}\right)-\Delta}\right] \\[3mm]
%
I_3 &= \frac{1}{4} \left(y_3\ln\left[\frac{\Delta}{2 y_3\left(1+\sqrt{1-\frac{\Delta}{y_3}}\right)-\Delta}\right] - \ln\left[\frac{\Delta}{2\left(1+\sqrt{1-\Delta}\right)-\Delta}\right]\right)\nonumber \\
&\quad + \frac{1}{2} \left(\sqrt{y_3\left(y_3-\Delta\right)}-\sqrt{1-\Delta}\right) \\[3mm]
%
%
I_4 &= \frac{1}{36}\Bigg(6y_3^{\frac{3}{2}} \ln\left[\frac{\Delta}{2 y_3\left(1+\sqrt{1-\frac{\Delta}{y_3}}\right)-\Delta}\right] - 6 \ln\left[\frac{\Delta}{2\left(1+\sqrt{1-\Delta}\right)-\Delta}\right] \nonumber\\
&\quad + (13 y_3 - \Delta)\sqrt{y_3-\Delta} - (13 - \Delta)\sqrt{1-\Delta} \Bigg)\\[3mm]
%
%
I_5 &= \frac{1}{8}C_{00}\left(\Delta\cdot\ln\left[\frac{2 y_3(1+\sqrt{1-\frac{\Delta}{y_3}})-\Delta}{2(1+\sqrt{1-\Delta})-\Delta}\right] + 2\sqrt{1-\Delta} - 2\sqrt{y_3(y_3-\Delta)}\right) \nonumber\\
&\quad + \frac{1}{12}\left(C_{01}+C_{10}\right)\left((1-\Delta)^{\frac{3}{2}}-(y_3-\Delta)^{\frac{3}{2}}\right)
\end{align}
with $\Delta$ used as a cut-off on $4p_\perp^2$ and $y_3 =\frac{(\sqg+\sgq)^2}{s^2}$.

\subsection{Strong Ordering Gluon Splitting}
\label{app:StrongOrderingGS}

The branching of a gluon splitting into a quark antiquark pair can only take place at the $3 \to 4$ level splitting. The generation of a gluon splitting takes place through an alternative form of phase space generation than the discussed $m_D$, $p_\perp$ and $E_n$ variables. Instead phase space is sampled in a triangular surface comparable to $m_D$ ordering, yet in this case using only one cutoff, the $Q^2$ generated at the $2\to 3$ level, to avoid the singular region of the gluon splitting antenna. The gluon splitting antenna is given by  
\begin{equation}
  \label{eq:9}
  a_{\bar{q}/qg}(y_1,y_2) =\frac{1}{m^2_{IK}}\left(\frac{(1-2y_1)}{2y_2}+ \frac{y^2_1}{y_2} +C_{00} + C_{01}y_1 + C_{10} y_2  \right).
\end{equation}
Because the integration surface is similar for all evolution types only depending on the cutoff $Q^2$ the integration is demonstrated for all types 
\begin{align}
&H=\frac{1}{2m^2_{IK}} \int_{Q_E^2}^{m^2_{IK}} \d s_2 \int_0^{m^2_{IK}-s_2} \d s_1 a_{\bar{q}/qg}(s_1,s_2) = \frac{m^2_{IK}}{2} \int_{y_E=\frac{Q_E^2}{m^2_{IK}}}^{1} \d y_2 \int_0^{1-y_2} \d y_1 a_{\bar{q}/qg}(y_1,y_2)\label{eq:AnIntGS} \\
&= \frac{1}{2}\left[\frac13 \ln\left(\frac{1}{y_E}\right) - \frac{13}{36} + \frac{y_E}{2} -
  \frac{y_E^2}{4} + \frac{y_E^3}{9} + \frac{\left(1-y_E\right)^2}{2}\left(
  C_{00} +
  \frac{C_{01}}{3}\left(1-y_E\right)
+ \frac{C_{10}}{3}\left(1+2y_E\right)\right)\right]
\nonumber
\end{align}
where the factor a half has been added for the sake of consistency with respect to the treatment of gluon emission. The factor $m^2_{IK}$ needs to be replaced by either $\sqg$ or $\sgq$ dependent on which antenna is being integrated. 

\subsection{Smooth Ordering Gluon Emission}

The phase space integral in the case of smooth ordering differs from strong ordering by allowing integration over the whole phase space region. The inclusion of a damping factor regulates the accessible region of phase space which generates a different phase space occupancy than in the case of strong ordering. A general form for smooth ordering integration of a gluon emission antenna is

\begin{align}
\frac{1}{16\pi^2 m^2_{IK}}\int_{0}^{m^2_{IK}} \d s_1\int_0^{m^2_{IK}-s_1} \d s_2 \frac{Q_{E_j}^2}{Q_{E_j}^2+Q_3^2}\, a_{g/IK}(s_1,s_2)\label{eq:AnIntGRSm}
\end{align}
Where we use the definition of \eqRef{eq:gluon_radiation} with $s_i=y_i m^2_{IK}$, $Q_3^2$ denotes the branching scale and $Q_{E_j}$ indicates the evolution variable used for gluon emission. We define the following integrals

\begin{align}
\mathcal{D}_s = \frac{1}{m^4_{IK}} \int_0^{m^2_{IK}}
\mbox{d}s_1\int_0^{m^2_{IK}-s_1} \mbox{d}s_2 \,
\frac{Q_{E_j}^2}{Q_{E_j}^2+Q_3^2}
\end{align}
\begin{align}
L_1 &=\mathcal{D}_s\frac{m^4_{IK}}{s_1s_2}\label{eq:L1}\\[3mm]
L_2 &=\mathcal{D}_s\frac{m^2_{IK}}{s_1}=\mathcal{D}_s\frac{m^2_{IK}}{s_2}\\[3mm]
L_3 &=\mathcal{D}_s\frac{s_1}{2s_2}=\mathcal{D}_s\frac{s_2}{2s_1}\\[3mm]
L_4 &=\mathcal{D}_s\frac{s_1^2}{2m^2_{IK}s_2}=\mathcal{D}_s\frac{s_2^2}{2m^2_{IK}s_1}\\[3mm]
L_5 &=\frac12\mathcal{D}_s \left[ C_{00} + C_{01}\frac{s_1}{m^2_{IK}}  + C_{10}\frac{s_2}{m^2_{IK}} \right]. \label{eq:L5}
\end{align}
So that, in these terms, the results read
\begin{equation}
\label{eq:42}
\frac{1}{16\pi^2 m^2_{IK}} \int_{0}^{m^2_{IK}} \d s_1 \int_0^{m^2_{IK}-s_1} \d s_2 \frac{Q_{E_j}^2}{Q_{E_j^2}^2+Q_3^2} \, a_{g/IK}   =
 \frac{1}{8\pi^2}\left(\sum_{i=1}^{5} K_i L_i  \right)
\end{equation}
where
\begin{equation}
\label{eq:43}
  K_1 = 1\, , \;\; K_2 = -2\, \;\; K_3 = 2, \;\; K_4 = -\delta_{Ig}-\delta_{Kg}, \;\; K_5 =1.
\end{equation}
We now turn to specific cases.

\subsubsection{Smooth mass ordering}

The only term from \eqRef{eq:AnIntGRSm} that requires specification is the damping factor
\begin{align}
1 - P_{\mrm{imp}} = \frac{Q^2_{E_j}}{Q_{E_j}^2+Q_3^2} = \frac{\min(s_1,s_2)}{\min(s_1,s_2)+\min(\sqg,\sgq)}.\label{eq:SmoothmDPimp}
\end{align}
The computation of the individual antenna parts will require separating the phase space triangle in two regions ($s_1>s_2$ and vice versa)  in order to make the damping factor definite. After summing over these two regions we obtain the following values for gluon emission contributions  
\begin{align}
L_1 &= 2\left[\ln(2)\ln\left(1+\frac{2}{y_3^2}\right)-\operatorname{Li_2}\left(-\frac{1}{y_3^2}\right)-\operatorname{Li_2}\left(\frac{2}{2+y_3^2}\right)+\operatorname{Li_2}\left(\frac{1}{2+y_3^2}\right)\right] \\[3mm]
L_2 &= -1 + \ln\left(2+\frac{2}{y_3^2}\right) + y_3^2 \ln\left(1+\frac{1}{y_3^2}\right) - \frac{1}{2} y_3^2 \ln\left(1+\frac{2}{y_3^2}\right) \ln(2) \nonumber\\
&\quad + \frac{1}{2}y_3^2\left[\operatorname{Li_2}\left(-\frac{1}{y_3^2}\right)+\operatorname{Li_2}\left(\frac{2}{2+y_3^2}\right)-\operatorname{Li_2}\left(\frac{1}{2+y_3^2}\right)\right] \\[3mm]
L_3 &= \frac{1}{8}\left( -3 + 2\ln\left(\frac{2+2 y_3^2}{y_3^2}\right)
+ 2y_3^2\ln\left(\frac{1+y_3^2}{2 y_3^2}\right) +
y_3^4\ln\left(1+\frac{2}{y_3^2}\right)\ln(2)\right. \nonumber\\
&\quad \left. - y_3^4\left[\operatorname{Li_2}\left(-\frac{1}{y_3^2}\right)+\operatorname{Li_2}\left(\frac{2}{2+y_3^2}\right)-\operatorname{Li_2}\left(\frac{1}{2+y_3^2}\right)\right]\right) \\[3mm]
%
%
L_4 &= \frac{1}{72}\left[ - 22 + 12\ln\left(\frac{2+2y_3^2}{y_3^2}\right) - 3 y_3^2 + 18 y_3^2\ln\left(\frac{1+y_3^2}{\sqrt{2}y_3^2}\right) - 3 y_3^4 + 9y_3^4\ln\left(\frac{2+2y_3^2}{y_3^2}\right) \right] \nonumber\\
&\quad + \frac{1}{24}y_3^6\ln\left(\frac{1+y_3^2}{y_3^2}\right) - \frac{1}{16}y_3^6\ln\left(\frac{2+y_3^2}{y_3^2}\right)\ln(2) \nonumber\\
&\quad + \frac{1}{16}y_3^6\left[\operatorname{Li_2}\left(-\frac{1}{y_3^2}\right) + \operatorname{Li_2}\left(\frac{2}{2+y_3^2}\right) - \operatorname{Li_2}\left(\frac{1}{2+y_3^2}\right)\right] \\[3mm]
L_5 &= \frac{1}{48}\bigg[4(3 C_{00} + C_{01} + C_{10}) + 3(8 C_{00} + C_{01} + C_{10})y_3^2 - 6(C_{01} + C_{10})y_3^4 \nonumber\\
&\quad \left. - 6y_3^2(1+y_3^2)(4 C_{00} + C_{01} + C_{10} - (C_{01} + C_{10})y_3^2)\ln\left(\frac{1+y_3^2}{y_3^2}\right)\right]
\end{align}
with $y_3^2 =\frac{2\min(\sqg,\sgq)}{m^2_{IK}}$.

\subsubsection{Smooth transverse momentum ordering}

In the case of smooth ordering for transverse momentum we find the following result for the ordering requirement
\begin{align}
1 - P_{\mrm{imp}} = \frac{Q^2_{E_j}}{Q_{E_j}^2+Q_3^2} = \frac{ \frac{s_1 s_2}{m^2_{IK}} }{\frac{s_1s_2}{m^2_{IK}} +\frac{\sqg \sgq}{s} }.\label{eq:SmoothpTPimp}
\end{align}
Where $m^2_{IK}$ should be replaced by $\sqg$ or $\sgq$ dependent on the dipole of integration. In combination with \eqRef{eq:AnIntGRSm} we find the following results for the partial gluon emission antenna parts
\begin{align}
L_1 &= \frac{1}{2}\ln^2\left(\frac{y_3^2}{2\left(1 + \sqrt{1 + y_3^2}\right) + y_3^2}\right) \\[3mm]
L_2 &= -2 - \sqrt{1+y_3^2} \cdot \ln\left(\frac{y_3^2}{2\left(1 + \sqrt{1 + y_3^2}\right) + y_3^2}\right) \\[3mm]
L_3 &= - \frac{1}{2} - \frac{1}{4}\sqrt{1+y_3^2} \cdot \ln\left(\frac{y_3^2}{2\left(1 + \sqrt{1 + y_3^2}\right) + y_3^2}\right) \\[3mm]
L_4 &= -\frac{13}{36}-\frac{1}{12}y_3^2-\frac{1}{24}(4+y_3^2)\sqrt{1+y_3^2} \ln\left(\frac{y_3^2}{2\left(1 + \sqrt{1 + y_3^2}\right) + y_3^2}\right) \\[3mm]
%
%
L_5 &= (C_{01} + C_{10}) \left(\frac{1}{12} + \frac{1}{4}y_3^2 + \frac{1}{8}y_3^2\sqrt{1+y_3^2}\ln\left(\frac{y_3^2}{2\left(1 + \sqrt{1 + y_3^2}\right) + y_3^2}\right)\right) \nonumber\\
&\quad + C_{00} \left(\frac{1}{4} - \frac{1}{16} y_3^2\ln^2\left(\frac{y_3^2}{2\left(1 + \sqrt{1 + y_3^2}\right) + y_3^2}\right)\right)
\end{align}
with $y_3^2 = 4 \frac{\sqg \sgq}{s\, m^2_{IK}} = \frac{4}{N_\perp}\frac{Q_3^2}{m_{IK}^2}$, so that 
$y_3^2 = 4 \sqg / s$ for $m^2_{IK}=\sgq$, and 
$y_3^2 = 4 \sgq / s$ for $m^2_{IK}=\sqg$. 

\subsection{Smooth Ordering Gluon Splitting}

Additionally we also need to consider the gluon splitting antenna function for smooth ordering. Similar to the strong ordering case, the separate generation of gluon splitting variables allows for a new choice for evolution variable and thereby a different phase space surface. As in the case of gluon emission we allow for integration over the whole phase space, using the damping factor to limit the accessible area. A general notation is the following

\begin{equation}
G=\frac{1}{2m^2_{IK}}\int_{0}^{m^2_{IK}} \d s_2\int_0^{m^2_{IK}-s_2} \d s_1 \frac{Q_{E_j}^2}{Q_{E_j^2}^2+Q_3^2}\, a_{\bar{q}/qg}(s_1,s_2), \label{eq:AnIntSmGS}
\end{equation}
with the definition for the gluon splitting antenna as in \eqRef{eq:9}
and the default choice of evolution variable for gluon splittings
being $Q_{Ej}^2=m_{q\bar{q}}^2=s_2$.

\subsubsection{Emissions ordered in $\mathbf{m_D}$, splittings 
  in $\mathbf{m_{q\bar{q}}}$}

With the gluon splitting antenna as defined in \eqRef{eq:9} and the
phase space integral \eqRef{eq:AnIntSmGS} we find the following result
for $Q_3^2 = N'\min(\sqg,\sgq)$:

\begin{align}
G=&\frac{1}{72 s_P^3} \left( -12 ( s_P^3 +3  ( -1 +2C_{00} +C_{10} ) y_3^2 s_P^2  + 3 ( -1 +2 C_{00} - 2C_{01} + 2 C_{10} +2C_{10}) y_3^4 s_P  \right. \nonumber \\
& +(-2-6C_{01}+3C_{10}) y_3^6 ) \,\mbox{arctanh}\left( \frac{s_P}{2y_3^2 +s_P}  \right)  \nonumber \\
&+s_P \left( ( -13+18C_{00}+6C_{01}+6C_{10}) s_P^2 + 3(-4 +12C_{00} - 6C_{01}+9C_{10}) y_3^2 s_P \right. \nonumber \\
&\left. \left. -6(2 + 6C_{01} - 3C_{10}) y_3^4 + 18 s_P^2 \ln \left( \frac{s_P+y_3^2}{y_3^2}  \right) 
\right) \right),
\end{align}
with $y_3 = \frac{N' \min(\sqg,\sgq)}{s}$ and $s_P=\max(\sqg,\sgq)/s$. Note that the gluon splitting antenna has been defined with the singularity in $y_2$ which determines the form of the damping factor. 

\subsubsection{Emissions ordered in $\mathbf{p_\perp}$, splittings
  in $\mathbf{m_{q\bar{q}}}$}

With the gluon splitting antenna defined in \eqRef{eq:9} and the phase
space integral \eqRef{eq:AnIntSmGS} we find the following result for
$Q_3^2 = N_\perp \sqg\sgq/ s$:

\begin{align}
G=&\frac{1}{72}\bigg( -13 + 18C_{00} +6C_{01}+6C_{10}+3(-4+12C_{00}-6C_{01}+9C_{10})y_3^2 \nonumber \\
&-6(2+6C_{01}-3C_{10})y_3^4+36y_3^2 \mbox{acoth}(1+2y_3^2) - 6(-2+y_3^2(6C_{00}(1+y_3^2) \nonumber\\
&+3C_{10}(1+y_3^2)^2-y_3^2(3+2y_3^2+6C_{01}(1+y_3^2))))\ln\left( 1+\frac{1}{y_3^2} \right)\bigg),
\end{align}
with $y_3^2 = N_\perp \frac{\sqg \sgq}{s\,m^2_{IK}} = Q_3^2/m_{IK}^2$, so that 
$y_3^2 = N_\perp \sqg / s$ for $m^2_{IK}=\sgq$, and 
$y_3^2 = N_\perp \sgq / s$ for $m^2_{IK}=\sqg$. 

\section{NLO Tune Parameters \label{app:tunes}}

\begin{table}[tp]
\scalebox{0.95}{%
\begin{tabular}{llll}
\\
\\
& NLO Tune & LO Tune &\\
Parameter & (Nikhef) & (Jeppsson 5) & Comment\\
! * alphaS
\\Vincia:alphaSvalue        &= 0.122  &= 0.139 &! alphaS(mZ) value
\\Vincia:alphaSkMu          &= 1.0    &= 1.0 &! Renormalization-scale prefactor
\\Vincia:alphaSorder        &= 2      &= 1 &! Running order
\\Vincia:alphaSmode         &= 1      &= 1 &! muR = pT:emit and Q:split
\\Vincia:alphaScmw          &= on    &= off &! CMW rescaling of Lambda on/off
\\
\\\multicolumn{3}{l}{
! * Shower evolution and IR cutoff}
\\Vincia:evolutionType      &= 1      &= 1 &! pT-evolution
\\Vincia:orderingMode       &= 2      &= 2 &! Smooth ordering
\\Vincia:pTnormalization    &= 4.     &= 4. &! QT = 2pT
\\Vincia:cutoffType         &= 1      &= 1 &! Cutoff taken in pT
\\Vincia:cutoffScale        &= 0.8    &= 0.6 &! Cutoff value (in GeV)
\\
\\\multicolumn{3}{l}{! * Longitudinal string fragmentation parameters}
\\StringZ:aLund             &= 0.40   &= 0.38 &! Lund FF a (hard fragmentation supp)
\\StringZ:bLund             &= 0.85   &= 0.90 &! Lund FF b (soft fragmentation supp)
\\StringZ:aExtraDiquark     &= 1.0    &= 1.0 &! Extra a to suppress hard baryons
\\
\\\multicolumn{3}{l}{! * pT in string breakups}
\\StringPT:sigma            &= 0.29  &= 0.275 &! Soft pT in string breaks (in GeV)
\\StringPT:enhancedFraction &= 0.01   &= 0.01 &! Fraction of breakups with enhanced pT
\\StringPT:enhancedWidth    &= 2.0    &= 2.0 &! Enhancement factor
\\
\\\multicolumn{3}{l}{! * String breakup flavour parameters}
\\StringFlav:probStoUD     &= 0.215   &= 0.215 &! Strangeness-to-UD ratio
\\StringFlav:mesonUDvector &= 0.45    &= 0.45 &! Light-flavour vector suppression
\\StringFlav:mesonSvector  &= 0.65    &= 0.65 &! Strange vector suppression
\\StringFlav:mesonCvector  &= 0.80    &= 0.80 &! Charm vector suppression
\\StringFlav:probQQtoQ     &= 0.083   &= 0.083&! Diquark rate (for baryon production)
\\StringFlav:probSQtoQQ    &= 1.00    &= 1.00 &! Optional Strange diquark suppression
\\StringFlav:probQQ1toQQ0  &= 0.031   &= 0.031 &! Vector diquark suppression
\\StringFlav:etaSup        &= 0.68    &= 0.68 &! Eta suppression
\\StringFlav:etaPrimeSup   &= 0.11    &= 0.11 &! Eta' suppression
\\StringFlav:decupletSup   &= 1.0     &= 1.0 &! Optional Spin-3/2 Baryon Suppression\\
\end{tabular}}
\caption{Parameters of the ``Nikhef'' NLO tune, compared to those of the
  ``Jeppsson 5'' LO tune.\label{tab:tunes}}
\end{table}

In \tabRef{tab:tunes} below, we list the perturbative and non-perturbative
fragmentation parameters for the Nikhef NLO tune of \Vc. For
reference, we compare them to the current (LO) default Jeppsson 5
tune, which was used for comparisons to LO \Vc in this paper.

\bibliography{antenna-nlo}

\end{document}